\documentclass{aa}  

\newcommand{\micro}{$\mu$m}

\newcommand{\MH}{H$_{2}$}
\newcommand{\B}{$\beta$}

\newcommand{\msun}{M$_{\odot}$}

\newcommand{\artemis}{ArT\'{e}MiS}
\newcommand{\CM}{cm$^{-2}$}
\newcommand{\R}{$R_{\rm flat}$}

\newcommand{\herschel}{$N_{\rm \mathcal{H}}$}
\newcommand{\hcrop}{$N_{\rm \mathcal{H}, crop}$}
\newcommand{\feather}{$N_{\rm \mathcal{F}}$}
\newcommand{\herschelI}{$I_{\rm \mathcal{H}}$(350)}
\newcommand{\featherI}{$I_{\rm \mathcal{F}}$(350)}

\usepackage{color}

\usepackage{graphicx}
\usepackage{rotating} 
\usepackage{longtable} 

\usepackage{listings}
\usepackage[varg]{txfonts}
\usepackage{natbib,twoopt}
\usepackage[]{hyperref}
\bibpunct{(}{)}{;}{a}{}{,}             
\makeatletter
  \newcommandtwoopt{\citeads}[3][][]{\href{http://adsabs.harvard.edu/abs/#3}%
    {\def\hyper@linkstart##1##2{}%
     \let\hyper@linkend\@empty\citealp[#1][#2]{#3}}}
  \newcommandtwoopt{\citepads}[3][][]{\href{http://adsabs.harvard.edu/abs/#3}%
    {\def\hyper@linkstart##1##2{}%
     \let\hyper@linkend\@empty\citep[#1][#2]{#3}}}
  \newcommandtwoopt{\citetads}[3][][]{\href{http://adsabs.harvard.edu/abs/#3}%
    {\def\hyper@linkstart##1##2{}%
     \let\hyper@linkend\@empty\citet[#1][#2]{#3}}}
  \newcommandtwoopt{\citeyearads}[3][][]%
    {\href{http://adsabs.harvard.edu/abs/#3}
    {\def\hyper@linkstart##1##2{}%
     \let\hyper@linkend\@empty\citeyear[#1][#2]{#3}}}
\makeatother

\graphicspath{{./}{/home/local/emannfor/PAPERS/Filaments/article/pictures/}}

\usepackage{graphicx}
\begin{document}

   \title{Comparison of \textit{Herschel} and \artemis\, observations of massive filaments}


   \author{E. Mannfors
          \inst{1}
          \and
          M. Juvela\inst{1}
          \and
          T. Liu\inst{2}	
          \and
          V.-M. Pelkonen\inst{3}
          }

   \institute{Department of Physics, P.O. box 64, FI- 00014, University of Helsinki, Finland\\
              \email{emma.mannfors@helsinki.fi}
              \and
              {Shanghai Astronomical Observatory, Chinese Academy of Sciences, 80 Nandan Road, Shanghai 200030, People’s Republic of China}
              \and
              {Institut de Ci\`{e}ncies del Cosmos, Universitat de Barcelona, IEEC-UB, Mart\'{i} i Franqu\'{e}s 1, E-08028 Barcelona, Spain}
             }
   \date{Received Month Day, 1996; accepted Mo dd, 1997}

 
  \abstract
   {Filaments are a fundamental part of the interstellar medium (ISM). Their morphology and fragmentation can give crucial information on the nature of the ISM and star formation. OMC-3 in the Orion A Cloud is a nearby, high-mass star-forming region, and therefore ideal to study massive filaments in detail.  }
   { We analyze how the inclusion of higher-resolution data changes the estimates of the filament properties, including their widths and fragmentation properties. We also test the robustness of filament fitting routines.  }
   {  \artemis\, and \textit{Herschel} data are combined to create high-resolution images. Column densities and temperatures are estimated with modified blackbody fitting. The nearby OMC-3 cloud (\textit{d} = 400\,pc) is compared to the more distant G202 and G17 clouds (\textit{d} = 760 and 1850\,pc, respectively). We further compare the OMC-3 cloud as it appears at \textit{Herschel} and \artemis\, resolution.   }
   { Column densities of dense clumps in OMC-3 are higher in combined \artemis\,and \textit{Herschel} data (FWHM$\sim$8.5\arcsec), when compared to \textit{Herschel}-only data (FWHM$\sim$20\arcsec). Estimated filament widths are smaller in the combined maps, and also show signs of further fragmentation when observed with the \artemis\, resolution. In the analysis of \textit{Herschel} data  the estimated filament widths are correlated with the distance of the field.     }
   { Median filament \textit{FWHM} in OMC-3 at higher resolution is 0.05\,pc, but 0.1\,pc with the \textit{Herschel} resolution, 0.3\,pc in G202 and 1.0\,pc in G17, also at the \textit{Herschel} resolution. It is unclear what causes the steep relation between distance and filament \textit{FWHM}, but likely reasons include the effect of the limited telescope resolution combined with existing hierarchical structure, and convolution of large-scale background structures within the ISM. Estimates of the asymptotic power-law index of the filament profile function \textit{p} is high. When fit with the Plummer function, the individual parameters of the profile function are degenerate,  while the FWHM is better constrained. OMC-3 shows negative kurtosis, and all but OMC-3 at the \textit{Herschel} resolution some asymmetry.   }


   \maketitle
%

\section{Introduction}

Filaments in molecular clouds (MCs) are a crucial component of star formation (SF) and are ubiquitous within the dense interstellar medium (ISM) \citep{Elmegreen1979,Schneider1979,Bally1987,Myers2009, Molinari2010, Andre2010, Menshchikov2010}. Rarely are star-forming filaments uniform structures, but are rather composed of denser clumps and cores within less dense gas  \citep{Wang2016}. Therefore, the properties of filaments must be characterized in order to understand their formation mechanisms and physical conditions, fragmentation, and the formation of stars.

The precise formation mechanism of filaments is still uncertain. Turbulence, magnetic energy, and gravity all play a role, but their relative importance varies \citep{Mattern2018,Liu2018b,Tang2019,Soam2019,Traficante2020}. Low-mass filaments, such as in the Polaris cloud, are dominated by large-scale turbulence but have a few gravitationally-bound regions \citep{Hartmann2007,Padoan2011,Arzoumanian2013,Andre2014,Smith2016}. Not surprisingly, gravity has a larger impact in denser clouds \citep{Kirk2015}, and filaments there are often aligned with the longer extent of their host clouds \citep{Andre2014}. Simulations have shown that the inclusion of the effects of magnetic fields results in more filamentary structures \citep{Hennebelle2013}. Contributions also come from shocks, from supernovae and feedback from massive stars \citep{Peretto2012} or cooling in the post-shock regions of large-scale colliding flows \citep{Padoan2007,Heitsch2008, Vazquez2011}.

The same processes which create filaments can also cause further fragmentation into clumps and cores, which with the right conditions can form stars. The mass and density of a filament affect the probability of SF, and the mass of the forming stars.  Nearby clouds ($d\leq 500$\,pc) include the extensively-studied Gould Belt clouds \citep{Andre2010}, and host low- to intermediate-mass SF. Within low-mass star-forming clouds observed with \textit{Herschel}, SF has been observed only within the densest filaments \citep{Andre2010,Andre2014,Konyves2015}. 
An extinction threshold for SF of $A_{\rm v} \sim$ 7$^{\rm mag}$ has been observed, possibly resulting from the shielding of cold gas from interstellar UV radiation \citep{Evans2009,Clark2014,Konyves2015}. 
The more massive, gravitationally supercritical, structures are preferentially located in the Galaxy's spiral arms \citep{Wang2016}. The highest-mass stars and clusters are formed in these dense structures \citep{Carey1998,Peretto2013,Andre2014,Dewangan2020}.

While there is much variation in other filamentary properties, a typical width of 0.1\,pc was observed in nearby, low-mass filaments within the Gould Belt using \textit{Herschel} \citep{Andre2014}, and later also in more distant clouds \citep{Arzoumanian2011, Juvela2012, Malinen2012, Palmeirim2013, Benedettini2015, Kirk2015, Kainulainen2016, GCCVII} and using other instruments \citep{Andre2016}. This has been interpreted to be caused by the change between supersonic and subsonic turbulent gas motions \citep{Padoan2001,Federrath2016} or due to the dissipation mechanism of magneto-hydrodynamic (MHD) waves \citep{HennebelleAndre2013,Hennebelle2013}. However, recent studies have called this typical width into question \citep{Smith2014, Panopoulou2017,Panopoulou2022}. Filament width has been found to be correlated with distance to the filament \citep{GCCVII,Panopoulou2022}.

The hierarchical nature of the ISM seems to extend even to the internal structure of (SF) filaments. Recent discoveries suggest that the internal structure of filaments may be quite complex. According to simulations, single filaments in column density maps may instead be a network of subfilaments in three-dimensional space \citep{Moeckel2015,Smith2016}. ALMA\footnote{Atacama Large Millimeter Array} molecular-line observations of dense gas tracers in the integral-shaped filament in Orion have detected over 50 velocity-coherent fibers within the wider filament \citep{Hacar2018}. \citet{Shimajiri2019} have also detected five fibers of length around 0.5\,pc in the massive filament NGC 6334.

While the \textit{Herschel} observatory enabled great strides in the study of filaments, ground-based instruments such as \artemis\footnote{\url{http://www.apex-telescope.org/instruments/pi/artemis/}\\ARchitectures de bolomètres pour des TElescopes à grand champ de vue dans le domaine sub-MIllimétrique au Sol} on the Atacama Pathfinder Experiment (APEX)\footnote{Based on observations with the APEX telescope under program ID 0101.F-9305(A). APEX is a collaboration between the Max-Planck-Institut fuer Radioastronomie, the European Southern Observatory, and the Onsala Observatory} telescope \citep{APEX} are able to observe their densest substructures. Furthermore, as most high-mass SF regions are located at distances $\geq$ 1\,kpc, higher resolution is necessary to observe them in as much detail as nearby, low-mass SF regions \citep{Hacar2018}. Recently ground-based observations with \artemis, as well as interferometers such as ALMA, have been used to study filament fragmentation in greater detail \citep[e.g.][]{Andre2016,Schuller2021}. However, ground-based bolometers usually miss extended emission due to atmospheric filtering, and interferometers such as ALMA do not detect large-scale structures.

In this paper, we study fragmentation and morphology of three dense Galactic filamentary clouds using \textit{Herschel} and \artemis\, observations. A more full picture of the entire region, from dense sub-filaments to the extended environment, and the higher angular resolution of the ground-based data, will help to answer questions of the structure and fragmentation of high-mass filaments, and the claimed correlation between filament parameters and the resolution of observations.

The paper is structured as follows: 
We discuss our observations, sources, and methods in Sect. \ref{sec:methodology}. Results are presented in Sect. \ref{sec:results} and discussed further in Sect. \ref{sec:discussion}. We test the sensitivity of Plummer fitting with various error sources in Appendix \ref{sec:app_simulations_Plummer}.

\section{Methodology \label{sec:methodology}  }

\subsection{Observations}

Our primary data are observations of three fields using the \textit{Herschel} space telescope and the APEX telescope \citep{HerschelObservatory}. \textit{Herschel} observations were made with two instruments: SPIRE \citep[Spectral and Photometric Imaging Receiver;][]{SPIRE} and PACS \citep[Photodetector Array Camera and Spectrometer;][]{PACS}, and accessed through the Herschel Science Archive\footnote{\url{http://archives.esac.esa.int/hsa/whsa/}}. The SPIRE data cover the wavelengths of 250, 350, and 500\,\micro\, \citep[resolution 18.2\arcsec, 24.9\arcsec, and 36.3\arcsec, respectively;][]{SPIREObsManual} and PACS 160\,\micro\, \citep[resolution 13.6\arcsec;]{PACSObsManual}. The SPIRE data have been cross-correlated with Planck intensities to achieve an absolute zero-point. PACS 70- and 100\,\micro\, data were not used. 
APEX observations (PI: M. Juvela, Program ID: 0101.F-9305(A), Project ID O-0101.F-9305A-2018)  were made with the the \artemis\, instrument (\citet{ARTEMIS}; 350 and 450\,\micro, resolution 8.5\arcsec\, and 9.4\arcsec, respectively, though 450\,\micro\, data were not used). The pixel size of the \artemis\, map is $\sim$0.8\arcsec.

\begin{table*} 
	\centering
	\caption{Observation IDs}
	\label{tbl:observations}	
	\begin{tabular}{l|llll p{0.18\textwidth}  ll}
		\hline\hline
		Field & (ra,dec)  & distance& & Full name  & Observation & Proposal & Map size \\
		 &  (\degr)  & (pc)& &  & ID\tablefootmark{a}    & ID  & (\degr) \\
		
		\hline
		G17 &  (275.16, -14.07) &  1850 & \textit{Herschel} & G017.69-00.15 & 1342218995   & KPOT smolinar  & $0.40 \times 0.40$\\
		 &  & & \artemis &  G16.96+0.27 & 21203  & 0101.F-9305(A)  & $0.18 \times 0.18$ \\

		G202  & (100.24, 10.59) & 760  &  \textit{Herschel} & G202.16+02.64 & 1342228342  & KPOT mjuvela   & $0.40 \times 0.40$\\
		 &  &  & \artemis &  G202.32+2.51 & 21418, 21428, 21433, 21710, 21716, 21717, 21731 &  0101.F-9305(A) & $0.19 \times 0.20$\\
		
		G208  &(83.84, -5.03)&  400 &  \textit{Herschel} & G208.63-20.36 & 1342218968  & KPGT pandre  & $0.40 \times 0.40$\\

		 &  & &  \artemis & G208.68-19.2 & 21417, 21427, 21432, 21709, 21714, 21730, 33165, 33166, 42699, 42702, 43737, 49523, 49524 &  0101.F-9305(A) & $0.22 \times 0.22$\\

		\hline
	\end{tabular}
	\tablefoot{ Central coordinates, distances, and observation IDs of our fields. \textit{Herschel} field G017 corresponds with \artemis\, field G16. The \textit{Herschel} data have been cropped from the original archive sizes to better correspond with \artemis\, data sizes. \\
\tablefoottext{a}{\textit{Herschel} observation ID and \artemis\, scan number(s) used. } }
\end{table*}

Four regions were observed with \artemis: G017.69-00.15/G016.97+00.28, G017.38+02.26, G202.16+02.64, and G208.63-20.36, of which three have \textit{Herschel} data: G017.69-00.15/G016.97+00.28, G202.16+02.64, and G208.63-20.36 (hereafter G17, G202, and G208). We list central coordinates, distances, observation IDs, and mapsizes of our fields in Table \ref{tbl:observations}. The G208 \artemis\, data were observed in May--August 2018 in on-the-fly scanning mode. Airmass was between 1.05 -- 1.95 and precipitable water vapor (pwv) between 0.3 -- 0.7\,mm, corresponding to an atmospheric opacity between 0.5--1.2\footnote{\url{https://www.apex-telescope.org/sites/chajnantor/atmosphere/transpwv/}}  at the elevation of the observations. Thirteen scans were taken, for a total integration time of 5.1\,h. Primary calibration frames were taken of Mars and Uranus, as well as of the secondary calibrator V883-ORI. Data reduction was performed using the IDL \artemis\, pipeline.\footnote{\url{https://www.apex-telescope.org/instruments/pi/artemis/data_reduction/}} The size of the G208 map is $\sim$0.2$\times$0.2\degr, with a RMS noise of 182\,MJy\,sr$^{-1}$ and maximum SNR of 69. We estimate calibration uncertainty of 30\% as in \citet{Andre2016,Schuller2021}.  

\begin{figure*}[h]
	\sidecaption
	\includegraphics[width=12cm]{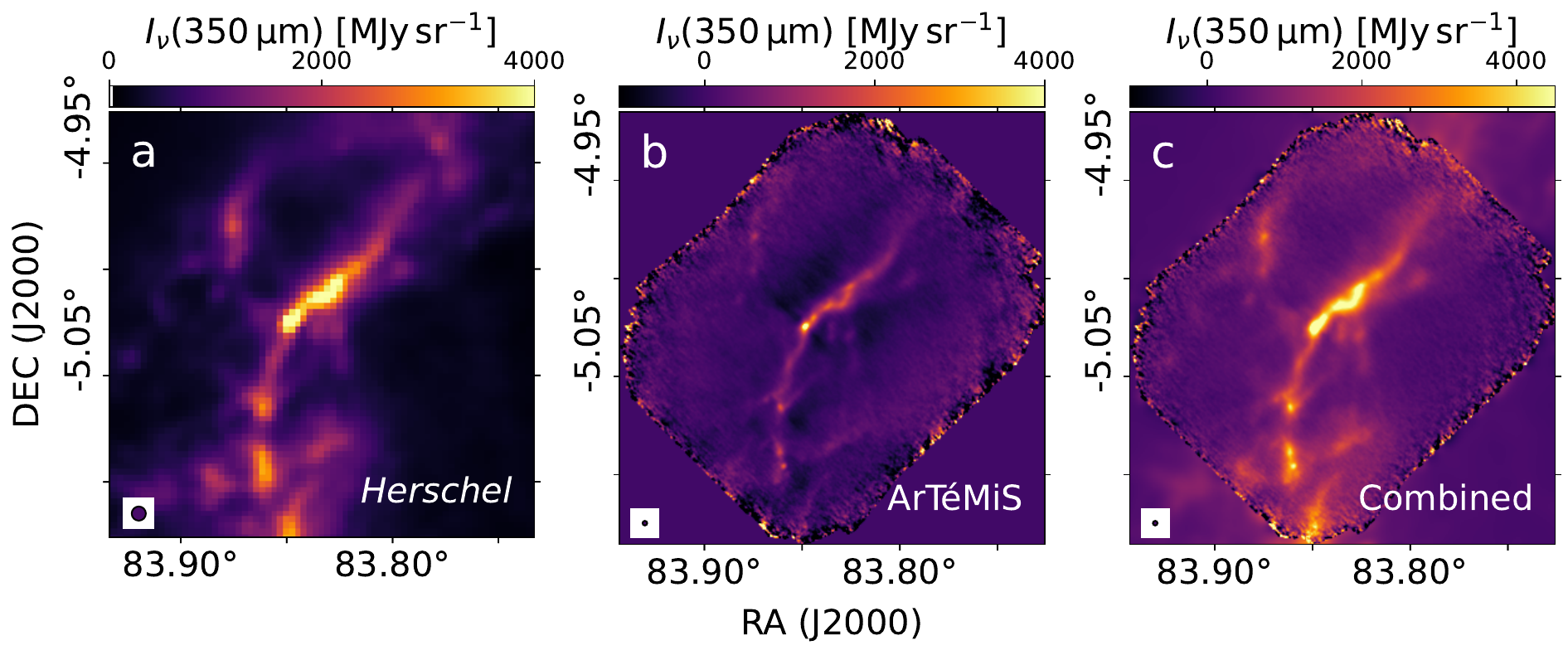}
	\caption{   G208 with \textit{Herschel} and \artemis.  (a) \textit{Herschel} 350\,\micro\, surface brightness map (\herschelI, resolution $\approx$ 20\arcsec) (b) \artemis\, 350\,\micro\, surface brightness map (c) The combined map obtained by feathering (\featherI). The beam is shown in the lower left corner of each image. The resolution of images (b) and (c) is around 10\arcsec \label{fig:feather} }
\end{figure*}

The \artemis\, observations of G208 are shown in Fig. \ref{fig:feather}b, and of the three other fields in Fig. \ref{fig:art_fields}. RMS noise for all four fields is listed in Table \ref{tbl:art_obs_pars}. 
\textit{Herschel} images are shown in Fig. \ref{fig:filament_aligned}, top row. In this paper we focus on the field G208 due to its higher signal-to-noise ratio (SNR), but also compare the G208 results to \textit{Herschel} data of G17 and G202.

G208 covers the Orion Molecular Cloud 3 (OMC-3), within the Orion A cloud in the northernmost part of the Orion MC complex.  We adopt a distance of $d\sim$ 400\,pc, within the estimates given in \citet{Grossschedl2018}. Though not within the most active region of the Orion MC \citep[e.g.][]{Suri2019}, OMC-3 is a region of active embedded SF \citep{Takahashi2013}. \cite{Chini1997} have found several outflows corresponding to protostars, including in MMS 6 near the center of our \artemis\, map. \cite{Megeath2012} have also found a number of dusty YSOs and protostars in the OMC-3 region.

Based on the Planck cold clumps catalog (resolution$\sim$5\arcmin) G17 has a column density of \textit{N}(\MH) $\approx 8.1 \times 10^{21}$\,cm$^{-2}$ and dust temperature $T \approx 13.3 \pm 5.9$\,K \citep{PlanckXXVIII}. The G202 region is part of the Monoceros OB1 cloud and has been studied extensively in \citet{Montillaud2019a,Montillaud2019b}. It has a column density of 1--3$ \times 10^{21}$\,cm$^{-2}$ within the Planck catalog  \citep{PlanckXXVIII}. G202 is made up of two possibly colliding filaments that feed SF throughout the cloud. According to the Planck satellite, G208 has a column density of $\approx 1.7 \times 10^{21}$\,cm$^{-2}$ \citep{PlanckXXVIII}. However, \textit{Herschel} observations show already much higher column density than that detected by Planck.

In the case of field G208, we used feathering to combine the lower-resolution \textit{Herschel} image with the higher-resolution \artemis\, data. Feathering is often used in interferometry, when lower-resolution single-dish data are used to complement interferometric observations that have higher angular resolution but lack information of low spatial frequencies. Feathering was performed with the uvcombine\footnote{\url{https://github.com/radio-astro-tools/uvcombine}} routine using the 350\,\micro\, images, and the results are shown in Fig. \ref{fig:feather}c. We find that uvcombine does not lose signal at intermediate spatial scales (Appendix \ref{sec:app_uvcombineTest}).

\subsection{SED and MBB fitting \label{sec:method_MBB}    }

To estimate dust (color) temperatures, \textit{Herschel} 160--500\,\micro\, observations were fit with a modified blackbody spectrum:
\begin{equation}
\label{eq:MBB}
I_{\rm \nu} = I_{\rm \nu 0} \dfrac{ B_{\rm \nu }(T)}{ B_{\rm \nu 0}(T) } \Big( \dfrac{\nu}{\nu_{\rm 0}}\Big)^{\rm \beta}, 
\end{equation}
where \textit{B}$_{\rm \nu}$ is the blackbody function and $I_{\rm\nu}$ the intensity at frequency $\nu$, $\nu_{\rm 0}$ is a reference frequency, and $\beta$ is the assumed value of the dust opacity spectral index. Equation \ref{eq:MBB} assumes that the emission is optically thin, as is the case in our fields with the possible exception of some dense cores at scales below the resolution of our data. 
We convolve the data using the convolution kernels provided in \citet{Aniano2011}.

A least-squares fit was performed on each pixel separately, and from this fit we derive the dust temperature. Column density \textit{N}(\MH) is calculated from the obtained temperature with
\begin{equation}
\label{eq:NH2}
N(H_{\rm 2}) =  \frac{ I_{\rm \nu} }{B_{\rm \nu}(T) \kappa_{\rm \nu}\mu m_{\rm H}},
\end{equation}
where $I_{\rm\nu}$ is the fitted intensity at frequency $\nu$, $\mu = 2.8$ is the mean molecular weight per free particle, $m_{\rm H}$ is the mass of a Hydrogen molecule, and dust opacity $\kappa$ is calculated as $\kappa_{\rm \nu} = 0.1 \Big(\frac{\nu}{ 1000 GHz}\Big)^{\beta}$cm$^{2}$\,g$^{-1}$ \citep{Beckwith1990}, where we use $\beta = 1.8$, consistent with values found in many dense clouds \citep{GCCVI} and which has been observed to be accurate in OMC-3 \citep{Sadavoy2016}.

\subsection{High-resolution \textit{N}(\MH) maps \label{sec:method_Palmeirim}    }

To perform blackbody fits, one must normally convolve and reproject the data to the lowest resolution (in our case, \textit{Herschel} PLW at $\sim$35\arcsec). We instead calculate column density maps at 20\arcsec\, resolution following the method described in \cite{Palmeirim2013}. In this method, the column density at the resolution of the 250\,\micro\, observations is calculated using the \textit{Herschel} bands from 160\,$\mu$m up to the wavelength indicated by the sub-indices of \textit{N}(\MH). In the following, an arrow refers to convolution to a certain FWHM, e.g. \textit{N}(\MH)$_{\rm 350\rightarrow 500}$ is the convolution of a column density map derived using 160--350\,\micro, convolved to the resolution of 500\,\micro. 
\begin{equation}
\begin{split}
N(\textrm{H}_{\rm 2})_{\rm P} = N(\textrm{H}_{\rm 2})_{\rm 500} + \Big(  N(\textrm{H}_{\rm 2})_{\rm 350} - N(\textrm{H}_{\rm 2})_{\rm 350\rightarrow 500} \Big) \\
+ \Big( N(\textrm{H}_{\rm 2})_{\rm 250} - N(\textrm{H}_{\rm 2})_{\rm 250\rightarrow350}\Big).
\end{split}
\end{equation}
For further details we refer the reader to Appendix A of \cite{Palmeirim2013}. Using the 350\,\micro\, feathered intensity map (Fig.~\ref{fig:feather}c) and temperatures estimated from \textit{Herschel} data, we calculate column density for the feathered map using Eq.~(\ref{eq:NH2}).

In this paper, we refer to the column density \textit{Herschel}-only G208 map with the symbol \herschel\, and to the feathered G208 map with the symbol \feather. The 350\,\micro\, intensity maps are marked with \herschelI\, and \featherI, respectively. The final resolution of the feathered map is 10\arcsec, and the resolution of the \textit{Herschel} column density maps are 20\arcsec. 

\subsection{Dense clump detection \label{sec:method_denseClumps}  }

According to Jeans' criterion, a self-gravitating clump is believed to fragment into cores of mass %
$$ M_{\rm Jeans} = \dfrac{\pi^{5/2}c_{\rm eff}^{3}}{6\cdot \sqrt{G^3 \rho_{\rm eff}}},$$
and size 
$$ \lambda_{\rm Jeans} = \sigma_{\rm tot} \Big(  \dfrac{\pi}{G\rho_{\rm eff}} \Big)^{1/2},$$
where \textit{G} is the gravitational constant, $\rho_{\rm eff}$ is the volume density, and the effective sound speed $c_{\rm eff}$ is replaced by the total velocity dispersion $\sigma_{\rm tot} = \sqrt{\sigma_{\rm TH}^2 + \sigma_{\rm NT}^2}$ \citep{Palau2014,Wang2014}. The nonthermal velocity dispersion $\sigma_{\rm NT}$ is found from molecular line observations,while the thermal velocity is described by  $\sigma_{\rm TH} = \sqrt{\dfrac{k_{\rm B}T_{\rm kin}}{\mu m_{\rm H}}}$. Here $k_{\rm B}$ is the Boltzmann constant and $T_{\rm kin}$ the kinetic temperature. We assume dust temperature and kinetic temperature are similar, as is the case in regions with density of $n_H \geq 10^{5}$ \citep{Goldsmith2001}. For a dust temperature of 20\,K, similar to G208, the thermal component is $\sigma_{\rm TH} \approx 0.27$\,km\,s$^{-1}$. Temperature for each substructure is calculated as the median of all pixels within the substructure. Density $\rho_{\rm eff}$ is estimated assuming the clump is a sphere with radius $R_{\rm eff} = \sqrt{r_{\rm max}\times r_{\rm min}}$, where $r_{\rm max}$ and $r_{\rm min}$ are the clump major and minor axes.

\citet{Suri2019} have analyzed Orion A in \element[][]{C}\element[][18]{O} in the CARMA-NRO Orion Survey (resolution 8\arcsec). From Fig. 1 of their paper we estimate velocity dispersion in OMC-3 to be $\sim$0.6--0.8\,km\,s$^{-1}$, rising to $\sim$1.2--1.4\,km\,s$^{-1}$ in the southern bright clump. We assume $\sigma_{\rm NT} \sim$ 0.8\,km\,s$^{-1}$ in the center of the filament.

\subsection{Filament analysis}
\begin{figure*}[h]
	\includegraphics[width=\linewidth]{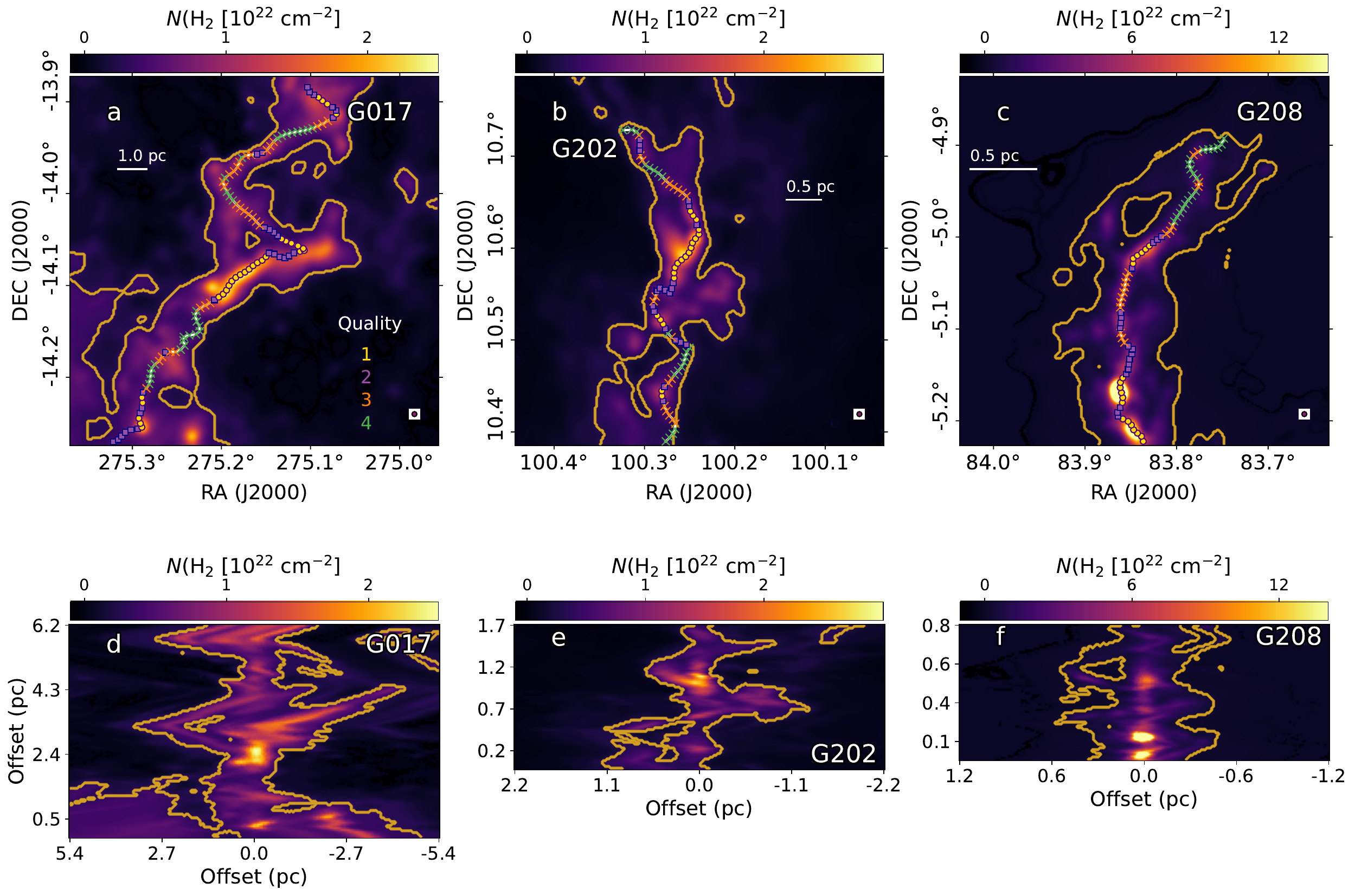}
	\caption{ Extracted filaments on the \textit{Herschel} column density maps.  (Top row): Column density images ($FWHM = 20$\arcsec) of fields G17, G202, and G208 showing the extracted filaments in each field. Filament quality is marked by color of symbols: \textit{q} = 1 (yellow circles), \textit{q} = 2 (purple circles), \textit{q} = 3 (orange crosses), and \textit{q} = 4 (green crosses). The yellow contour shows the limit of $A_{\rm V}= 3$\,mag. (Bottom row): Extracted filament profiles, in which the filament center is at the center of the image  \label{fig:filament_aligned} }
\end{figure*}

The main filaments in each field were traced in the column density maps by eye. The filament profiles (perpendicular to the local filament direction) were extracted at even steps corresponding to one beam, 20\arcsec\, in the case of \textit{Herschel} column densities and 8.5\arcsec\, in the case of the \feather\, map. Each field has between 50 and 150 extracted profiles, depending on filament length (Table \ref{tbl:properties}). Filament paths are shown in Fig. \ref{fig:filament_aligned}. Subfilaments in G17 and G202 complicate the filament selection. We relate extinction to column density by \textit{N}(\MH)$/A_{\rm v}  = 10^{21}$\,cm$^{-2}$\,mag$^{-1}$ \citep{Sadavoy2012}, and we use a threshold extinction $A_{\rm v} = $ 3$^{\rm mag}$ to eliminate background emission when calculating filament line masses. This $A_{\rm v} \geq $ 3 mask is twice the mean extinction of the background in G208 \herschel\, and is larger than the standard deviation in all fields. The $A_{\rm v} < $ 3 component is not omitted during radial profile fitting.

The radial profiles of filament column density are generally well-represented by a Plummer-like function \citep{Arzoumanian2011,Andre2014}. We modify this function to include a linear background and convolve the model profile according to the angular resolution of the observed data, 
\begin{equation}
\label{eq:PlummerBG}
Y   =   \textrm{Conv} \Bigg[ N_{\rm 0} \cdot \Bigg( 1.0+\Bigg(\frac{r+\Delta r}{R_{\rm flat}}\Bigg)^{2} \Bigg)^{0.5-0.5\cdot p} \Bigg]   + a + b\cdot r,
\end{equation}
where \textit{r} is the offset from the filament center, $N_{\rm 0}$ is the filament's central column density, \R\, the radius of the flat inner region, and \textit{p} is the asymptotic power-law exponent. The term $\Delta r$ allows the center of the fitted filament to shift to better match the observations when the filament is not perfectly centered on the path that was traced by eye.

We also fit the profiles with asymmetric models, where the two sides of the filament have independent $R_{\rm flat}$ and $p$ parameters, though assuming the same values of \textit{a} and \textit{b} on both sides of the filament. Unless otherwise noted, results refer to the asymmetric fits. Effects of noise, sky background, background component model, distance to the source, fitting area, and convolution are explored in Appendix \ref{sec:app_simulations_Plummer}.

The Plummer \R\, and \textit{p} parameters give the FWHM of the fitted profile: 
\begin{equation}
	\label{eq:FWHM}
	FWHM = 2\cdot |R|\cdot \sqrt{   2^{2/(p-1)}-1      }
\end{equation}
which is more robust compared to the individual values of \R\, and \textit{p} due to their partial degeneracy.

We calculated for each profile a SNR value by dividing the peak value of the filament segment with the noise level obtained from a region with the lowest emission (Table \ref{tbl:centers_of_empty_regions}). Profiles were assigned quality flags from \textit{q} = 1 in the highest SNR quartile to \textit{q} = 4 in the lowest SNR quartile. Individual profiles with poor quality flags ($q>2$) were rejected from further analysis. We use quartiles as opposed to direct SNR values so that we have different qualities within each field, with sufficient number of datapoints to facilitate comparison.

\section{Results  \label{sec:results}  }
In the following section, we first present general filament properties such as line masses and profile asymmetry. We then focus on Plummer profile fitting, and compare the results derived for \herschel\, and \feather\, data. Finally, we analyze fragmentation of the regions by studying clumps and cores, and with wavelet analysis.

\subsection{Filament properties}
Lengths of the filament spines and filament masses are listed in Table \ref{tbl:properties}. 
Total filament mass is calculated using the pixels within the $A_{\rm v} \geq $ 3 mask: 
\begin{equation}
\label{eq:Mass}
M_{\rm fil} = \Sigma_{i,j} \Big( N(\mathrm{H}_{\rm 2})_{i,j} \Big)\cdot (\Delta x)^{2} \cdot m_{\rm H}\cdot\mu_{\rm H_{2}},
\end{equation}
where \textit{N}(\MH)$_{i,j}$ is the column density, and $\Delta x$ is the physical size of the pixel. Line mass, or the mass per length, was then estimated as the average over the filament after removal of the background. Using this criterion, G202 has a low line mass of 103\,\msun\,pc$^{-1}$, whereas other filament sections have line masses above 150\,\msun\,pc$^{-1}$. To accurately compare \feather\, and \herschel\, filament properties, we include \hcrop, the G208 \herschel\, field cropped to the size of the \feather\, map. Mean \textit{N}(\MH) is the same within the \feather\, and \hcrop\, maps, though of the two \feather\, has higher maximum \textit{N}(\MH). This is due to the higher resolution of the \feather\, map, where the same high density is more diluted in the \hcrop\, map.

Global gravitational stability of a filament can be estimated using the critical line mass of an isothermal filament \citep{Ostriker1964}: 
\begin{equation}
\label{eq:MLCrit}
\Bigg( \dfrac{M}{L}\Bigg)_{\rm crit, TH} = \dfrac{2 \sigma_{\rm TH}^2}{G},
\end{equation}
where $\sigma_{\rm TH}$ is the 1-D thermal velocity dispersion, defined in Sect. \ref{sec:method_denseClumps} and \textit{G} is the gravitational constant. Temperature used to calculate $\sigma_{\rm TH}$ is the median over the $A_{\rm v} \geq 3$ region. We first assume the full filament is thermally supported, and thus use only thermal velocity dispersion. At $T\sim 20$\,K, critical line masses are around 30--40\,\msun\,pc$^{-1}$, showing that all the filaments are gravitationally unstable without support from magnetic or turbulent energy.

We again use \element[18][]{C}\element[][]{O} velocity dispersion from \citet{Suri2019} to estimate $\sigma_{\rm NT}$ \citep{FiegePudritz2000b} for G208, taking for the full OMC-3 region a value of $\sigma_{\rm NT}\sim 0.6$\,km\,s$^{-1}$. \citet{Montillaud2019b} studied G202 in several molecular lines, and from the region corresponding to our filament, we estimate  $\sigma_{\rm NT}\sim 0.7$\,km\,s$^{-1}$ from their \element[18][]{C}\element[][]{O} data. For G17 we assume a similar value, $\sigma_{\rm NT}\sim 0.7$\,km\,s$^{-1}$, due to the lack of molecular line data for the region. Critical line masses $M_{\rm l,crit, TH+NT}$ now range from 200 to 270\,\msun\,pc$^{-2}$. G17, the full G208 \herschel\, field, and the \feather\, map have line mass greater than critical line mass, but with the addition of turbulent support the other fields may be gravitationally stable. We however note that the estimate for G17 is only a general estimate.

\begin{table*}
	\centering
	\caption{Filament properties.} 
	\label{tbl:properties}
	\begin{tabular}{llllllllll}
		\hline\hline

			Field  & $\langle N$(\MH)$\rangle$ &  max($N$(\MH)) & \textit{T} & $M$  & $L$ & $M_{\rm line}$   & $M_{\rm l,crit, TH}$   & $M_{\rm l,crit, TH+NT}$     \\ 
 		& ($10^{22}$\,cm$^{-2}$) & ($10^{22}$\,cm$^{-2}$) & (K) & ($M_{\rm \odot}$) & (pc) & ($M_{\rm \odot}$\,pc$^{-1}$)  & ($M_{\rm \odot}$\,pc$^{-1}$)  & ($M_{\rm \odot}$\,pc$^{-1}$) \\

		\hline

		G17 & 0.6 $\pm$ 0.4 & 2.5 $\pm$ 1.2 & 24.1 $\pm$ 3.4 & 6443 $\pm$ 3226 & 20.7  $\pm$ 0.6 & 310 $\pm$ 155 &  39.6 $\pm$ 19.8  &  267 $\pm$ 134 \\ 
		G202 & 0.6 $\pm$ 0.4 & 3.0 $\pm$ 1.5 & 20.0 $\pm$ 1.0 & 555 $\pm$ 279 & 5.4  $\pm$ 0.4 & 103 $\pm$ 52 &  32.9 $\pm$ 16.6  &  261 $\pm$ 132 \\ 
		\herschel & 1.8 $\pm$ 3.7 & 60.5 $\pm$ 30.2 & 23.3 $\pm$ 4.5 & 696 $\pm$ 358 & 2.9  $\pm$ 0.4 & 243 $\pm$ 128 &  38.3 $\pm$ 19.7  &  206 $\pm$ 106 \\ 
		\hcrop & 1.7 $\pm$ 1.6 & 9.9 $\pm$ 4.9 & 21.8 $\pm$ 2.7 & 233 $\pm$ 120 & 1.3  $\pm$ 0.2 & 173 $\pm$ 91 &  35.9 $\pm$ 18.5  &  203  $\pm$ 105 \\ 
		\feather & 1.7 $\pm$ 1.9 & 22.9 $\pm$ 11.5 & 22.6 $\pm$ 3.5 & 280 $\pm$ 144 & 1.3  $\pm$ 0.2 & 208 $\pm$ 110 &  37.3 $\pm$ 19.2  &  205  $\pm$ 106 \\ 

		\hline
	\end{tabular}
	\tablefoot{Mean column density, maximum \textit{N}(\MH), temperature \textit{T}, mass \textit{M}, length \textit{L}, and line mass $M_{\rm line}$ for the four filaments. Column densities are the mean over the $A_{\rm v} \geq $ 3$^{\rm mag}$ region. We assume an uncertainty in temperature of 1\,K. Also assuming an error in $\kappa$ of 50\%, we calculate a 50\% uncertainty in \textit{N}(\MH). Masses are derived using $N_{\rm P}$. Assuming a 50\,pc uncertainty in distance, the uncertainties in \textit{M} and $M_{\rm line}$ are approximately 50\% due to propagation of uncertainty.  $M_{\rm l,crit, TH}$ is the critical line mass assuming only thermal support, whereas $M_{\rm l,crit, TH+NT}$ includes also turbulent support. We assume an uncertainty in $\sigma_{\rm NT}$ of 0.2\,km\,s$^{-1}$. G208$_{\rm crop}$ (\herschel) refers to the \herschel\, map cropped to the dimensions of the \feather\, map. }
\end{table*}

\subsection{Filament skewness and kurtosis}

\begin{figure*}[h]
	\sidecaption
	\includegraphics[width=12cm]{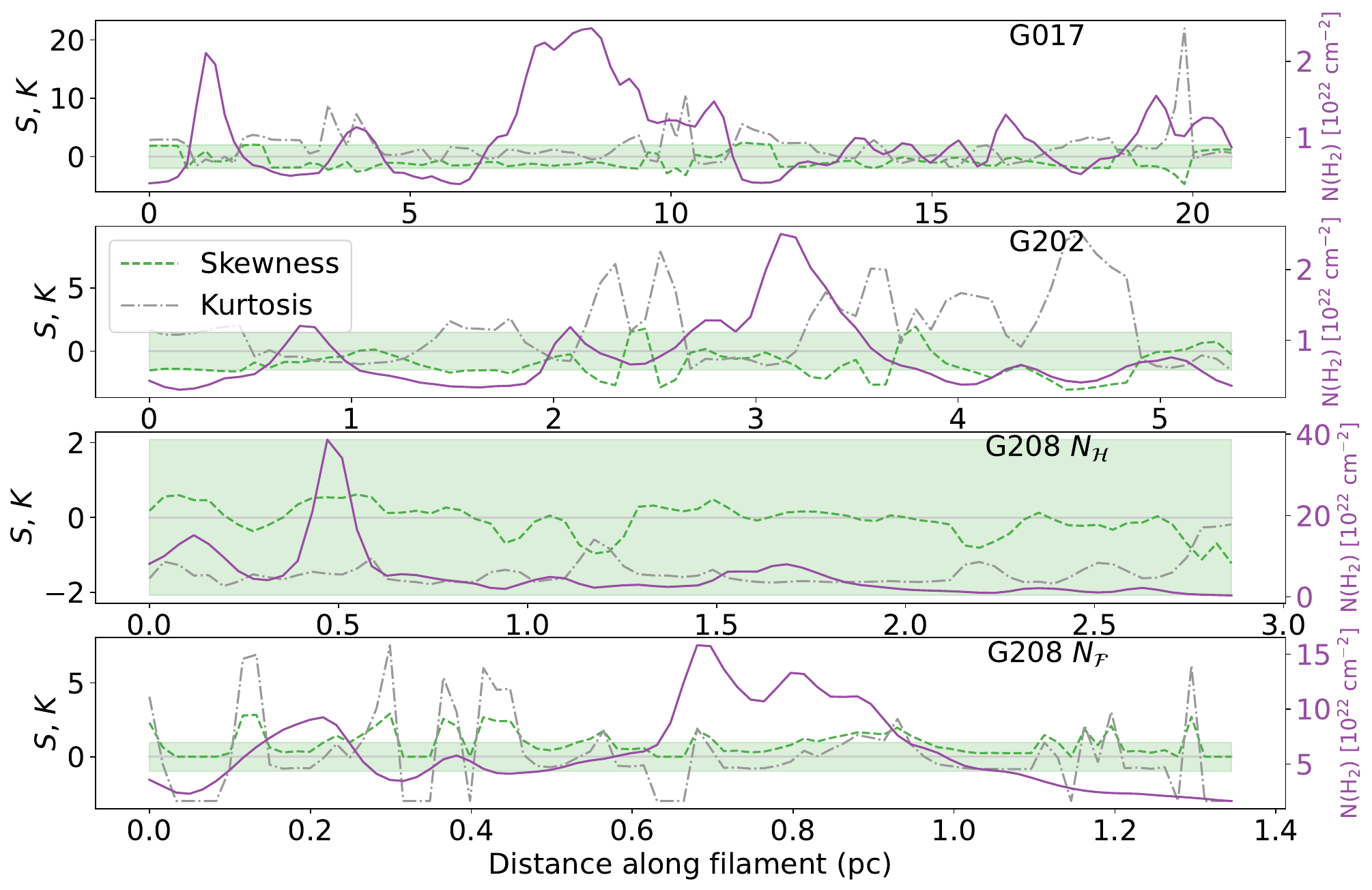}
	\caption{ Skewness (green dashed line) and kurtosis (gray dash-dotted line) along the the filament spine of G17, G202, and G208 \herschel\, and \feather. The right-hand y-axis shows column density along the filament spine (purple solid line). The faint gray horizontal line marks \textit{K} = \textit{S} = 0. \textit{S} which falls outside of the green highlighted region is significant (see Sect. \ref{sec:discussion_Plummer_symmetry}).  \label{fig:skewLine} }
\end{figure*}

We plot Fisher-Pearson coefficient of skewness \textit{S}, Fisher kurtosis \textit{K}, and \textit{N}(\MH) along the filament spine in Fig. \ref{fig:skewLine}. Column density is computed as the mean of the pixels in a 3$\times$3 grid surrounding the spine center, with pixel sizes for \feather\, and \herschel\, being 0.76" and 6", respectively. \textit{S} and \textit{K} are calculated from the Plummer fits from which we have subtracted the linear background. \textit{S} describes the direction of the asymmetry of the filament, where positive \textit{S} indicates a left-leaning profile with a stronger tail to the right (i.e. leaning toward the positive RA axis).Skewness is calculated as\footnote{scipy.stats.skew (bias=True)} 
\begin{align}	
	\label{eq:skewness}
	S &= \dfrac{m_{\rm 3}}{m_{\rm 2}^{3/2}},\\
	m_{\rm i} &= \dfrac{1}{N}\sum^{N}_{n=1}\Big( x[n] - \bar{x} \Big)^{i},
\end{align}
where $m_{\rm i}$ is the biased sample \textit{i}th central moment, and $\bar{x}$ the sample mean. \textit{K} describes the shape of the profile compared to a Gaussian, therefore a higher \textit{K} describes a filament with stronger tails while a Gaussian has \textit{K} = 0. Kurtosis is calculated as\footnote{scipy.stats.kurtosis (fisher=True)} 
\begin{align}
	\label{eq:kurtosis}
	K &= \dfrac{\mu_{\rm 4}}{\sigma^4} - 3.0,
\end{align}
where $\mu_{\rm 4}$ is the fourth central moment and $\sigma$ the standard deviation.  Pearson correlation coefficients between column density, skewness, and kurtosis are shown in Table \ref{tbl:Pearson_coefficients_NH2_S_K}.

\begin{table*}
	\centering
	\caption{Mean and 1-$\sigma$ standard deviation of \textit{S} and \textit{K} in the fields.} 
	\label{tbl:skewNess}
	\begin{tabular}{l|lllll}
		\hline\hline
		& G017 & G202 & G208 \herschel & G208 \hcrop  & G208 \feather \\
		
		\hline
		 $\langle S\rangle$ & -0.8 $\pm$ 1.3 &  -1.1 $\pm$ 1.1 &  -0.1 $\pm$ 0.4 &  -0.0 $\pm$ 0.4   & 0.9 $\pm$ 0.8 \\
		$\langle  K \rangle$ & 1.6 $\pm$ 2.9 &  1.8 $\pm$ 2.8 & -1.5 $\pm$ 0.3 &  -1.5 $\pm$ 0.2  & 0.1 $\pm$ 2.5 \\

		\hline
	\end{tabular}
	
\end{table*}

 For a test sample of 1000 numpy random normal distributions of 1000 values each, mean values are $\langle S\rangle  = -0.002 \pm  0.08, \langle  K \rangle = -0.03 \pm 0.15.$ In comparison, we list mean values from our data in Table \ref{tbl:skewNess}.  The 1-$\sigma$ spread in \textit{S} is consistent with a normal distribution in G208 \herschel\, and \hcrop, but slightly higher in other fields. G17 and G202 show negative \textit{S}, G208 \feather\, positive. G17 and G202 show strong tails (i.e. a high \textit{K}) when compared to a Gaussian, but this difference is less significant in G208 \feather. There is no statistically significant correlation between \textit{N}(\MH) and \textit{S} or \textit{K}.  The high peak in \textit{K} in G208 \feather\, corresponds to the region of highest column density.

\begin{table*}
	\centering
	\caption{Pearson correlation coefficients between column density, skewness \textit{S}, and kurtosis \textit{K}.} 
	\label{tbl:Pearson_coefficients_NH2_S_K}
	\begin{tabular}{l|ll|ll|ll}
		\hline\hline 
		Field & \multicolumn{2}{l|}{ \textit{N}(\MH) \& \textit{S} } & \multicolumn{2}{l|}{ \textit{N}(\MH) \& \textit{K} } & \multicolumn{2}{l}{ \textit{S} \& \textit{K} } \\ 
		& \textit{r} & \textit{p}-value & \textit{r} & \textit{p}-value & \textit{r} & \textit{p}-value\\
		\hline

		G017  &  -0.15  &  0.102  &  -0.17  &  0.071  &  -0.34  &  $\leq 0.005$ \\ 
		G202  &  0.1  &  0.392  &  -0.21  &  0.074  &  -0.71  &  $\leq 0.005$ \\ 
		G208 $N_\mathcal{H}$  &  0.56  &  $\leq 0.005$  &  -0.13  &  0.278  &  -0.6  &  $\leq 0.005$ \\ 
		 G208 \hcrop  &  0.43  &  $\leq 0.005$  &  -0.36  &  $\leq 0.005$  &  -0.59  &  $\leq 0.005$ \\ 
		G208 \feather  &  0.08  &  0.457  &  0.04  &  0.692  &  0.97  &  $\leq 0.005$ \\

		\hline
	\end{tabular}
	\tablefoot{Pearson correlation coefficients \textit{r} and their \textit{p}-values between column density   \textit{N}(\MH), skewness \textit{S} and kurtosis \textit{K} along the filament spine.  }
\end{table*}

\subsection{Plummer fitting}
\begin{figure}[h]
	\sidecaption
	\resizebox{\hsize}{!}{\includegraphics{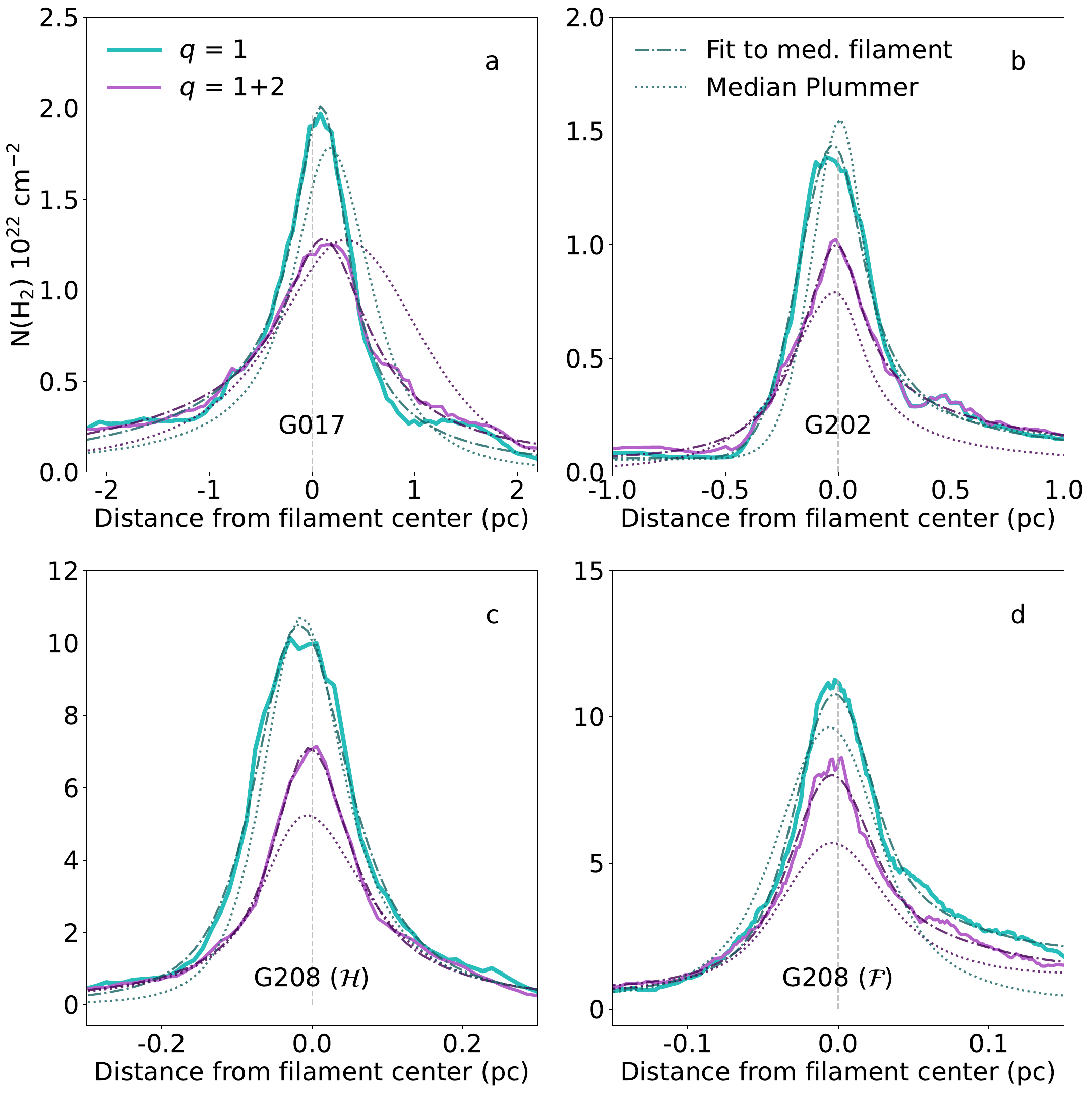}}
	\caption{Filament profiles in G17 (a), G202 (b), G208 \herschel\, (c), and G208 \feather\, (d). The median filament profile is shown with solid lines, for quality 1 filaments (blue) and quality 1 and 2 (purple) profiles. The dash-dotted line shows the Plummer fit to the median filament profile. The dotted line shows the median Plummer profile, by taking the median of each individual Plummer parameter calculated.  \label{fig:mean_filament_profiles} }
\end{figure}

Filament column density profiles were fitted with the Plummer model. We plot the median profiles for each quality bin and the derived Plummer model in Fig. \ref{fig:mean_filament_profiles} for asymmetric fits. Plummer profiles are fitted out to 10\arcmin, which corresponds to $r\sim$ 5\,pc, 2\,pc, and 1\,pc from filament center for fields G17, G202, and G208 \herschel, respectively. Due to the smaller extent of the \artemis\, map, it can only be fitted to a distance of 1.3\arcmin\, from filament center ($\sim$ 0.16\,pc). For comparison with \feather\, data, we also fit the \herschel\, data to the same extent and refer to this also as \hcrop. 
Median derived plummer values are shown in Appendix \ref{sec:app_Plummer_pars} (Table \ref{tbl:PlummerResults}, Fig. \ref{fig:derived_violins}).

We derive mean \textit{FWHM} values of 0.1--1.0\,pc for asymmetric \textit{q} = 1 fits using the three \textit{Herschel} fields, but only  $\sim$0.05\,pc using G208 \feather\, data. Power-law exponent of the Plummer profile \textit{p} is generally below 5. We further compare G208 \hcrop\, and \feather\, fits in the next section.

\begin{figure*}[h]
	\sidecaption
	\includegraphics[width=\linewidth]{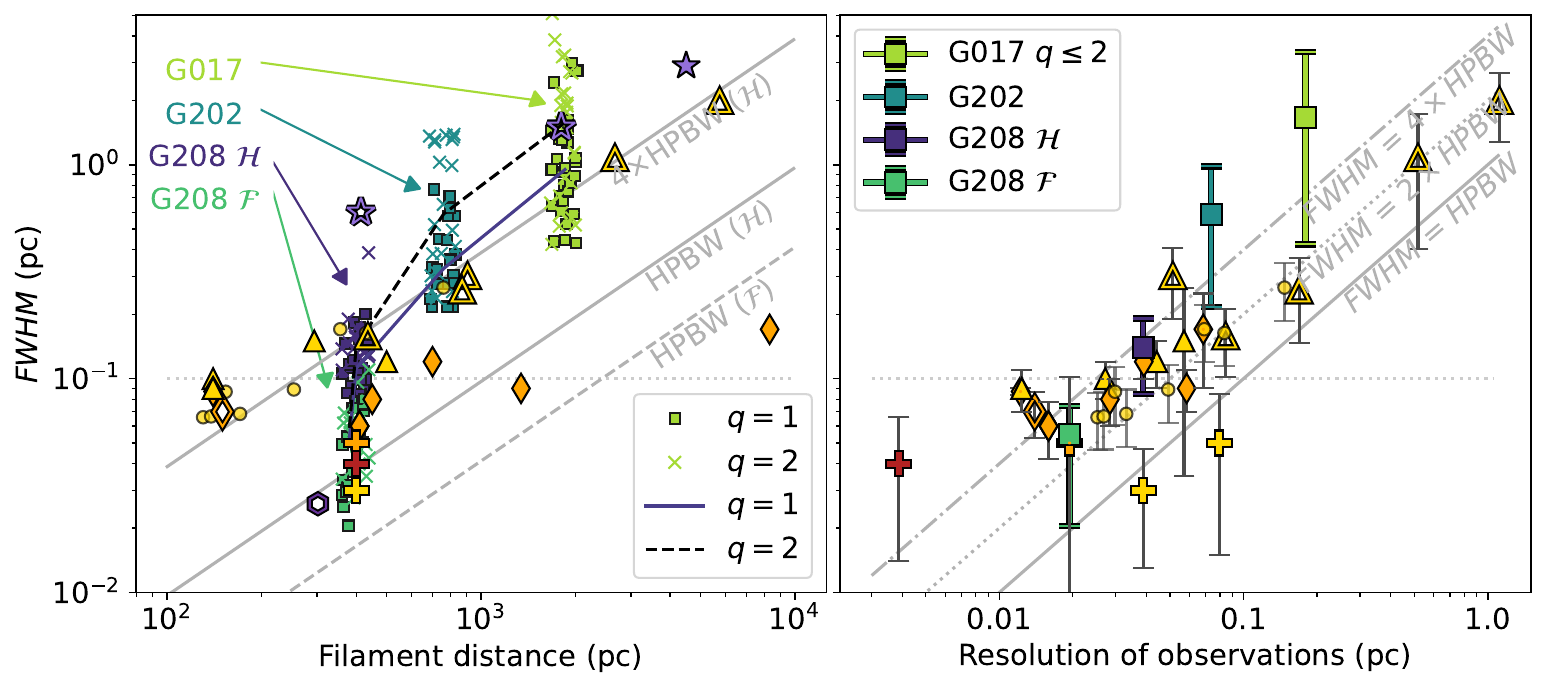}
	\caption{The observed dependence of \textit{FWHM} on filament distance and observational resolution.  (Left:) \textit{FWHM} from asymmetric fits plotted against the distance of the field of each \textit{q} = 1--2 profile (squares and crosses, respectively).  For  readability, the distance values include small added jitter. Median of each distribution is plotted with a solid line (\textit{q} = 1) and dashed line (\textit{q} = 2). Grey lines show the proposed relation of $FWHM = 4\times HPBW$ \citep{Panopoulou2022}, as well as the \textit{Herschel} (solid line) and \artemis\, (dashed line) beamsizes. 	(Right:) The median value of each $q$ = 1--2 dataset as a function of observational resolution in pc (solid squares).  The errorbars represent the standard deviation of the \textit{FWHM} widths in each dataset. Grey lines show relations of $FWHM = [1,2,4]\times HPBW$.  In both frames we also include values from the literature (see text for references): \textit{Herschel} continuum (yellow triangles), other continuum (orange diamonds), mean deconvolved filament widths from Fig. 1 of \citet{Panopoulou2022} (yellow circles), and values from \citet{Juvela2023} with solid plusses (LR and HR in yellow, AR with orange, and MIR with red). The errorbars represent either the published uncertainty or a 30\% uncertainty on the \textit{FWHM}. On the left frame only, we show additional literature values from CO  (purple stars), and NH$_{\rm 3}$ (purple hexagon) observations. Solid symbols represents deconvolved widths, those without beam deconvolution are shown with outlines. 	 \label{fig:distance_vs_FWHM} } 
\end{figure*}

Figure~\ref{fig:distance_vs_FWHM} shows the fitted $FWHM$ values as the function of the field distance (frame a) and observational resolution in parsecs (frame b) for all three \textit{Herschel} fields and the \feather\, map. Including only the \textit{Herschel} filament profiles of quality 1 and 2, we derive a relation between distance in kpc and \textit{FWHM} in parsecs of: 
$$ FWHM_{\langle q\rm = 1+2 \rangle} \approx 0.65 \textrm{ pc}\times (d/\textrm{kpc}), $$
and, including \artemis\, and all \textit{Herschel} data, a relation between instrumental resolution $HPBW$ and \textit{FWHM} of:
$$ FWHM_{\langle q\rm = 1+2 \rangle} \approx 6.7 \times\textrm{ }HPBW\textrm{[pc]},  $$  where \textit{HPBW} and \textit{FWHM} are in pc.

We see a dependence of \textit{FWHM} on distance and spatial resolution in our data, slightly steeper than the $4\times HPBW$ relation proposed by \citet{Panopoulou2022}. Our fit is based on only four maps (including two observations of the G208 field), not sufficient to draw any universal conclusions. A correlation between distance or spatial resolution and derived \textit{FWHM} is also seen in the literature, albeit with a more shallow slope than what we observe. In Fig.~\ref{fig:distance_vs_FWHM} we include \textit{Herschel} continuum observations from \citet{Palmeirim2013,Schisano2014,Zhang2020b,Andre2022,Zhang2022,Li2023}, and continuum observations using other telescopes from  \citet{Hill2012,Salji2015b,Federrath2016b,Kainulainen2016,Howard2019,Zavagno2020,Schuller2021}. Though continuum and molecular lines do not necessarily trace the same filament, we include CO observations from \citet{Zheng2021, Guo2022, Yamada2022}, and NH$_{\rm 3}$ observations  from \citet{Chen2022} in the right-hand frame.  
\textit{FWHM} values estimated with continuum observations (yellow and orange markers in Fig.~\ref{fig:distance_vs_FWHM}) seem to follow the $2\times HPBW$ relation. Tentatively, it appears that in continuum observations \textit{FWHM} converges to a value of $\sim$0.05--0.1\,pc when resolution is better than 0.3\,pc. To test this possible convergence would require a larger sample of data at high resolution.

\subsection{Comparing Plummer FWHM and column density}
\begin{figure}[h]
	\resizebox{\hsize}{!}{\includegraphics{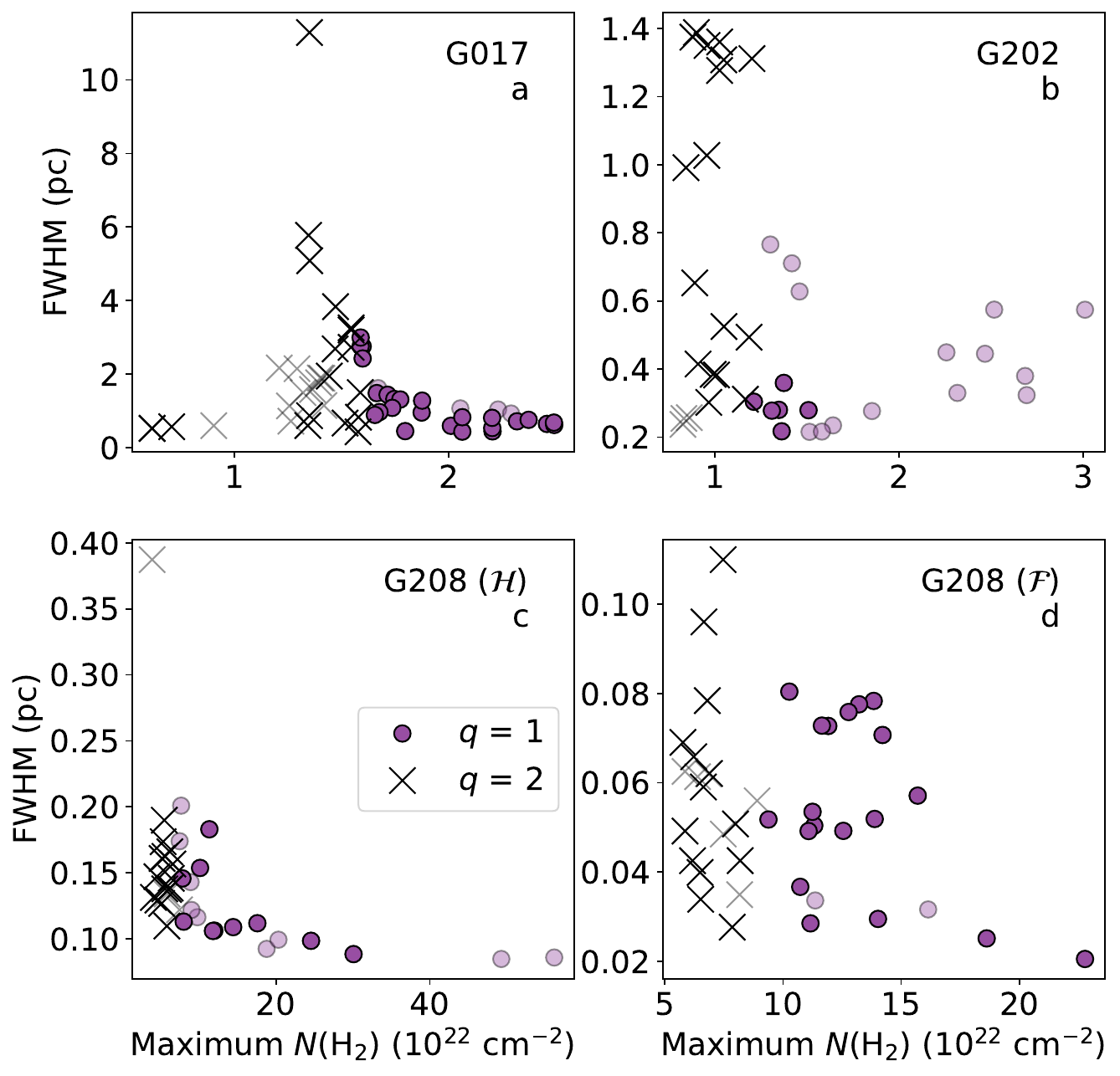}}
	\caption{  Plummer FWHM plotted against the column density at the filament spine for \textit{Herschel} fields G17 (a), G202 (b), and G208 (c), and the \feather\, map (d), for filaments of quality 1 (purple circles) and 2 (black crosses).  Transparent symbols show profiles for which $p > 10$. \label{fig:compare_FWHM_NH2} }
\end{figure}

FWHM values for each \textit{q} = 1--2 filament profile are plotted against \textit{N}(\MH) in Fig. \ref{fig:compare_FWHM_NH2}. Pearson correlation was calculated between these two parameters, and the \textit{r}- and \textit{p}-values are listed in Table \ref{tbl:Pearson_coefficients_FWHM_NH2}. In \textit{q} = 1 profiles, significant correlation is found in G17 and G208 \herschel. In \textit{q} = 2 profiles, correlation is generally not significant. This is possibly caused by low column density profiles, in which fits can return high FWHM values depending on background fluctuations. However, as we study only \textit{q} = 1--2 profiles, column density in these cases is significantly higher than the background. In fits with Plummer \textit{p $\leq 10$}, correlation is less significant, due in part to the lower number of datapoints. By including only those profiles with $p \leq 10$ in G202, we detect a (non-significant) anticorrelation between \textit{FWHM} and \textit{N}(\MH); conversely by including all profiles field G202 shows a positive correlation.

\begin{table} 
	\centering
	\caption{Pearson correlation coefficients between Plummer FWHM and peak column density.} 
	\label{tbl:Pearson_coefficients_FWHM_NH2}
	\begin{tabular}{llllllllll}
		\hline\hline 

		Quality flag & \multicolumn{2}{l}{\textit{q} = 1} & \multicolumn{2}{l}{\textit{q} = 2}\\
		
		\hline
		Field & $r$ & $p$-value & $r$ & $p$-value \\ \hline

		\multicolumn{4}{c}{All profiles} \\
		G017 & -0.682  & 0.000  & 0.233  & 0.224 \\ 
		G202 & 0.240  & 0.323  & 0.160  & 0.527 \\ 
		G208 ($\mathcal{H}$) & -0.610  & 0.006  & -0.447  & 0.063 \\ 
		G208 ($\mathcal{F}$) & -0.411  & 0.064  & -0.187  & 0.429 \\ 

		\multicolumn{4}{c}{(Plummer) $p \leq 10.0$} \\
		G017 & -0.684  & 0.000  & 0.213 & 0.397 \\ 
		G202 & -0.104  & 0.845  & -0.158  & 0.546 \\ 
		G208 ($\mathcal{H}$) & -0.585  & 0.075  & 0.165  & 0.557 \\ 
		G208 ($\mathcal{F}$) & -0.426 & 0.069  & -0.078  & 0.790 \\ 

		\hline
	\end{tabular}
	\tablefoot{Pearson correlation coefficients \textit{r} and their \textit{p}-values for Plummer parameters in fits to quality 1 and 2 filamentary profiles. The lower section includes only fits with mean Plummer $p \leq 10$.  }
\end{table}

\subsection{Comparison between \feather\, and \herschel\, in field G208 \label{sec:results_feather_herschel_comparison} }

\begin{figure}[h]
	\includegraphics[width=\linewidth]{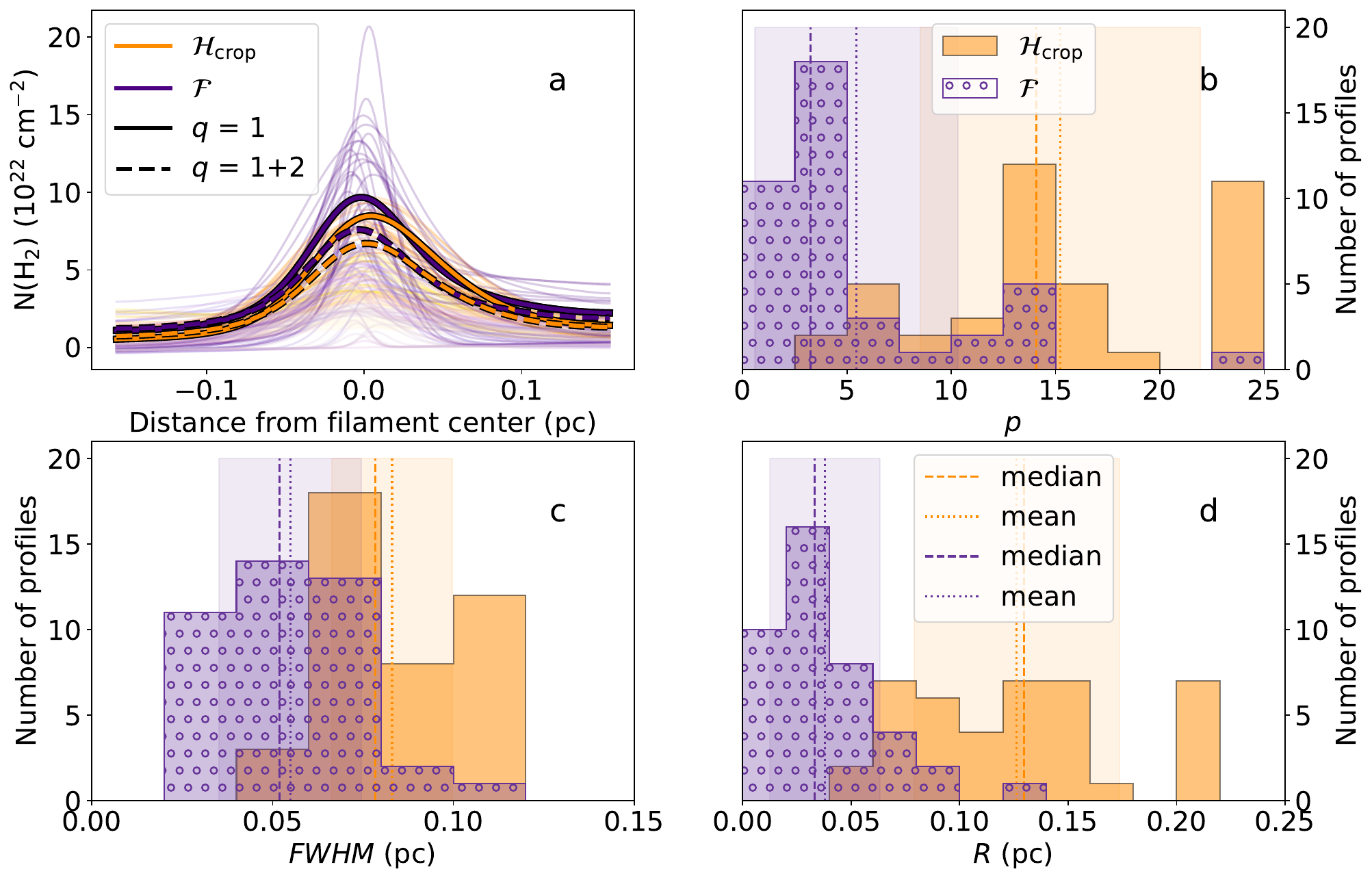}
	\caption{Comparison of G208 \feather\, and \hcrop\, filament shapes. (Frame a): Profiles of \herschel\, (orange) and \feather\, (purple) \textit{q} = 1--2 filaments. Median profiles are plotted with solid and dashed lines for \textit{q} = 1 and \textit{q} = 2 profiles, respectively.  (Frames b-d): histograms of \textit{p} (b), \R\, (d), and \textit{FWHM} (c) for all \textit{q} = 1--2 filaments in field G208. Medians and means for each distributions are plotted in dashed and dotted lines, respectively. The shaded areas correspond to the central 1-$\sigma$ regions of the distributions. \label{fig:compare_PlummerPars} }
\end{figure}

We plot the mean asymmetric Plummer profiles and the distributions of $R_{\rm flat}$, \textit{p}, and \textit{FWHM} in Fig. \ref{fig:compare_PlummerPars}, using the \feather\, and \hcrop\, maps. 
T-tests comparing the two sets are shown in Table \ref{tbl:tTest_featherHerschel}. The differences in \textit{FWHM}, \R\, and \textit{p}  are all significant. Fitting the Plummer profile to the \hcrop\, map, using the \artemis\, pixel size, results in higher values of \textit{p} but also lower \textit{FWHM} than by fitting the full \herschel\, map. However, restricting analysis to those profiles with $p\leq$ 5.0 results in average \textit{FWHM} values of 0.10 $\pm$ 0.01 and 0.05 $\pm$ 0.02 for \hcrop\, and \feather, respectively.

\begin{table}
	\centering
	\caption{T-test results between Plummer parameters computed for G208 \hcrop\, and \feather. }
	\label{tbl:tTest_featherHerschel}
	\begin{tabular}{lllll}
		\hline\hline
		Parameter &  \hcrop & \feather & t-score & p-value \\ 
		\hline
		 $R_{\rm L}$  &  0.11 $\pm$ 0.05  &  0.03 $\pm$ 0.04  &  -7.35  &  $\leq$ 0.005  \\ 
		 $R_{\rm R}$  &  0.16 $\pm$ 0.08  &  0.02 $\pm$ 0.04  &  -7.35  &  $\leq$ 0.005  \\ 
		 $\langle R\rangle$  &  0.13 $\pm$ 0.05  &  0.03 $\pm$ 0.03  &  -10.41  &  $\leq$ 0.005  \\ 
		 $p_{\rm L }$  &  11.37 $\pm$ 9.84  &  2.99 $\pm$ 5.99  &  -5.27  &  $\leq$ 0.005  \\ 
		 $p_{\rm R }$  &  17.45 $\pm$ 9.76  &  2.43 $\pm$ 7.54  &  -5.09  &  $\leq$ 0.005  \\ 
		 $\langle p\rangle$  &  14.09 $\pm$ 6.72  &  3.26 $\pm$ 4.87  &  -7.44  &  $\leq$ 0.005  \\ 
		 $\Delta r$  &  -0.0 $\pm$ 0.01  &  0.01 $\pm$ 0.01  &  1.73  &  0.087  \\ 
		 $FWHM$  &  0.08 $\pm$ 0.02  &  0.05 $\pm$ 0.02  &  -6.89  &  $\leq$ 0.005  \\ 

		\hline
	\end{tabular}
	\tablefoot{  \textit{q} $\geq$ 2 filament profiles are used in this analysis. $\langle R\rangle$  and $\langle p\rangle$ are the median of values for the left (L) and right (R) sides of the fit.  The t-test assumes identical variance.  }
\end{table}

\begin{figure}[h]
	\resizebox{\hsize}{!}{\includegraphics{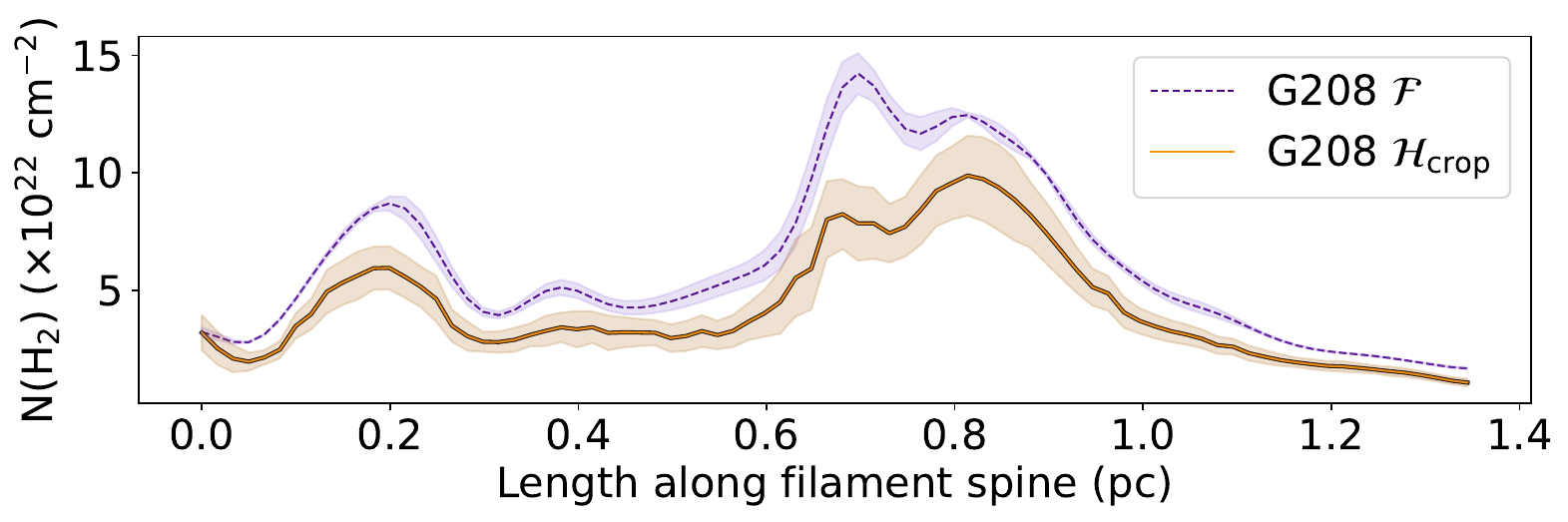}}
	\caption{Comparison of derived \textit{N}(\MH) in G208.  \textit{N}(\MH) along the filament crest for the \feather\, (purple dashed line) and \hcrop\, (orange solid line).  The \hcrop\, map is at 20\arcsec\, resolution, and \feather\, has been convolved to the same 20\arcsec\, resolution.   \label{fig:NH2_comparison} }
\end{figure}

Column densities using \herschel\, and \feather\, data along the filament crest are shown in Fig. \ref{fig:NH2_comparison}. There is difference of about 30\% in column densities, consistent with the \artemis\, calibration uncertainty. 

\subsection{Fragmentation \label{sec:fragmentation}  }
 We study fragmentation of the filaments in G208 using wavelet decomposition as well as by using the extracted clumps. In this analysis the noisy edges are removed from the \feather\, map, and we use the \hcrop\, map (Fig.~\ref{fig:clumps}).

\subsubsection{Clumps}

Dense clumps were identified in the \feather\, and \hcrop\, column density maps using the astrodendro package\footnote{\url{http://www.dendrograms.org/}} \citep{dendrograms}. The dendrogram algorithm describes the data as a hierarchy of structures of progressively smaller size. In order to find the densest clumps, we use the leaves of the dendrogram. We require fitted structures have a minimum column density of $2\times RMS$, where $RMS$ is the standard deviation within a relatively empty region (the orange circle in Fig.~\ref{fig:clumps}).  Five clumps are found in the \hcrop\, map and sixteen in the \feather, though three of the \feather\, clumps are likely due to noise at the map edges, and are excluded from further analysis.  The ellipses for these clumps are shown in Fig. \ref{fig:clumps} and their properties are compared in Fig.~\ref{fig:compare_clump_chars} and Table~\ref{tbl:Jeans}. 	

Though \feather\, clumps are smaller, they have higher median \textit{N}(\MH) and mass per area, though derived clump mass is generally lower than Jeans mass. Figure \ref{fig:G208_SED} shows the spectral energy distributions (SEDs) of the four densest clumps in each field from 160 to 500\,\micro. Based on MBB fits assuming constant \B\, = 1.8, clumps in all but G202 are generally warm, at $T\sim 18-20$\,K.

We  calculated the distances between the clumps. Mean separation is ({0.28 $\pm$ 0.14)\,pc for the \hcrop\, map and (0.13 $\pm$ 0.06)\,pc for the \feather. The mean clump separation in the \feather\, map is close to the estimated effective Jeans length of 0.1\,pc, while mean separation in the \hcrop\, map is approximately double the effective Jeans length. It is therefore likely that many of the clumps in OMC-3 are gravitationally unstable with the possibility for future star-formation.

We searched for YSOs from the \citet{Megeath2012} Spitzer catalog, finding 23 YSOs which overlapped with the \hcrop\, clumps. Of these, \citet{Megeath2012} have classified nine as D-class, or pre-main-sequence stars with a disk. The rest are classified as P-class, or protostars. Ten of these YSOs also overlap with clumps in the \feather\, map, with all six classified as protostars. 
All \hcrop\, clumps are spatially associated with at least one YSO. 

\begin{figure}[h]
	\resizebox{\hsize}{!}{\includegraphics{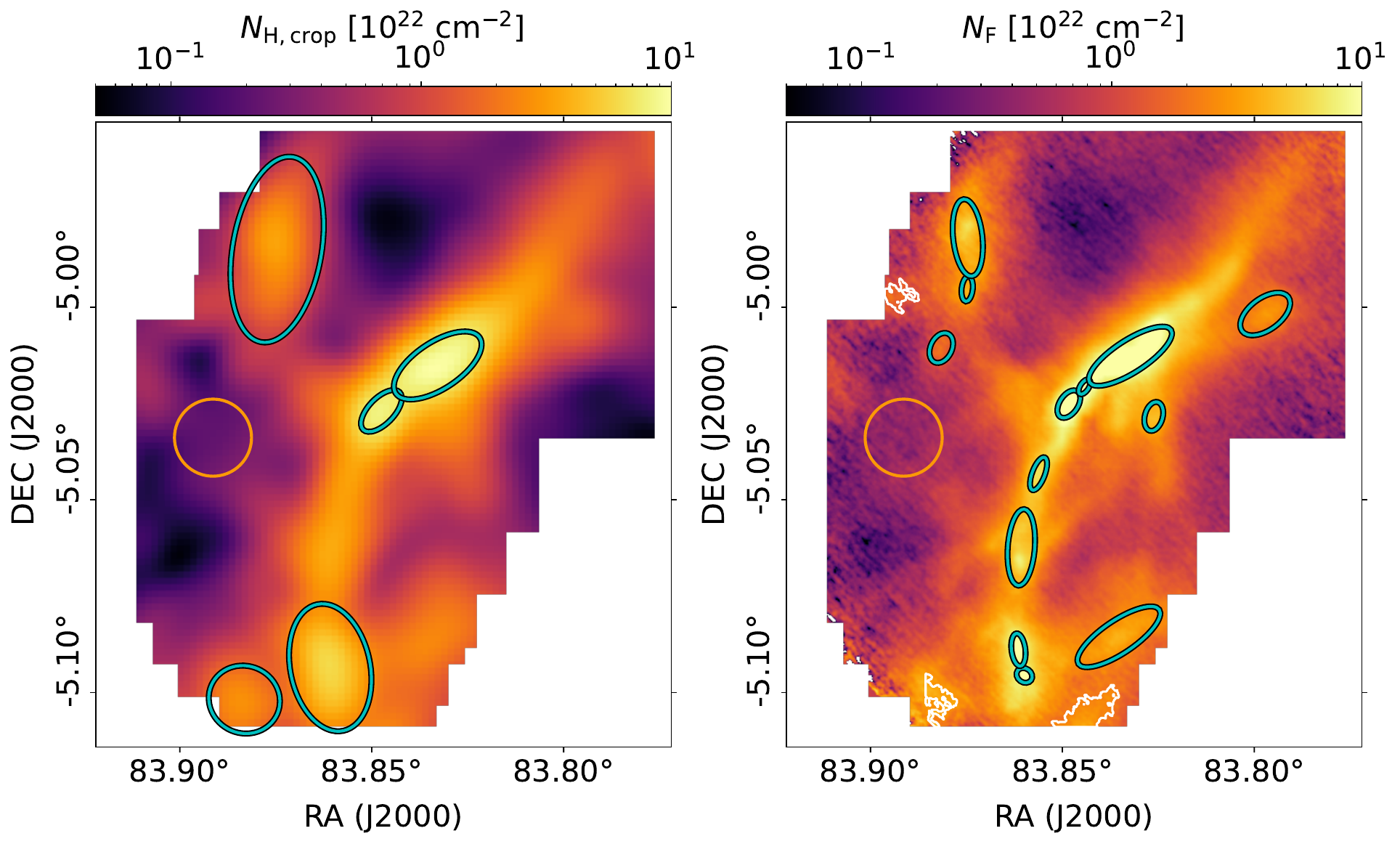}}
	\caption{   Clump ellipses (cyan) plotted on top of the \hcrop\, (left) and \feather\, (right) column density maps. The three clumps removed from analysis are drawn with white outlines on the \feather\, map. The orange circle shows the region used for determining the RMS noise. 	\label{fig:clumps} }	
\end{figure}

\begin{figure}[h]
	\resizebox{\hsize}{!}{\includegraphics{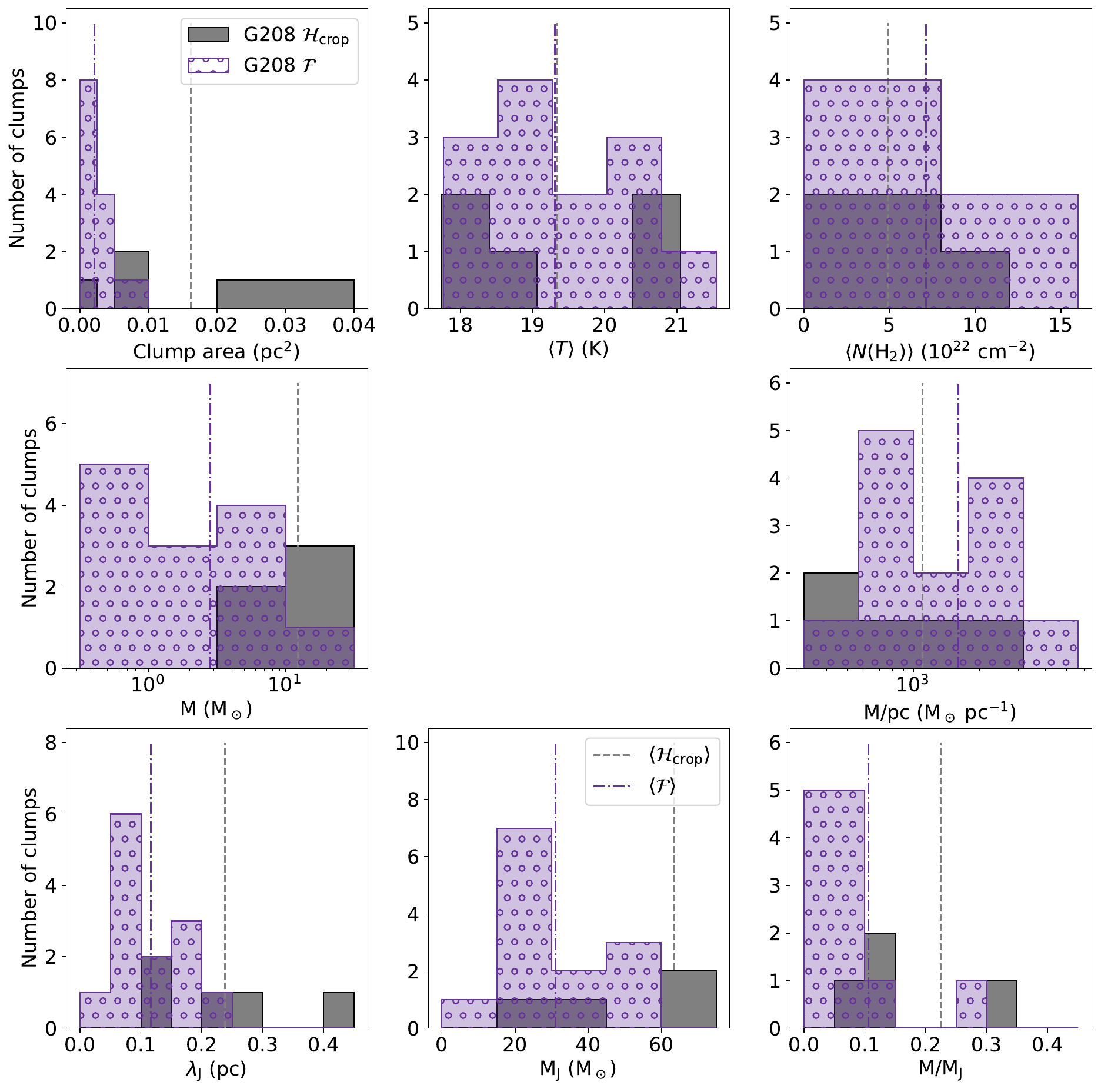}}
	\caption{   Comparison of clump properties found in \hcrop\, (gray) and \feather\, (purple dotted) column density maps. The corresponding median values are plotted with gray dashed and purple dash-dotted lines for \hcrop\, and \feather, respectively.  \label{fig:compare_clump_chars} }
\end{figure}

\begin{table*}
	\centering
	\caption{Clump characteristics in G208.} 
	\label{tbl:Jeans}
	\begin{tabular}{lllllllllll}
		\hline\hline 
		& $\langle T \rangle$ & $\langle N$(\MH) $\rangle$     & $\langle R_{\rm eff}\rangle$   &   $\langle M \rangle$ &  $\langle \lambda_{\rm J}\rangle$ &  $\langle M_{\rm J} \rangle$ & $\langle M/area \rangle$  & $\langle s \rangle$ \\ 
		& (K)  &  ($10^{22}$\,cm$^{-2}$) & (pc)  &  (\msun)  &  (pc) & (\msun) & (\msun\,pc$^{-2}$) & (pc)  \\ 
	
		\hline

		\hcrop  & 19.3 $\pm$ 1.3 &  4.9 $\pm$ 2.8 &  0.03 $\pm$ 0.01 & 12.3 $\pm$ 6.6 & 0.24 $\pm$ 0.12 & 63.5 $\pm$ 31.3 & 1099 $\pm$ 623 & 0.3 $\pm$ 0.1 \\   
		\feather  & 19.3 $\pm$ 1.1 &  7.1 $\pm$ 4.5 &  0.01 $\pm$ 0.01 & 2.8 $\pm$ 3.1 & 0.12 $\pm$ 0.05 & 31.1 $\pm$ 14.4 & 1598 $\pm$ 999 & 0.1 $\pm$ 0.1 \\   

		\hline
	\end{tabular}
	\tablefoot{ Dust temperature, column density, clump radius, and clump mass (columns 1-4) are estimated from MBB fits to \textit{Herschel} 160--500\micro\, data. Effective radius is calculated as $R_{\rm eff} = \sqrt{a\times b}$, where \textit{a} and \textit{b} are the clump major and minor axes. Columns 5-6 list the effective Jeans' length $\lambda_{\rm J}$ and effective Jeans' mass $M_{\rm J}$, where non-thermal   velocity dispersion is estimated from \citet{Suri2019}, $\langle \sigma_{\ NT}\rangle \approx 0.8$\,km\,s$^{-1}$. Due to the similar mean temperatures, $\langle \sigma_{\ TH} \rangle$ = 0.24\,km\,s$^{-1}$ and $\langle \sigma_{\ tot}\rangle$ = 0.83\,km\,s$^{-1}$ for both \feather\, and \hcrop\, clumps. Column 7 lists the average mass surface density, and column 8 the mean separation between neighboring clumps (\textit{s)}. }
\end{table*}

Clump profiles in the column density maps were fitted with power-laws $r^{-\alpha}$ following the methodology of \citet{Shirley2000}. We derive mean major-axis power-law indices of $\alpha = $ 1.54 $\pm$ 0.96 for G208 \hcrop, and $\alpha = $ 1.21 $\pm$ 0.74 for G208 \feather. Minor-axis power-law indices are smaller, with 0.97 $\pm$ 0.59 for \hcrop\, and 0.75 $\pm$ 0.69 for \feather.

\begin{figure}[h]
	\resizebox{\hsize}{!}{\includegraphics{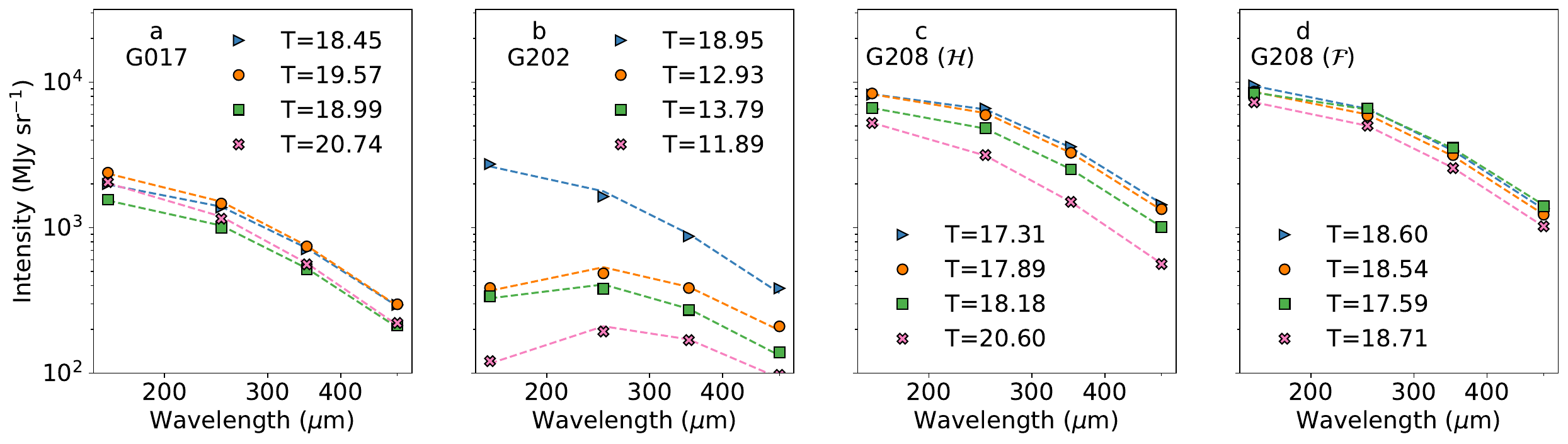}}
	\caption{SEDs of the four brightest clumps in each field, at 160--500\,\micro. The dashed lines represent MBB fits with constant \B =1.8 for each clump. The best-fit temperature for each clump is listed in the legend.\label{fig:G208_SED} }
\end{figure}

\subsubsection{Wavelet decomposition \label{sec:wavelets}   }
We performed wavelet filtering to identify structures at various spatial scales, as in \citet{Mattern2018}. Column density maps are decomposed into scale maps $X_{\rm i}$ at scales from 0.01--0.4\,pc, where we require that the minimum scale used is larger than the pixel size of the image. The maximum scale used is the scale at which no distinct structures are visible in the map after convolution. For this analysis, we study only the $A_{\rm v}\geq 3^{\rm mag}$ regions. 
We require that each structure at level \textit{i} overlaps with a structure at levels \textit{i}+1 and \textit{i}-1. At the lowest and highest levels, we require the structure to only overlap at levels \textit{i}+1 or \textit{i}-1, respectively. This method is explained in detail in \citet{Kainulainen2014}. We then find clumps in these decomposed images using Dendrograms, taking each leaf as an individual structure.
Derived properties of these structures are listed in Table \ref{tbl:wavelet}.

Mass of each structure is calculated with 
$$M = \langle N(H_{\rm 2})\rangle \cdot \Omega d \mu m_{\rm H},$$
where $\langle N(H_{\rm 2}\rangle$ is the mean \textit{N}(\MH) of the region, calculated as the mean over all pixels in the clump footprint, $\Omega$ the solid angle, \textit{d} the distance, and $\mu m_{\rm H}$ the mass of the Hydrogen molecule.  Hydrogen number density \textit{n}$_{\rm H}$ is
\begin{equation}
\label{eq:H_number_density}
n(H_{\rm 2}) = \langle N(H_{\rm 2})\rangle \times \pi \times \Big(\dfrac{a\times b}{V}\Big),
\end{equation}
where \textit{a} and \textit{b} are the major and minor axes of the structure, $V$ is the volume of the structure, assuming a prolate spheroid shape, $V = \frac{4}{3}\pi\times  a\times  b\times  \min(a,b)$. 
We calculate median separation \textit{s} between substructures.

Not surprisingly, the number of structures increases toward smaller scales, though not in a linear fashion. Though the total mass increases with spatial scale, the highest masses are around scales of 3.0, 1.0, and 0.75\,pc for G17, G202, and G208, respectively. This may simply be due to the smaller number of structures at the largest scales. Both mean volume and column densities decrease toward larger scales, not surprising as more of the diffuse ISM is included in the structures. 
In G208 \herschel\, \textit{N}(\MH) is at its peak at scales of 0.05\,pc. In G208 \feather\, the median separation between structures in all but the largest two levels are smaller than the Jeans length, whereas in the \herschel\, G208 field, median separations are all larger.

\section{Discussion \label{sec:discussion}    }

\subsection{  Cloud properties }
The \textit{Herschel} Gould Belt survey has observed a range of nearby star-forming regions, with filament line masses in the range 1--204\,\msun\,pc$^{-1}$ \citep{Arzoumanian2011,Palmeirim2013,Arzoumanian2019}. We find similar line masses in all fields. Quiescent clouds generally have lower line masses, such as  those observed with CO molecular line transitions in the Perseus MC \citep{Guo2022} or the quiescent Musca cloud \citep{Kainulainen2016}, or for a sample of nearby ($d< 500$\,pc) filaments imaged with \textit{Herschel} \citep{Arzoumanian2019}.
We derive line masses of $\sim$ 170--300\,\msun\,pc$^{-1}$ in OMC-3, similar to those derived by \citet{Schuller2021} and \citet{Li2022b}, also for OMC-3. 
Similar line masses are also derived in the high-mass SF region G345.51+0.84 with ALMA and \textit{Herschel} \citep{Pan2023} and in extragalactic sources, such as the N159W-North filament in the Large Magellanic Cloud (LMC) observed with ALMA \citep{Tokuda2022}.
Our derived line masses of 100--300\,\msun\,pc$^{-1}$ (Table \ref{tbl:properties}) are under half of that derived for the Nessie filament \citep{Mattern2018} and a fifth of those derived for a sample of large-scale filaments within the ATLASGAL survey \citep{Ge2022}.

Our derived column densities for G208 \feather, with maximum \textit{N}(\MH) over $10^{23}$\,\CM , are similar to that of \citet{Schuller2021} (Fig. \ref{fig:compare_schuller}). Within the densest regions of the filament, the column densities derived by \citet{Schuller2021} are higher by up to $10^{23}$\,\CM, corresponding to a difference of 56\% of total column density. The differences in derived \textit{N}(\MH) are likely due to two factors: calibration uncertainty and different feathering methods used in this work and in \citet{Schuller2021}, as well as a different combination of maps used to derive temperature and column density. The significantly higher column densities in the \citet{Schuller2021} map correspond to dense cores, which also have higher surface brightness than in our data (Fig. 1 in \citet{Schuller2021} compared to Fig.~\ref{fig:feather} in this paper). 
Similar column densities to those in G208 have been detected in a sample of 11 IRDCs and in the high-mass star-forming NGC 6334 complex using \textit{Herschel} and \artemis\, data \citep{Andre2016,Peretto2020}. 
Column densities of $\sim$ 2$\times 10^{22}$\,\CM\, have been observed in two IRDCs using \artemis, LABOCA\footnote{Large APEX BOlometer CAmera}, and \textit{Herschel} data \citep{Miettinen2022}. Slightly higher column densities of $\sim$ 5$\times 10^{22}$\,\CM\, were observed using \artemis\, and \textit{Herschel} SPIRE/HOBYS data of the Galactic \ion{H}{II} region RCW 120  \citep{Zavagno2020}, similar to the mean values over the whole G208 filament.

\begin{figure}[h]
	\resizebox{\hsize}{!}{\includegraphics{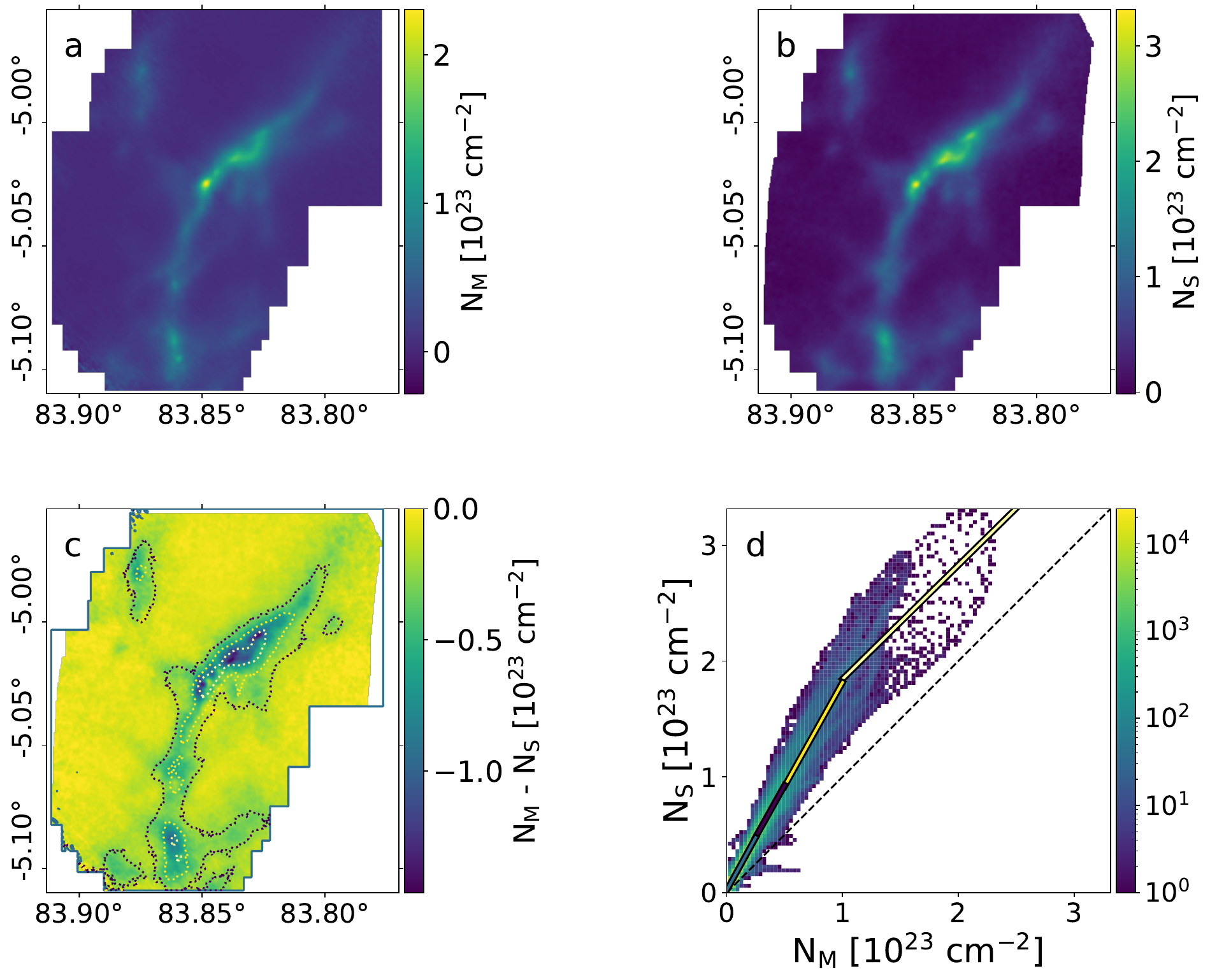}}
	\caption{Comparison between column densities derived in this work ($N_{\rm M}$) and in \citet{Schuller2021} ($N_{\rm S}$) using a combinations of \textit{Herschel} and \artemis\, observations. Frame a: this work. Frame b: \citet{Schuller2021}, APEX project 098.F-9304. Frame c: difference between this work and that of Schuller. The contours correspond to $N_{\rm M} = (0, 0.25, 0.5, 1.0)\times 10^{23}$\,\CM\, Frame d: comparison plot between \textit{N}(\MH) values. The black dashed line represents a 1-1 relation, the colored lines the linear fits between $N_{\rm M}$ and $N_{\rm S}$, with colors corresponding to the contours in frame c. \label{fig:compare_schuller} }
\end{figure}

\subsection{Plummer fitting}

\subsubsection{Symmetry of the filaments  \label{sec:discussion_Plummer_symmetry}  }

To study what are normal values of skewness (\textit{S}) and kurtosis (\textit{K}), we have simulated filaments with the symmetric Plummer parameters of our \textit{q}=1 filament segments (listed in Table \ref{tbl:PlummerResults_sym}). We include a powerlaw background and an additional Gaussian large-scale component to simulate hierarchical structure. Asymmetric Plummer fits are performed to these simulated filaments, and from a total of 1280 filament profiles we have calculated mean \textit{S} and \textit{K}. As the generated filament profiles are symmetric, the values in Table \ref{tbl:simulated_SK} give lower limits of \textit{S} which correspond to significant asymmetry. 
\begin{table}
	\centering
	\caption{Skewness and kurtosis calculated for simulated filaments with the symmetric Plummer parameters of our \textit{q}=1 profiles.}
	\label{tbl:simulated_SK}
	\begin{tabular}{lll}
		\hline\hline
		Field & \textit{S} & \textit{K}\\
		\hline
		G17 & 1.72$\pm$0.27 & 2.55$\pm$1.02 \\  
		G202 & 1.21$\pm$0.29 & 0.71$\pm$0.65 \\ 
		G208 \herschel & 2.02$\pm$0.06 & 3.55$\pm$0.30 \\ 
		G208 \feather & 0.97$\pm$0.01 & -0.51$\pm$0.02 \\ 		
		\hline
	
	\end{tabular}
	\tablefoot{The field refers to the field from which Plummer parameters were taken. Values are the mean $\pm$ standard deviation. }
\end{table}

When comparing the derived values of \textit{S} (Table \ref{tbl:skewNess}), all filaments have mean values lower than the limits set by the simulation. However, as seen in Fig. \ref{fig:skewLine}, G202 and G208 \feather\, have regions in which skewness is outside of these limits. Strong external directional forces (e.g. from SF) likely affect the OMC-3 filament. \citet{Zheng2021} also find an abundance of asymmetric filaments within the Orion A cloud with CO observations, with most filament segments in OMC-3 being asymmetric in their paper. Likewise, \citet{Peretto2012} find multiple asymmetric filament profiles in the Pipe nebula using \textit{Herschel} observations, which they interpret to be due to compression flows from the  Sco OB2 association.  The multiple filament segments and colliding filaments in G17 and G202 also likely raise \textit{S} in these filaments, and it is unclear whether the segments of significant asymmetry in these two filaments are caused by external forces or filament collision.

\subsubsection{Filament widths \label{sec:discussion_PlummerFitting_width}   }

We find mean filament FWHM widths of $\sim$ 1.0, 0.3, 0.1, 0.08, and 0.05\,pc for fields G17, G202, G208 \herschel, G208 \hcrop, and G208 \feather, respectively. As shown in Fig. \ref{fig:distance_vs_FWHM}, distance to the field and instrumental resolution seem to affect derived width. This is likely due to the increasing resolution of nearer fields: at larger distances the observed filament includes more of the extended structure. \citet{Juvela2012} found that at the nominal \textit{Herschel} resolution, assuming 0.1\,pc widths, filament parameters could be recovered reliably only up to a distance of 400\,pc. The artemis resolution should thus double this threshold. A dependence on distance has been found in multiple studies \citep{GCCVII,Panopoulou2022, Andre2022}.  \citet{Panopoulou2022} suggest a relation between distance of $FWHM_{\rm filament} \approx 4\times HPBW,$ shallower than that that found among our sample but steeper than a fit to continuum \textit{FWHM} values from Fig.~\ref{fig:distance_vs_FWHM}, which finds a relation of $FWHM \approx 1.8\times HPBW.$
\citet{Andre2022}, who also find a slight dependence between distance and width, suggest that a characteristic width exists, but is unresolved at distances $>$ 1\,kpc.

Molecular line observations from the literature also show a potential correlation between resolution and \textit{FWHM} although molecular lines do not necessarily trace the same filaments as continuum data. Simulations by \citet{PriestleyWhitworth2020} find that filaments detected in CO are several times larger than the same filaments detected in continuum observations, but narrower when detected using dense gas tracers such as HCN and N$_{\rm 2}$H$^{+}$. Observations of the Orion regions show a similar discrepancy between molecular line and continuum observations \citep{Shimajiri2023a}.  
Further, slightly different fitting processes (including the level of background subtraction, deconvolution, and filament extraction methods) can affect derived filament widths.

Though G208 \herschel\, shows the characteristic 0.1\,pc width, the same field observed with the resolution of \artemis\, shows widths of only $\sim$0.05\,pc. This is similar to results found by \citet{Smith2016}, \citet{Schuller2021}, and \citet{Panopoulou2022}. Due to the multiscale nature of the dense ISM, it is not surprising that filament-like structures can be observed at many scales. Though dense fibers have been detected in the Southern OMC-3 region in N$_{\rm 2}$H$^{+}$ \citep{Hacar2018}, further high-resolution continuum data are needed to quantify the internal structure of filaments at the scale of the fibers.  While fibers are not visible in the MIR extinction data of \citet{Juvela2023}, these observations are limited by 8\,\micro\, absorption saturating toward the densest filaments.  
 \citet{Schuller2021} have compared their observations with those of of \citet{Hacar2018}. They find that while N$_{\rm 2}$H$^{+}$ generally seems to correlate with the densest region of the filaments visible in the continuum, N$_{\rm 2}$H$^{+}$ can be destroyed by interactions with CO and by high-energy radiation. This can narrow filament profiles observed with molecular lines.

\citet{Andre2022} have suggested that the apparent relation of distance to \textit{FWHM} can be mostly explained due to the convolution of a Plummer-like filament combined with background noise fluctuations. However, assuming an intrinsic width of 0.1\,pc, the mean \textit{FWHM} values of G202 should be only $\sim$0.13\,pc according to their Figure 3b.

This conclusion of course also depends on the relative strengths of the background components and the filament, as well as on the sizes of the background fluctuations.
The power-law exponent of the background fluctuations in the full \textit{Herschel} fields is around -2, shallower than that found in \textit{Herschel} SPIRE observations of the Polaris Flare \citep{MivilleDeschenes2010}. However, \citet{MivilleDeschenes2010} have both a larger map size, as well as lower resolution of $\sim$30\arcsec, possibly explaining this difference in power-law exponent. Whether the background is described by an exponent of -2 or -3 does not significantly affect derived filament \textit{FWHM}. Performing the simulations in Sect. \ref{sec:app_simulations_Plummer_distance}, using a modeled background with a power-law slope of -2 results in \textit{FWHM} values of 22.3 $\pm$ 5.5 pixels, and assuming a slope of -3 we derive FWHM = 25.1 $\pm$ 10.4 pixels, identical within the uncertainties. Both power-law backgrounds  increase the derived \textit{FWHM} from the original value of 20.78\,pixels. 

The relative strengths of the background components are difficult to identify, and depend on the positions of the filament segment in relation to the wider cloud. Comparing the standard deviation across the whole background (defined as the $A_{\rm v} < 3.0^{\rm mag}$ region) with the maximum intensity of each filament segment results in SNR values from 10--100. However, this method does not necessarily take into account all background structures, especially those on scales larger than the observations and thus the "true" SNR may be lower than that calculated. Simulations in Sect.~\ref{sec:app_simulations_Plummer_BG_strength} show the effects of lower SNR on filament widths. Though the mean \textit{FWHM} values do not significantly change, the uncertainty increases. 
Further simulations in Sect.~\ref{sec:app_simulations_Plummer_distance}, which include a wide Gaussian component with an ideal narrow filament, result in \textit{FWHM} values which increase by up to 30\% at 2\,kpc from the original width. Similar simulations not including this wide component result in an increase of \textit{FWHM} of up to 10\%. \textit{SNR} also affects derived widths by a further 5\% if it is decreased from 100 to 10. These simulations show both how large-scale structures can artificially widen filament profiles, but also that detected filament widths slightly increase as a function of distance even if one studies only an ideal narrow filament.

As the simulations of Sect.~\ref{sec:app_simulations_Plummer_distance} show, convolution and background fluctuations, combined with the multiscale nature of the ISM, will affect derived filament widths.
The chosen definition of a filament further complicates analysis. In C$^{18}$O data, \citet{Suri2019} find tens of filament segments in the OMC-3 region.  By segmenting these data in another way, derived mean filament widths can change significantly.

In \citet{Juvela2023}, we have compared four of the densest filament segments using \textit{Herschel} and \artemis\, emission, as well as Spitzer extinction. When comparing only the dense filament segments, also \textit{Herschel} data show a relatively low width of 0.05--0.1\,pc. This differs significantly from the results derived across the entire OMC-3 filament and may be explained by the consistently high column densities of the filament profiles in \citet{Juvela2023}. In this present paper, \textit{FWHM} is also lower in more dense filament segments.  It seems that, when looking only at the regions of OMC-3 with the highest filament strength compared to the background, derived width does not depend on resolution. Whether this would hold in G17 or G202 is unknown. 
Further, both resolution ($\sim$2\arcsec) and SNR in the mid-IR Spitzer data were higher than the \artemis\, resolution and SNR.

A wide range of filament widths have been found in the literature.  Using \textit{Herschel} SPIRE observations for a sample of filaments, \citet{Schisano2014} found mean filament widths between 0.1--2.0\,pc, increasing with distance to the filament. Using \artemis\, and \textit{Herschel}, \citet{Schuller2021} find similar widths to our \feather\, field of 0.06\,pc in OMC-3. Similar values of $\sim$0.07\,pc are detected by \citet{Kainulainen2016} using dust extinction mapping. Various combinations of JCMT SCUBA-2, \artemis, and \textit{Herschel} data results in \textit{FWHM} $\sim$0.08--0.12\,pc \citep{Hill2012,Salji2015b,Howard2019,Zavagno2020}. \textit{FWHM} values as high as 0.3\,pc have been detected in the Vela C cloud at a distance of $\sim$900\,pc \citep{Li2023} using \textit{Herschel} data alone. Similarly high values of 0.17\,pc are detected in the central molecular zone at a distance of 8.3\,kpc using ALMA and \textit{Herschel} data \citep{Federrath2016}.

\subsubsection{Parameters $p$ of the Plummer fits}

Out derived median power-law indices using asymmetric fits are $p\sim$2--5, though with some outliers reaching up to 25. However, values of \textit{p} over 5 are unlikely to be true, and are probably due to the degeneracy between the \R\, and \textit{p} values. Large \textit{p} may be caused when a filament has a steep profile, and the shape of the convolved profile is dominated by the convolution by the beam as opposed to emission from the filament. In addition, \citet{Juvela2023} find that dips in column density in profile tails can cause large values of \textit{p}.   However, large \textit{p} values do not affect the shape of the derived profile. For example, the asymmetric fits to G202 and G208 \feather\,in Fig. \ref{fig:asymmetric_vs_symmetric_profile} both have fitted $p = 25$. 

In symmetric fits to \textit{Herschel} data, the median indices $p$ are below 4. Values from the literature are generally lower, with \textit{p}$\sim$ 2.0$\pm$0.4 found in Taurus \citep{Palmeirim2013}, \textit{p}$\sim$ 2.6$\pm$11\,\% in Musca \citep{Kainulainen2016}, \textit{p}$\sim$ 2.2$\pm$0.3 in a sample of nearby filaments \citep{Arzoumanian2019}, and \textit{p}$\sim$ 1.5--2 in Hi-GAL filaments \citep{Andre2022}. \citet{Schuller2021} also find \textit{p} = 1.7--2.3 in OMC-3.

\subsection{Comparison between feathered and \textit{Herschel}-only fits  \label{sec:discussion_feather_herschel_comparison} }

We find significant differences between \herschel\, and \feather\, data. \textit{FWHM} of profiles is over two times larger in \herschel\, data, directly proportional to the data resolution.  This is a difference of $\sim (2-3)\sigma$ in \textit{q} = 1--2 filaments. This is true for both symmetric and asymmetric fits. This difference is similar to that found recently by e.g. \citet{Smith2016} and \citet{Panopoulou2022}, as well as \citet{Schuller2021} in OMC-3. Fragmentation also shows increased substructure with the inclusion of higher-resolution \artemis\, data, both through clump detection and wavelet analysis. With even higher-resolution data, such as from ALMA or JWST, smaller cores and fibers could possibly be detected.

\subsection{Comparison of various Plummer fitting methods}
In the following, we study how  the results depend on different methods of Plummer fitting. We compare fits using symmetric and asymmetric Plummer functions, examine the differences between the fits to individual filament cross sections and to the mean filament profile. Finally, we study how adding offset $\Delta r$ to the Plummer formula (Eq. \ref{eq:PlummerBG}) changes results. 

\subsubsection{Symmetric and asymmetric fits   }

Though we used asymmetric Plummer fits in our study, Fig.~\ref{fig:asymmetric_vs_symmetric} shows a comparison to symmetric fits. In a symmetric fit, both sides of the Plummer function have the same \R\, and \textit{p} values. 
Especially noticeable is the spread in \textit{p} for all but G208  \herschel. We plot as an example one \textit{q} = 1 filament with the highest offset for each field in Fig. \ref{fig:asymmetric_vs_symmetric_profile}. Values for the fits, including $\chi^{2}$ values, are listed in Table \ref{tbl:asymmetric_vs_symmetric}.

Comparing the results of symmetric and asymmetric fits for our whole sample from Tables \ref{tbl:PlummerResults}-\ref{tbl:PlummerResults_sym}, we see that generally they do not differ by more than 1$\sigma$. Median \textit{FWHM} does not differ significantly between the two fits. Though the median \textit{FWHM} of G208 \herschel\, filaments is approximately 0.1\,pc, neither the other fields nor the \feather\, map show a characteristic width. 

\begin{table*}
	\centering
	\caption{Values of the fits in Fig.~\ref{fig:asymmetric_vs_symmetric_profile}.   }
	\label{tbl:asymmetric_vs_symmetric}
	\begin{tabular}{l|ll|ll|ll|ll}
	\hline
	\hline
		
		Field & \multicolumn{2}{l}{\R\, [pc]} & \multicolumn{2}{l}{$p$} & \multicolumn{2}{l}{$FWHM$ [pc]} & \multicolumn{2}{l}{$\chi^{2}$} \\ 
		& S & A & S & A & S & A & S & A \\ 
		\hline
		G017.69-00.15 & 5.30 & 2.31 & 25.00 & 5.59 & 2.585 & 2.749 & 0.045 & 0.021 \\ 
		G202.16+02.64 & 1.30 & 1.29 & 25.00 & 25.00 & 0.633 & 0.628 & 0.020 & 0.019 \\ 
		G208.63-20.36 & 0.04 & 0.05 & 2.66 & 2.65 & 0.099 & 0.106 & 0.098 & 0.064 \\ 
		G208.63-20.36 & 0.08 & 0.05 & 25.00 & 13.92 & 0.041 & 0.034 & 0.644 & 0.634 \\ 
		\hline
	\end{tabular}
	\tablefoot{The headers S and A refer to symmetric and asymmetric fits, respectively. }
\end{table*}

\begin{figure*}[h]
	\sidecaption
	\includegraphics[width=12cm]{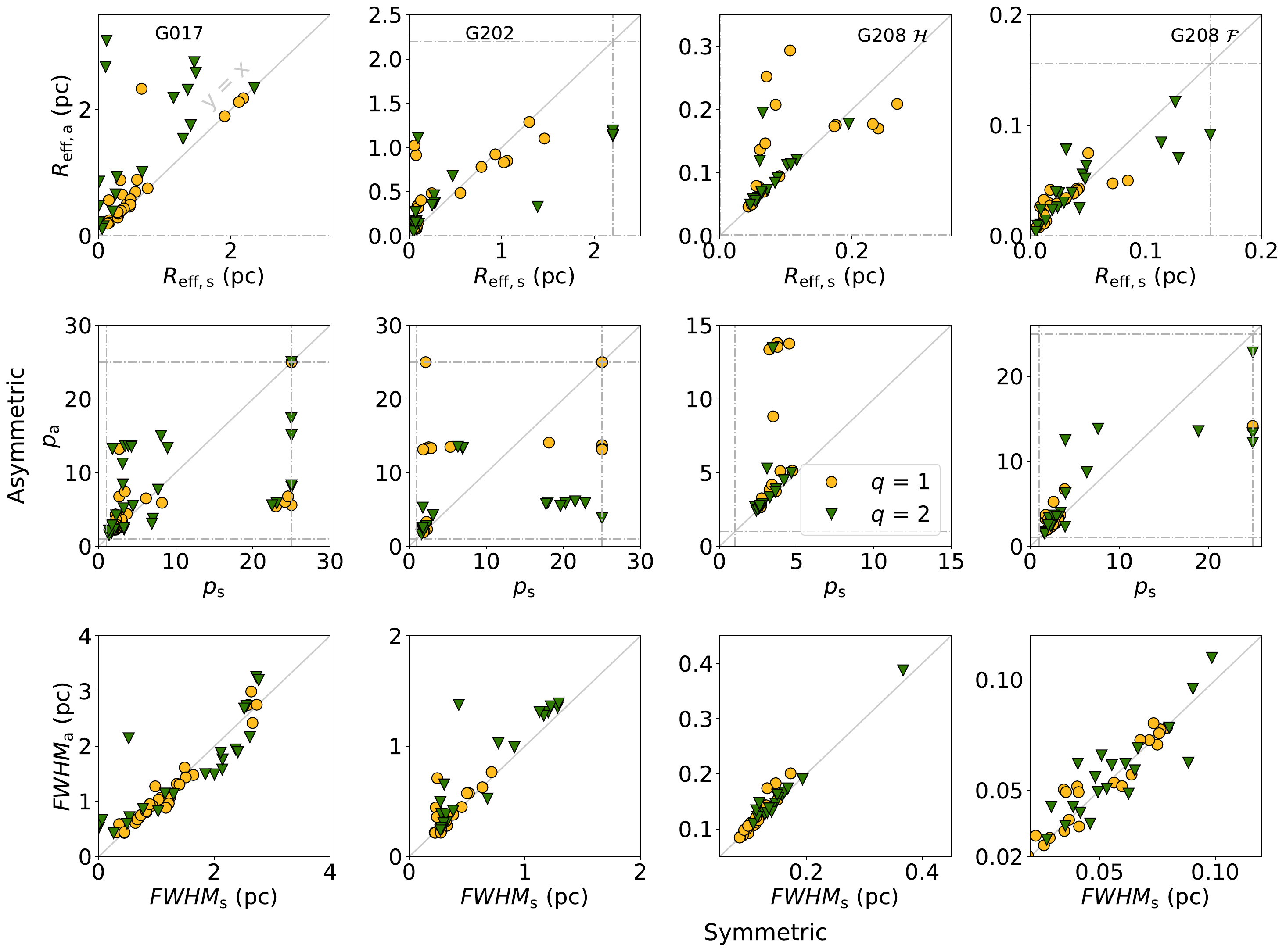}
	\caption{Comparison of symmetric and asymmetric Plummer fits for all fields.  \R\, (top row), \textit{p} (middle), and \textit{FWHM} (bottom row) for symmetric (x-axes) and asymmetric (y-axes) fits.  \textit{q} = 1 profiles are plotted with yellow circles and \textit{q} = 2 profiles with green triangles. The gray solid line shows a 1:1 relation and the gray dashed lines the allowed upper and lower limits of the Plummer parameters.  \label{fig:asymmetric_vs_symmetric} }
\end{figure*}

\begin{figure*}[h]
	\sidecaption
	\includegraphics[width=12cm]{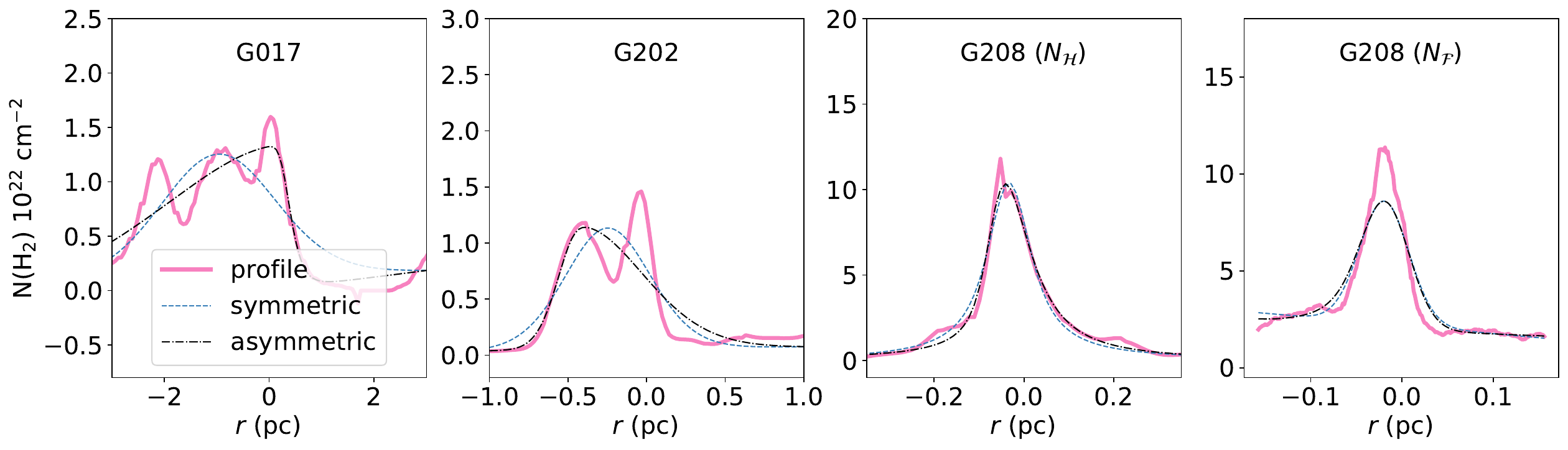}
	\caption{ Profiles of the \textit{q} = 1 filament with the largest offset in each of the four fields. The pink solid line corresponds to the observed profile, the blue dotted line to the symmetric Plummer fit and the black dashed line to the asymmetric Plummer fit. \label{fig:asymmetric_vs_symmetric_profile} }
\end{figure*}

\subsubsection{ Analysis of the mean profile or analysis of multiple filament profiles }
Many studies \citep[e.g.][]{Arzoumanian2011,GCCVII} of filament widths concentrate on the analysis of the mean or median profile. We have compared fits to the median profile and the median of all the fits to individual cross sections in Fig. \ref{fig:compare_medians2}. Median values of \textit{p} in G208 \feather\, are higher when estimated from the mean profile, but slightly lower in the \textit{Herschel} fields. \R\, and \textit{FWHM} are remarkably consistent regardless of fitting methods used. The correlation between the two fits is better in profiles of higher quality, not surprising as a stronger background or noise will introduce more uncertainty into the fits. Similar correlation also with the inclusion of MIR data was found in \citet[][Fig.~3d]{Juvela2023}. 

More information about the distribution of values is accessible by fitting individual profiles instead of just their average. As shown in Fig. \ref{fig:derived_violins}, the distribution of individual parameters may be quite wide, even at higher quality flags ($q\leq 2$). In contrast, the parameter \textit{p} is more robust when fitting to the median profile.

\begin{figure*}[h]
	\sidecaption
	\includegraphics[width=12cm]{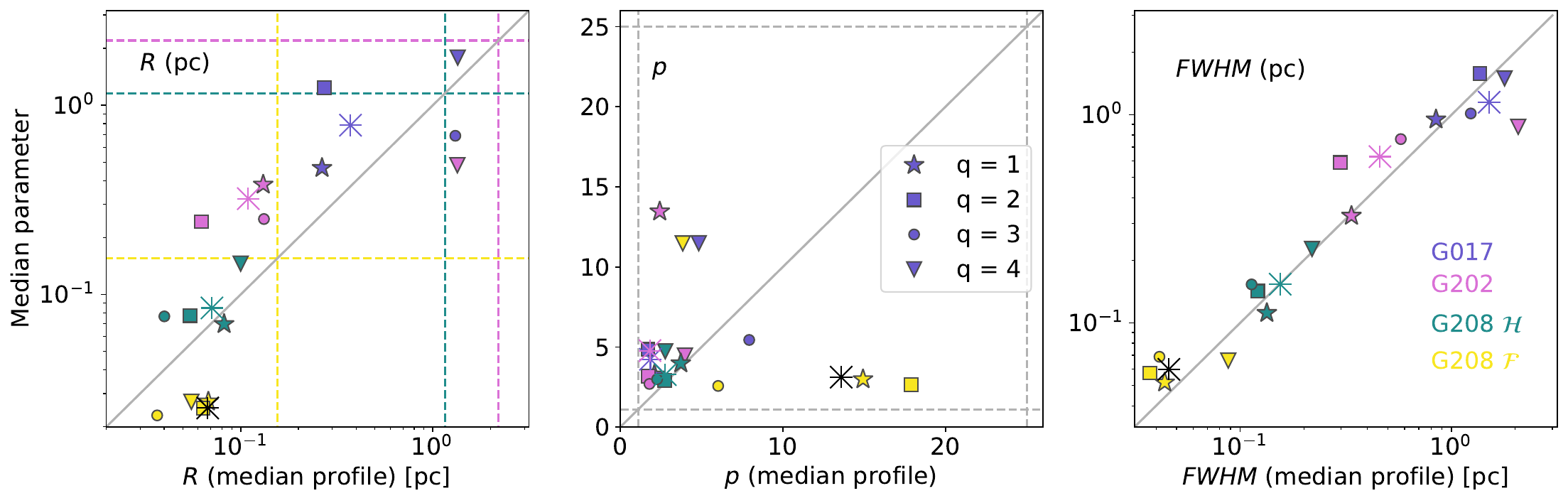}
	\caption{ (x-axis) Fit to the median profile plotted against (y-axis) the median values calculated separately for each filament profile. The gray solid line represents a relation of $y=x$. Quality flags are represented by a star, square, circle, and triangle, in order of decreasing quality. The fit to the entire field, regardless of quality flag, is shown by an asterisk. The dashed lines represent upper and lower bounds to the fits. In frame 1, the color of the line corresponds to the field. \label{fig:compare_medians2} }
\end{figure*}

\subsubsection{Comparison of fits with and without offset}
We have compared fits with and without $\Delta r$, the parameter which allows the filament peak to shift. \textit{FWHM} calculated without the offset term is plotted against FWHM derived using the offset in Fig. \ref{fig:offset_vs_none}. All four fields show good correlation between the values for the \textit{q} = 1--2 filaments, though generally \textit{FWHM} tends to be somewhat higher in fits with offset. Greater variation is seen in lower-quality filaments, especially in G17 and G208 \feather.

\begin{figure*}[h]
	\sidecaption
	\includegraphics[width=12cm]{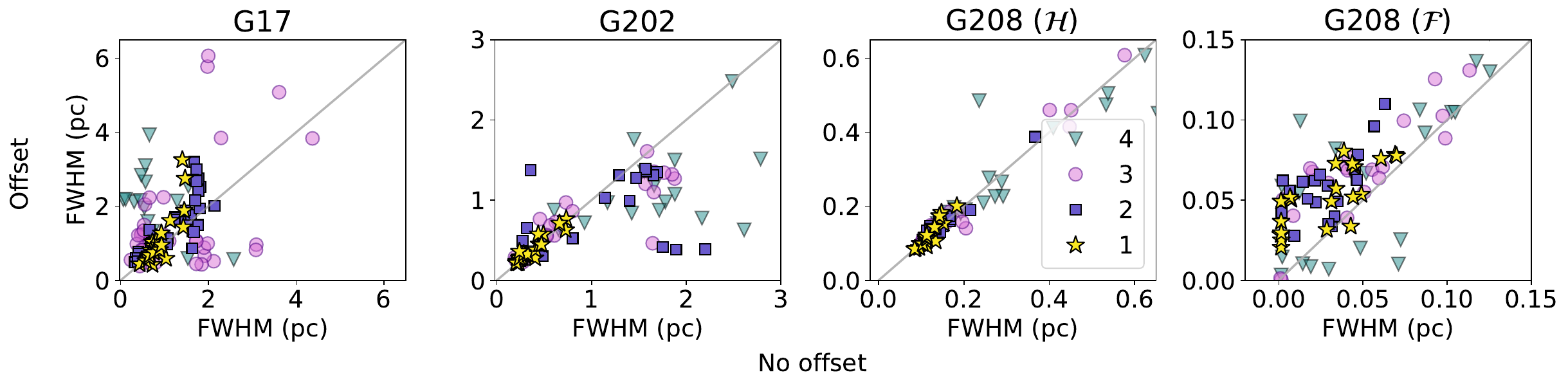}
	\caption{  \textit{FWHM} for all four datasets, calculated using the Plummer formula (Eq. \ref{eq:PlummerBG}) with no term for offset (x-axis) and including $\Delta r$ (y-axis). The gray line plots a 1:1 relation.  \label{fig:offset_vs_none} }
\end{figure*}

\subsection{Clump powerlaws}
We have fitted clump profiles with powerlaws of $r^{-\alpha},$ and find mean powerlaw indices of  $\langle \alpha\rangle = $ 1.21 $\pm$ 0.74 for G208 \feather, and $\langle \alpha\rangle =$ 1.54 $\pm$ 0.96 for G208 \hcrop. The \hcrop\, values are similar to those found in an analysis of Class I and 0 protostellar cores using the SCUBA instrument at the James Clerk Maxwell Telescope \citep[JCMT; ][$\langle \alpha\rangle =  1.48 \pm 0.35$]{Shirley2000}, whereas \feather\, clumps are somewhat shallower.

\subsection{Fragmentation length}

We estimate effective Jeans lengths of $\sim$0.1\,pc in G208, similar to the typical width for interstellar filaments, consistent with \citet{Arzoumanian2013}. Median separation between dense clumps in the \feather\, map is also a similar distance, suggesting that thermal gravitational fragmentation is sufficient to explain the fragmentation at these scales.

As in \citet{Mattern2018}, more structures are found at smaller spatial scales, though in their paper the most clumps are found in the second scale. In our fields, all \textit{Herschel} fields have a similar number of structures at the lowest two levels, whereas G208 \feather\, has more structures at level 1 (corresponding to 0.025\,pc). It is not unexpected that the median column and volume densities decrease toward larger scales, as was also found in Nessie.  However, the mean \textit{N}(\MH) in G208 \herschel\, stays constant until $\sim$0.5\,pc, in G202 until 0.1\,pc, and  in G17 until $\sim$0.5\,pc, at scales larger than the beam size. In G208 \herschel, the dependence between column density and spatial scale is approximately linear.

We do not detect fragmentation in G202 and G208 \herschel\, above 1.0\,pc. However, this is likely due to the limit of our map sizes, as opposed to a physical upper limit of fragmentation. In comparison, with a mapsize of 1$^\circ \times 20$\arcsec\, at a distance of 3.5\,kpc, \citet{Mattern2018} find fragmentation on scales from 0.1--10\,pc in Nessie. These larger fragmentation scales likely exist in our fields, but above the resolution of the data. 

In our data, median separation \textit{s} is negatively correlated with median density. A similar anticorrelation is detected in \citet{Mattern2018}, though OMC-3 has higher densities by a factor of $10^2$ compared to Nessie. \citet{Palau2014} have also observed an increase of fragmentation (which would lead to smaller separation) with higher density.
The slope of the \citet{Mattern2018} relation matches that found in G17 and G202. Orion shows a steeper relation between density and \textit{s}. 
Mean separation between substructures in all but G208 \feather\, is larger on all scales than the Jeans length. In \feather\, $\langle s \rangle \leq 0.1$\,pc at scales below 0.05\,pc. Both the analysis of dense clumps as well as wavelet decomposition show increased fragmentation at the \artemis\, resolution.

\subsection{Parameter uncertainties}

Due to uncertainties in the background, the Plummer fitting routine can fit the data with a range of parameter values. In order to study uncertainties due to the fitting routine, we performed a simulation of a filament plus a background sky that was generated based on a powerlaw power spectrum. We ran 100 fits to a map of size 100$\times$100\,pix (a total of 10000 fits), calculating the mean, median, and standard deviation of the Plummer parameters. For each run, the random fluctuations of the sky varied, but the power-law index (\textit{k} = -2) did not, and no white noise was added to the filament itself. The results of the runs are shown in Table \ref{tbl:errors_symmetric} and \ref{tbl:errors_asymmetric} for symmetric and asymmetric fits, respectively.

We find uncertainty due to background fluctuations on \R\, of 4.0\%, on \textit{p} of 2.7\%, and on \textit{FWHM} of 3.6\% for asymmetric fits. 
Uncertainty on symmetric fits is similar, at 4.2\%, 2.5\% for \R\, and \textit{p}, respectively, but slightly lower (3.0\%) for $FWHM$.

We studied sources of error in Plummer fitting further in Appendix \ref{sec:app_simulations_Plummer}. 
As we include only those filament profiles with high quality flags in our analysis, the error due to SNR variations will likely not be large. Only in field G17 is SNR low enough to affect derived parameters (Table \ref{tbl:SNR_qualityFlag}). Simulations of varying background fluctuation strength show that uncertainty increases with a stronger sky; this, too, should not be a problem in our sources. A stronger source of uncertainty can arise from large-scale structure similar to the large Gaussian structure in Appendix \ref{sec:app_simulations_Plummer_distance}, which can artificially widen the extracted filament by up to 30\% at 2\,kpc. This was discussed in more detail in Sect.~\ref{sec:discussion_PlummerFitting_width}.  
The model used to fit filament background can affect derived parameters. In this paper, we assume a linear background. We have tested fits with a polynomial background in Sect. \ref{sec:app_simulations_Plummer_BG_data}, finding no significant difference between linear and polynomial background fits in OMC-3. The possibility of a strong background component can results in uncertainty up to 50\% in \textit{FWHM}.

The Plummer fits included 1D beam convolution as part of the fitting procedure. In Appendix~\ref{sec:app_compasion1D2D} we test this against the more correct way of convolving the predictions of the fitted model in 2D. In this case the convolved values become dependent also on the neighboring filament cross sections. It is likely that the spread in derived Plummer values would decrease with the use of 2-dimensional convolution, however in regions with a distinct filament the difference between one- and two- dimensional convolution is small. Thus, in practice the difference between the 1D and 2D approaches does not cause significant differences in the parameter estimates.

Temperature variations along the line of sight can bias estimates of opacity spectral index $\beta$  \citep{Shetty2009,Malinen2011}. The submillimeter $\kappa$ and $\beta$ are also dependent on the assumed sizes and optical properties of dust grains \citep{Ormel2011}. In the dense cores and filaments, the line-of-sight temperature variations combined with the uncertainty of the dust properties can cause the column densities to be underestimated by even up to a factor of ten \citep{Malinen2011}. Because these effects depend the column density, they can have systematic effects also on the filament profiles that are derived via modified blackbody fits. This will again affect more the individual profile parameters, while the FWHM estimates will be more robust \citep{Juvela2023}.

\section{Conclusions}
We have analyzed the main OMC-3 filament using continuum \textit{Herschel} (\herschel) and \artemis\, data. The \textit{Herschel} and \artemis\, data have been combined to provide a higher-resolution image (\feather).  Filament morphology as well as fragmentation have been studied. We have further observed two other fields with \textit{Herschel}, G17 and G202, located at larger distances of 1.8\,kpc and 760\,pc, respectively. 
\begin{enumerate}
	\item Line masses of the examined filaments are in the range 103--310\,\msun\,pc$^{-2}$. Assuming only thermal support, all filaments are gravitationally unstable. With the addition of turbulence, G202 and G208 \feather\, are stable against gravity. 	All filaments show some asymmetry, but only G202 and G208 \feather\, has significant asymmetry across most of the filament. 
	\item The relation between filament \textit{FWHM} and instrumental resolution can be fit with by $ FWHM \approx 6.7 \times\textrm{ }HPBW\textrm{[pc]},  $  suggesting that telescope resolution can affect derived properties. We also find an increase in \textit{FWHM} with distance in simulated ideal filaments (Appendix \ref{sec:app_simulations_Plummer_distance}). However, in the densest filament segments of OMC-3 studied in \citet{Juvela2023}, with the highest signal relative to the background, filament \textit{FWHM} is relatively independent of resolution.  
	\item We do not detect significant correlation between filament \textit{FWHM} and column density. 
	\item Widths of \feather\, filaments in OMC-3  are $\sim$ 0.05\,pc, half of the 0.1\,pc typical width observed in many \textit{Herschel} filaments. Meanwhile, \textit{FWHM} of \herschel\,  in OMC-3 filament segments is  $\sim$0.1\,pc. Hierarchical structure within the ISM will result in filament-like structures visible at many scales and may in part explain the dependence on the resolution of the observations. Further contributions to this relation likely come from convolution of large-scale structures within the ISM. Models show that large-scale fluctuations in the background can increase derived \textit{FWHM}.

	\item Compared to \textit{Herschel}, \artemis\, probes denser structures, which have higher column density and mass per area but smaller physical size. 	Most of the clumps detected within OMC-3 are not gravitationally bound, with the exception of two dense clumps detected in the \feather\, map. Effective Jeans length in OMC-3 is $\sim$0.1\,pc. 
	\item A higher number of structures are visible at small scales in the \feather\, map, and the median separation of \feather\, structures at scales $\leq $ 0.05\,pc is below the Jeans length. In the \herschel\, map, the median separation between substructures at small scales is $\sim$0.1--0.2\,pc. Furthermore, median separation for clumps detected in the \feather\, data of (0.1$\pm$0.1)\,pc is within $1 \sigma$ of the Jeans length, but \herschel\, clumps have median separation of 0.3\,pc. 
	\item We have analyzed how the derived Plummer parameters depend on the way the fits are done. 
	\begin{enumerate}
		\item Asymmetric fits, in which the two sides of the filament profile are fit with different \R\, and \textit{p} values, were compared with symmetric fits. Mean values of the parameters do not generally differ by more than $1 \sigma$. Although individual \R\, and \textit{p} parameters can have large variation up to a factor of 2, the derived \textit{FWHM} is fairly similar. 
		\item Major differences in Plummer parameters are not visible when fitting the mean filament profile, compared to fitting individual filament cross sections. This difference increases in cases where the filament is weak compared to the background. Once again, \textit{FWHM} is robust against changes in \R\, and \textit{p}. 	
		\item Allowing the peak of the filament to shift along the x-axis results in slightly higher \textit{FWHM}. The effect is less than 20\% in \textit{q} = 1 filaments but can rise to almost a factor of two in low-quality filaments ($q>2$).
	\end{enumerate}
\end{enumerate}

\begin{acknowledgements}

E.M. is funded by the University of Helsinki doctoral school in particle physics and universe sciences (PAPU). MJ acknowledges support from the Research council of Finland grant No. 348342.\\

\textit{Software:} APLpy \citep{aplpy}, Matplotlib \citep{matplotlib}, Astropy \citep{astropy}, Numpy \citep{numpy}, uvcombine , Montage \citep{montage,montage_wrapper}, scipy \citep{scipy}

\end{acknowledgements}

\bibliographystyle{aa} 
\bibliography{/home/local/emannfor/PAPERS/EM} 

\begin{appendix}

\section{Simulations of Plummer fitting  \label{sec:app_simulations_Plummer} }

We tested the accuracy of Plummer fitting and derived parameters by varying the SNR, background, and distance (and therefore resolution) of simulated filaments. The initial parameters used to create the Plummer profile in these simulations are listed in Table \ref{tbl:par0}.

\begin{table}
	\caption{Initial Plummer parameters used in these simulations. }
	\label{tbl:par0}
	\begin{tabular}{lll}
		\hline
		
		Parameter & Unit \\
		\hline
		\hline
		$^\alpha N_{\rm 0}$ & & 1.0 \\
		\R & (pix) & 6.0 \\
		\textit{p}  & & 2.0  \\
		 \textit{a} & & 0.05  \\
		 \textit{b} & (pix) & 2$\times 10^{-4}$ \\
		 $^\beta$\textit{c} & (pix$^2$) & -2$\times 10^{-4}$   \\
		 $^\gamma \Delta x$ & (pix) & 0.0  \\
		 $^\delta$ \textit{FWHM} & (pix) & 20.78 \\

		\hline
	\end{tabular}
	\tablefoot{ Initial asymmetric Plummer profile models assume $R_{\rm L} = R_{\rm R} = R$ and  $p_{\rm L} = p_{\rm R} = p$. \\ $^\alpha$Sect. \ref{sec:app_simulations_Plummer_BG_model} uses $N_{\rm 0} = 4.0.$ \\ $^\beta$The variable \textit{c} is used in the polynomial background of the Plummer model in Section~\ref{sec:app_simulations_Plummer_BG_model}. \\  $^\gamma$Offset $\Delta x$ is not used in the tests in Sections~\ref{sec:app_simulations_Plummer_SNR} or \ref{sec:app_simulations_Plummer_BG_strength}. \\ $^\delta$ \textit{FWHM} is calculated from the values of \R\, and \textit{p} using Eq.~(\ref{eq:FWHM}).}
\end{table}

\subsection{SNR variation  \label{sec:app_simulations_Plummer_SNR}  }

Noise was added to the Plummer model using
\begin{align*}
Y_{\rm 0} &= Y  +  \dfrac{\textrm{max}(Y)}{SNR}N(0,1)
\end{align*}
where \textit{N}(0,1) are normally distributed random numbers, and \textit{Y} is the (symmetric) Plummer function (Eq. \ref{eq:PlummerBG}). 
We tested SNRs of 2, 5, 10, 25, 50, and 100. This Plummer model was convolved using a 1-D convolution. We fit the Plummer function to $Y_{\rm 0}$ using scipy's leastsq function. The derived values of \textit{FWHM}, \textit{R}, and \textit{p} are shown in Fig. \ref{fig:plummer_fit_SNR}. Already at SNR 25 the least squares fitting routine recovers the Plummer filament well, with mean estimated \textit{FWHM} being within 1\% of the \textit{FWHM} of the original profile. 

\begin{figure}[h]
	\includegraphics[width=\linewidth]{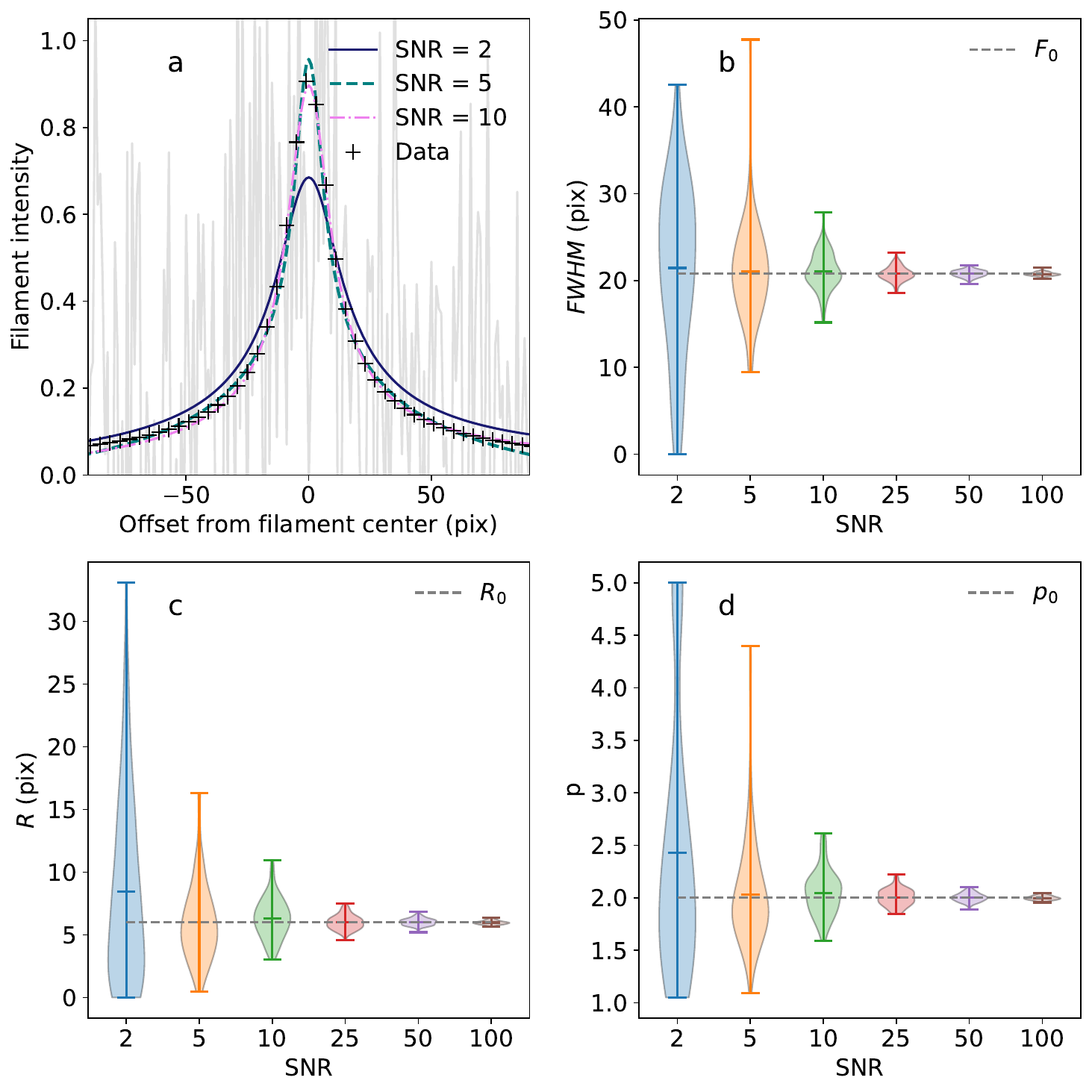}
	\caption{ Effect of varying SNR on Plummer filament fitting. (a) Profile showing the original Plummer without noise (black plusses), the fit with SNR = 2 (black solid line), SNR = 5 (teal dotted line), and SNR = 10 (dash-dotted line). The simulated filament profile assuming \textit{SNR} = 2 is plotted with faint gray. (b--d) violin plots of derived \textit{FWHM}, \textit{R}, and \textit{p} with increasing SNR. The dotted line corresponds to the original Plummer parameter, \textit{R} = 6.0 and \textit{p} = 2.0.   \label{fig:plummer_fit_SNR} }
\end{figure}

\subsection{Background variation  \label{sec:app_simulations_Plummer_BG}  }
We have tested the effect of the background on the derived Plummer model. In Sect. \ref{sec:app_simulations_Plummer_BG_strength} we test how the relative strength of the filament compared to the background sky affects derived Plummer parameters. In Sect. \ref{sec:app_simulations_Plummer_BG_model} we compare Plummer fits to a simulated filament with a linear or polynomial background component, and in Sect. \ref{sec:app_simulations_Plummer_BG_data} we test a polynomial fit on our four fields.

\subsubsection{Fluctuations of the background sky  \label{sec:app_simulations_Plummer_BG_strength}  }
We studied the effect of background sky fluctuations on derived filamentary properties. We assume that the background has a spatial power spectrum that is a powerlaw with an exponent of -2 (similar to the background powerlaw shape in OMC-3), and convolve it using a 2-D convolution. We varied the ratio of the filament peak divided by rms value of the background fluctuations, calling this property $SNR_{\rm filament}$. The Plummer filament itself does not vary (except for the random effects due to white noise corresponding to \textit{SNR }= 100, as presented in Sect. \ref{sec:app_simulations_Plummer_SNR}). We fit the Plummer model using the same method as in Sect. \ref{sec:app_simulations_Plummer_SNR}. Two resulting 1D profiles, along with background sky, are shown in Fig. \ref{fig:plummer_fit_BG}b-c, frame b showing a filament with a relatively strong background and frame c showing a relatively weak background. An example of a simulated filament is shown in Fig. \ref{fig:plummer_fit_BG}a.

Even with background fluctuations which are 10\% of the filament peak, the mean of derived Plummer values stays fairly constant, though in a filament with lower SNR (Sect. \ref{sec:app_simulations_Plummer_SNR}), the background fluctuations would undoubtedly play a stronger role. Further, the exponent of the background powerlaw does not significantly affect the derived parameters. For filaments with $SNR_{\rm filament}\approx 10$, the \textit{FWHM} was 22.33 $\pm$ 5.53 pixels assuming \textit{k} = -2 and 25.07 $\pm$ 10.36 pixels assuming \textit{k} = -3. 

\begin{figure*}[h]
	\sidecaption
	\includegraphics[width=12cm]{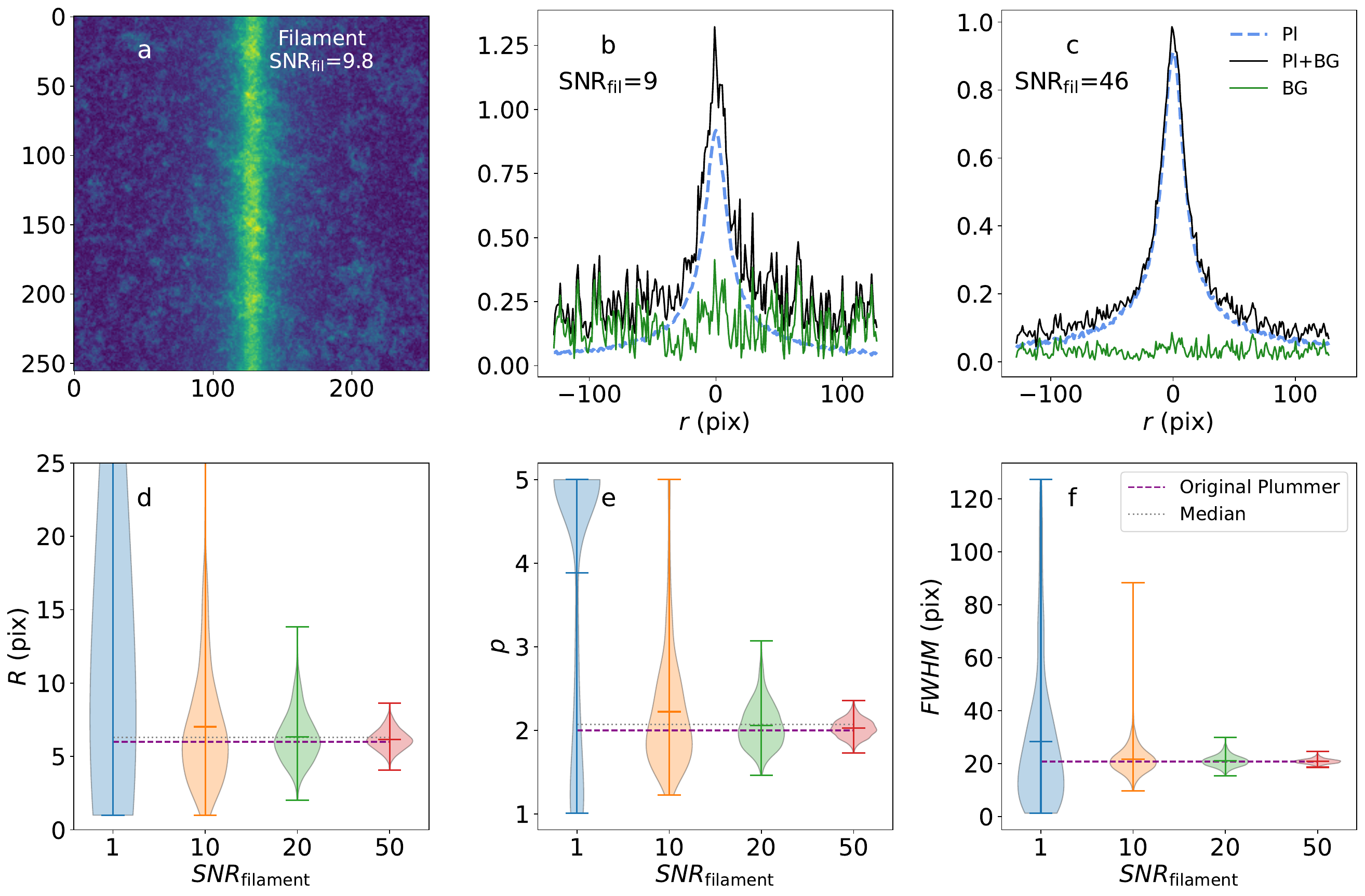}
	\caption{ Simulation of the effect of varying background level on Plummer filament fitting. (a): The generated filament on top of the background. (b,c) The results of the fit for two different levels of background noise, with the original Plummer filament plotted with blue dashed lines, the background with green solid lines, and the sum of the Plummer and background with black. Frame b shows a stronger background w.r.t. the filament.  (d,e,f) Violin plots of \textit{R} (frame d). \textit{p} (frame e), and \textit{FWHM} (frame f) showing the results of the Plummer fit. Background noise decreases when moving right. The horizontal lines show the original Plummer parameters (dashed lines), the median for our simulation (dotted line).  \label{fig:plummer_fit_BG} }
\end{figure*}

\subsubsection{Background model  \label{sec:app_simulations_Plummer_BG_model}  }

In addition to fluctuations of the background sky, the Plummer formula (Eq. \ref{eq:PlummerBG}) assumes a certain shape of the background component.  Which model is used to fit the background component can possibly also affect derived parameters. We have simulated two models. Both have \textit{SNR} = 100, and fairly weak background sky fluctuations ($SNR_{\rm filament} \approx 17$). The Plummer components ($N_{\rm 0}$, \R, \textit{p}, and $\Delta r$) are the same in both tests. We only vary the model of the  background component used to create the filament: in the first a linear model ($a + b r$) as in  in Eq.~(\ref{eq:PlummerBG}), and in the second a second-order polynomial  ($a + b r + c r^2$). The strength of this background component is also varied. We initially set \textit{a}, \textit{b}, and \textit{c} as in Table \ref{tbl:par0}, and then multiply \textit{b} and \textit{c} by \textit{m} = 0.1, \textit{m} = 1, and \textit{m} = 2. The parameter \textit{a} is kept constant.

To test how assuming a certain background model affects Plummer fits, both filaments were fit with a Plummer model which assumed a linear background. The first filament is thus correctly fitted with an appropriate background model. The second filament is incorrectly fitted with a linear background. We then compare the results of these fits, seeing how derived parameters differ.

Figure \ref{fig:compare_BG_model} shows the inputs and results of this simulation. Frames \textit{b} and \textit{e} show both the background sky fluctuation (dotted lines), and the background component (dash-dotted lines) for each fit. Median derived $N_{\rm 0}$, \R, \textit{p}, and \textit{FWHM} are listed in Table \ref{tbl:compare_linear_polynomial_BG}. The discrepancy between the two fits increases as the polynomial factor \textit{c} increases. $N_{\rm 0}$ is overestimated in most fits, possibly due to additional emission from the sky fluctuations.  At \textit{m} = 0.1, the derived values for both models are fairly similar. However, at \textit{m} $\geq$ 1, \R\, is significantly overestimated in filaments with a polynomial background component. As the polynomial grows larger, the shape of the filament becomes dominated by this component. Generally, \textit{p} has a wider range when fitting a linear profile, and is slightly underestimated. Meanwhile, if $m \geq$ 1, values of \textit{p} are overestimated in the polynomial profile. \textit{FWHM} is more consistent, but is slightly overestimated in the polynomial filament. In general, simulating a filament with a strong polynomial background creates another, more diffuse filamentary structure on top of the Plummer model, leading to increased width in fits.

\begin{table}
	\centering
	\caption{Results of simulations comparing fits to a Plummer filament with a linear or polynomial background. \label{tbl:compare_linear_polynomial_BG} }
	\begin{tabular}{llllll}	
		\hline\hline
		Variable & Original value & Linear BG & Polynomial BG \\ 
		& & component & component \\
		
		\hline
		
		\multicolumn{3}{l}{\textit{m} = 0.1} \\ 
		$N_{\rm 0}$ & 4.0 & 4.1$\pm$0.6 & 4.3$\pm$0.5 \\ 
		\R & 6.0 & 6.4$\pm$4.8 & 5.7$\pm$4.1 \\ 
		$p$ & 2.0 & 2.0$\pm$1.2 & 1.8$\pm$1.0 \\ 
		$FWHM$ & 20.8 & 21.7$\pm$3.8 & 23.2$\pm$4.0 \\ 
		\hline
		\multicolumn{3}{l}{m = 1} \\ 
		$N_{\rm 0}$ & 4.0 & 4.1$\pm$0.6 & 4.0$\pm$0.2 \\ 
		\R & 6.0 & 6.3$\pm$4.7 & 15.1$\pm$5.2 \\ 
		$p$ & 2.0 & 1.9$\pm$1.2 & 2.9$\pm$0.7 \\ 
		$FWHM$ & 20.8 & 21.5$\pm$3.9 & 31.7$\pm$3.8 \\ 
		\hline
		\multicolumn{3}{l}{m = 2} \\ 
		$N_{\rm 0}$ & 4.0 & 4.1$\pm$0.6 & 3.7$\pm$0.1 \\ 
		\R & 6.0 & 6.3$\pm$4.7 & 41.1$\pm$3.8 \\ 
		$p$ & 2.0 & 1.9$\pm$1.2 & 9.6$\pm$0.9 \\ 
		$FWHM$ & 20.8 & 21.6$\pm$3.9 & 34.9$\pm$1.7 \\ 

		\hline

	\end{tabular}
	\tablefoot{Values listed are the median $\pm$ standard deviation of the derived values. }
\end{table}

\begin{figure*}[h]
	\includegraphics[width=\linewidth]{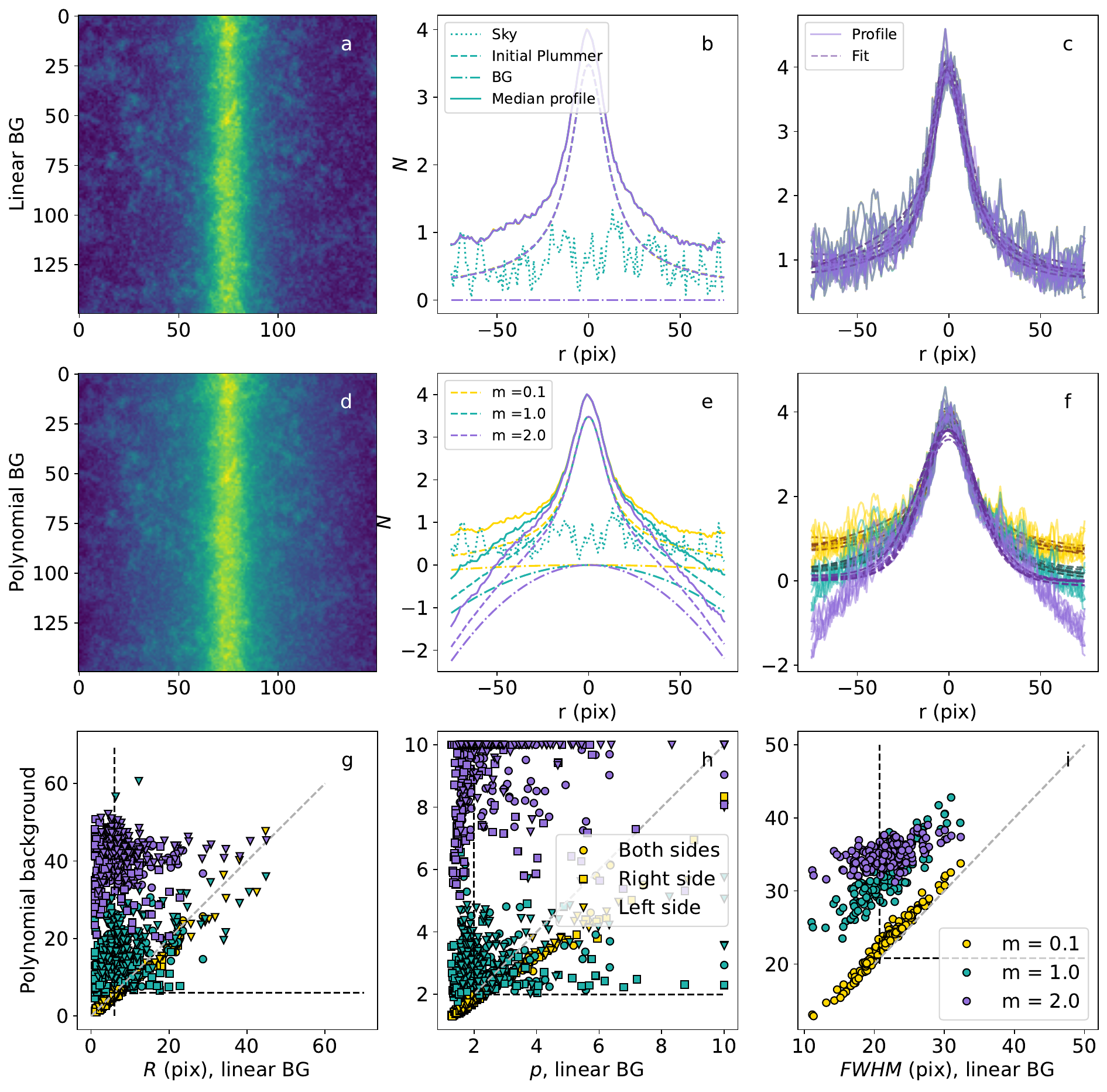}
	\caption{ Results of simulations comparing filaments with a linear or polynomial background component. (Top row): A simulated filament with a linear background. (Middle row): A simulated filament with a polynomial background. Both filaments also have random sky fluctuations with powerlaw index \textit{k} = -2. Frame a/d: The simulated filament, \textit{m} = 1.0. Frame b/e: Components of the filament at different values of \textit{m}: The sky (dotted line), initial Plummer (dashed line), the initial background component (dash-dotted lines), and the median profile (solid line). Yellow corresponds to the flattest background component (\textit{m} = 0.1), turquoise to \textit{m} = 1.0, and purple to the strongest background component (\textit{m} = 2.0). Frame c/f: Individual profiles (solid lines) and Plummer fits (dashed lines) to selected profiles in each map. Colors correspond to different strengths as in the previous frames. (Bottom row): \R\, (g), \textit{p} (h), and \textit{FWHM} (i) calculated for the profiles with a linear background (x-axis) and a polynomial background (y-axis). Colors are as in frame e.  The vertical and horizontal dashed lines mark the initial parameter value.   \label{fig:compare_BG_model} }
\end{figure*}

\subsubsection{Comparing fits to real data using a polynomial background model  \label{sec:app_simulations_Plummer_BG_data}  }
We have tested a Plummer fit to our four fields, but assuming a polynomial background component. As priors in our fit, we use the derived parameters from Table \ref{tbl:PlummerResults}, and initially assume the polynomial term \textit{c} = 0.0. Median profiles for \textit{q} = 1--2 profiles are shown in Fig. \ref{fig:mean_filament_profiles_POLY}, and median values in Table \ref{tbl:PlummerResults_POLY}. (For linear fits, the profiles are shown in Fig. \ref{fig:mean_filament_profiles}.) The profile widths in G17 differ by up to 70\% in \textit{q} = 2 filament segments. This discrepancy is only about 20\% in G202. The parameter \textit{p} is lower in polynomial fits to G17, but significantly higher in G202.  

In contrast, OMC-3 can be fit equally well by a linear and polynomial background. Median values in G208 \feather\, are nearly the same in both fits. $N_{\rm 0}$ in G208 \herschel\, differs by no more than 4\%. \R\, differs by 20\% (0.01\,pc) and \textit{p} by less than 9\%. The differences between the two fits are all within the 1-$\sigma$ standard deviation. In fits using \textit{Spitzer} extinction, as well as \textit{Herschel} and \artemis\, data of filament segments in OMC-3, \citet{Juvela2023} also do not see notable difference in \R, \textit{p}, or \textit{FWHM}.

\begin{figure*}[h]
	\sidecaption
	\includegraphics[width=12cm]{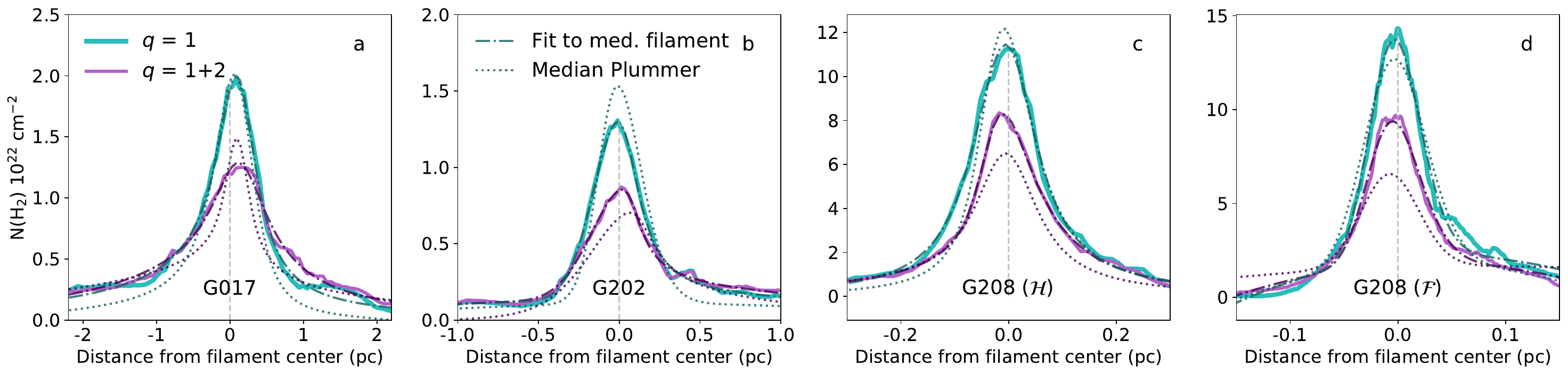}
	\caption{ Same as Fig. \ref{fig:mean_filament_profiles}, but for fits using a polynomial background.  \label{fig:mean_filament_profiles_POLY} } 
\end{figure*}

\begin{table*}
	\centering
	\caption{Median Plummer parameters calculated assuming a linear or polynomial background. }
	\label{tbl:PlummerResults_POLY}
	\begin{tabular}{llllllllll}
		\hline\hline
		Field & Qual & $N_{\rm 0, l}$  & $N_{\rm 0, p}$ & $R_{\rm flat, l}$  & $R_{\rm flat, p}$ & $p_{\rm l}$  & $p_{\rm p}$  &  $FWHM_{\rm l}$  & $FWHM_{\rm p}$   \\  
		& & ($10^{22}$\,cm$^{-2}$) & ($10^{22}$\,cm$^{-2}$) &  (pc) &  (pc) & & &  (pc) &  (pc) \\ 
		\hline
		G017 & 1 & 2.1$\pm$0.4 & 2.2$\pm$0.2 & 0.48$\pm$1.39 & 0.30$\pm$0.73 & 4.19$\pm$8.86 & 3.15$\pm$9.05 & 0.87$\pm$0.72 & 0.66$\pm$0.31 \\ 
		G017 & 2 & 1.2$\pm$0.3 & 2.0$\pm$0.0 & 1.03$\pm$1.67 & 0.04$\pm$0.53 & 5.69$\pm$9.03 & 1.58$\pm$2.17 & 1.29$\pm$0.76 & 0.32$\pm$0.44 \\ 
		\\
		
		G202 & 1 & 1.6$\pm$0.6 & 1.5$\pm$0.6 & 0.17$\pm$0.62 & 0.55$\pm$0.60 & 3.28$\pm$10.90 & 25.00$\pm$10.82 & 0.31$\pm$0.32 & 0.31$\pm$0.18 \\ 
		G202 & 2 & 0.8$\pm$0.2 & 0.7$\pm$0.2 & 0.18$\pm$0.76 & 0.37$\pm$0.80 & 2.11$\pm$7.61 & 5.60$\pm$9.92 & 0.61$\pm$0.46 & 0.51$\pm$3.25 \\ 
		\\
		
		G208 (\herschel) & 1 & 13.7$\pm$10.4 & 13.9$\pm$10.3 & 0.04$\pm$0.11 & 0.05$\pm$0.12 & 2.75$\pm$7.53 & 2.71$\pm$7.47 & 0.11$\pm$0.03 & 0.11$\pm$0.03 \\ 
		G208 (\herschel) & 2 & 6.8$\pm$1.4 & 7.1$\pm$1.3 & 0.05$\pm$0.18 & 0.05$\pm$0.09 & 2.57$\pm$2.61 & 2.35$\pm$3.72 & 0.15$\pm$0.09 & 0.14$\pm$0.05 \\ 
		\\
		
		G208 (\feather) & 1 & 15.5$\pm$3.3 & 15.5$\pm$3.3 & 0.09$\pm$0.03 & 0.09$\pm$0.03 & 25.00$\pm$2.71 & 25.00$\pm$2.20 & 0.05$\pm$0.01 & 0.04$\pm$0.01 \\ 
		G208 (\feather) & 2 & 7.6$\pm$1.5 & 7.6$\pm$1.5 & 0.08$\pm$0.03 & 0.08$\pm$0.03 & 25.00$\pm$7.59 & 25.00$\pm$4.48 & 0.04$\pm$0.01 & 0.04$\pm$0.01 \\ 

		\hline
	\end{tabular}
	\tablefoot{Median$\pm$ standard deviation of $N_{\rm 0}$, \R, \textit{p}, and \textit{FWHM} calculated for linear (subscript \textit{l}) and polynomial (subscript \textit{p}) fits. \R\, and \textit{p} are the median of both sides of the Plummer fit. Note: The discrepancy in \textit{q} = 2 linear-fit values between this table and Table \ref{tbl:PlummerResults} is due to the values in Table \ref{tbl:PlummerResults} being the value within the combined \textit{q} = 1 and 2 profiles.  }
\end{table*}

\subsection{Distance variation\label{sec:app_simulations_Plummer_distance} }

We further studied the effect of distance on derived Plummer parameters using a simulated filament, once again with SNR = 100 and relatively weak background fluctuations, and \textit{FWHM} = 20.78\,pixels (corresponding to a pixel size of 4.8$\times10^{-3}$\,pc assuming a physical FWHM width of 0.1\,pc). To simulate hierarchical structure within the ISM, we further added a wide Gaussian component with $\sigma = 125$\,pixels (0.6\,pc), half the original mapsize. This component is fainter and significantly wider than the main Plummer filament.  We also added eight small spherical Gaussian clumps of width $\sigma = 5-12$\,pixels to simulate smaller-scale hierarchical structure. The clumps are described by the formula $\exp\Big(   -4\ln (2) \times \frac{ (x-x_{\rm 0})^2 + (y-y_{\rm 0})^2 }{ \sigma^2 }  \Big),$ where \textit{x} and \textit{y} are the coordinates of each pixel, $x_{\rm 0}$ and $y_{\rm 0}$ the central pixel coordinates of the clump, and $\sigma$ the width of the Gaussian clump. The peak intensity of each clump is 1.0 before regridding. 

The original filament was convolved to distances of 250\,pc, 500\,pc, 1\,kpc, 2\,kpc, and 4\,kpc. The map was also reprojected to simulate increasing pixel size, thus, the size of the map in pixels also decreases with increased distance. The background sky, filament, noise, wide component, and clumps were generated at the beginning of the simulation and thus do not change during one run. 
The results of this simulation are shown in Fig. \ref{fig:plummer_fit_distance_simulation}. The first row shows the filament at different distances, the second row the profile (with all components) and fitted Plummer model, and the bottom violin plots of for the estimated $R$, $p$, and $FWHM$. Given an assumed intrinsic width of 0.1\,pc, we estimate a relation between FWHM and distance of $FWHM \approx 1.7\times 10^{-4}\cdot  d\textrm{ [pc]} + 0.10,$ (Fit 1)
 a shallower relation that that derived in this paper. When we also allow the (fitted) intrinsic width to vary, this results in a fit of $FWHM \approx 1.1\times 10^{-4}\cdot  d$ + 0.25 (Fit 2). This larger initial width is because of the blending of the large-scale Gaussian background structure with the Plummer filament.

We further perform this simulation with and without large-scale structures or small clumps, and by assuming \textit{SNR} = 10 and \textit{SNR} = 100, for a sample of 20 random realizations. We list the results of these tests in Table~\ref{tbl:D_FWHM_fits_simulations}. 
Varying the \textit{SNR} between 10 and 100, but keeping the other parameters the same, changes the \textit{FWHM} at 2\,kpc by only (1.4$\pm$0.7)\%. Whether Gaussian clumps are included or not changes \textit{FWHM} at 2\,kpc by (7.7$\pm$1.8)\%, and in fact small-scale structures slightly compensate for the increase in \textit{FWHM} due to distance. In contrast, the difference in \textit{FWHM} at 2\,kpc is around (42.8$\pm$2.8)\% larger when large-scale background structures are included. However, even in tests without large-scale structure or clumps, \textit{FWHM} can increase by up to 3$\times FWHM_{\rm True}$ at 2\,kpc. In the final three columns of Table~\ref{tbl:D_FWHM_fits_simulations} we list the slope, y-intercept value (i.e. the \textit{FWHM} at \textit{d} = 0\,pc), and calculated \textit{FWHM} at 2\,kpc for Fit 2. Though the initial simulated Plummer filament has a width of 0.1\,pc, in fits which include large-scale structures, the fitted intrinsic width is 0.2--0.3\,pc. In contrast, the four fits which do not include these large Gaussian structures have fitted intrinsic widths of 0.12-0.13\,pc.  

\begin{table*}
	\centering
	\caption{  The derived distance-\textit{FWHM} relation and \textit{FWHM} at 2\,kpc for the different simulations.   }
	\label{tbl:D_FWHM_fits_simulations}
	\begin{tabular}{lll|lll|lll}
		\hline\hline
		\textit{SNR} & Contains  & Contains &  \multicolumn{3}{c|}{Fit 1}    &  \multicolumn{3}{c}{Fit 2}  \\
		 &  clumps? &  large-scale  & slope & intercept   &   $\Delta F$ &  slope & intercept   &    $\Delta F$ \\
		 &   &   structure? & [$\times 10^{-4}$ pc] & [pc]    &   [pc] &   [$\times 10^{-4}$ pc] &  [pc]    &  [pc] \\
		\hline

		100 & \checkmark & \checkmark & 1.67$\pm$0.13 &  0.10 & 4.3$\pm$0.3 &  1.11$\pm$0.13 & 0.25$\pm$0.02 & 4.8$\pm$0.3\\  
		10 & \checkmark & \checkmark & 1.64$\pm$0.10 &  0.10 & 4.3$\pm$0.2 &  1.14$\pm$0.09 & 0.24$\pm$0.02 & 4.7$\pm$0.2\\  
		100 & - & \checkmark & 1.85$\pm$0.13 &  0.10 & 4.7$\pm$0.3 &  1.18$\pm$0.13 & 0.29$\pm$0.02 & 5.2$\pm$0.3\\  
		10 & - & \checkmark & 1.80$\pm$0.15 &  0.10 & 4.6$\pm$0.3 &  1.15$\pm$0.17 & 0.28$\pm$0.02 & 5.1$\pm$0.3\\  
		100 & - & - & 1.03$\pm$0.04 &  0.10 & 3.1$\pm$0.1 &  0.93$\pm$0.05 & 0.13$\pm$0.00 & 3.1$\pm$0.1\\  
		10 & - & - & 1.02$\pm$0.05 &  0.10 & 3.0$\pm$0.1 &  0.91$\pm$0.06 & 0.13$\pm$0.01 & 3.1$\pm$0.1\\  
		100 & \checkmark & - & 0.97$\pm$0.04 &  0.10 & 2.9$\pm$0.1 &  0.90$\pm$0.04 & 0.12$\pm$0.01 & 3.0$\pm$0.1\\  
		10 & \checkmark & - & 0.97$\pm$0.06 &  0.10 & 2.9$\pm$0.1 &  0.90$\pm$0.06 & 0.12$\pm$0.01 & 3.0$\pm$0.1\\  

		\hline
	\end{tabular}
	\tablefoot{ Fit 1 assumes an intrinsic 0.1\,pc width, Fit 2 leaves the intrinsic width as a free parameter. Values are the mean $\pm$ standard deviation of a sample of 20 random realizations. Increase in \textit{FWHM} is defined as $\Delta F = FWHM_{\rm (d=2\,kpc)}$ / $FWHM_{\rm True}$, where $FWHM_{\rm True}= 0.1$\,pc.   }
\end{table*}

It is clear from these simulations that at least part of the distance relation could be due to the presence of hierarchical and large-scale  structures. Furthermore, as real filaments such as those observed in the main paper are more complex than these ideal, straight filaments they may be further artificially widened. This is a likely explanation for the steeper relation found for our data, when compared with these simulations.

\begin{figure*}[h]
	\includegraphics[width=\linewidth]{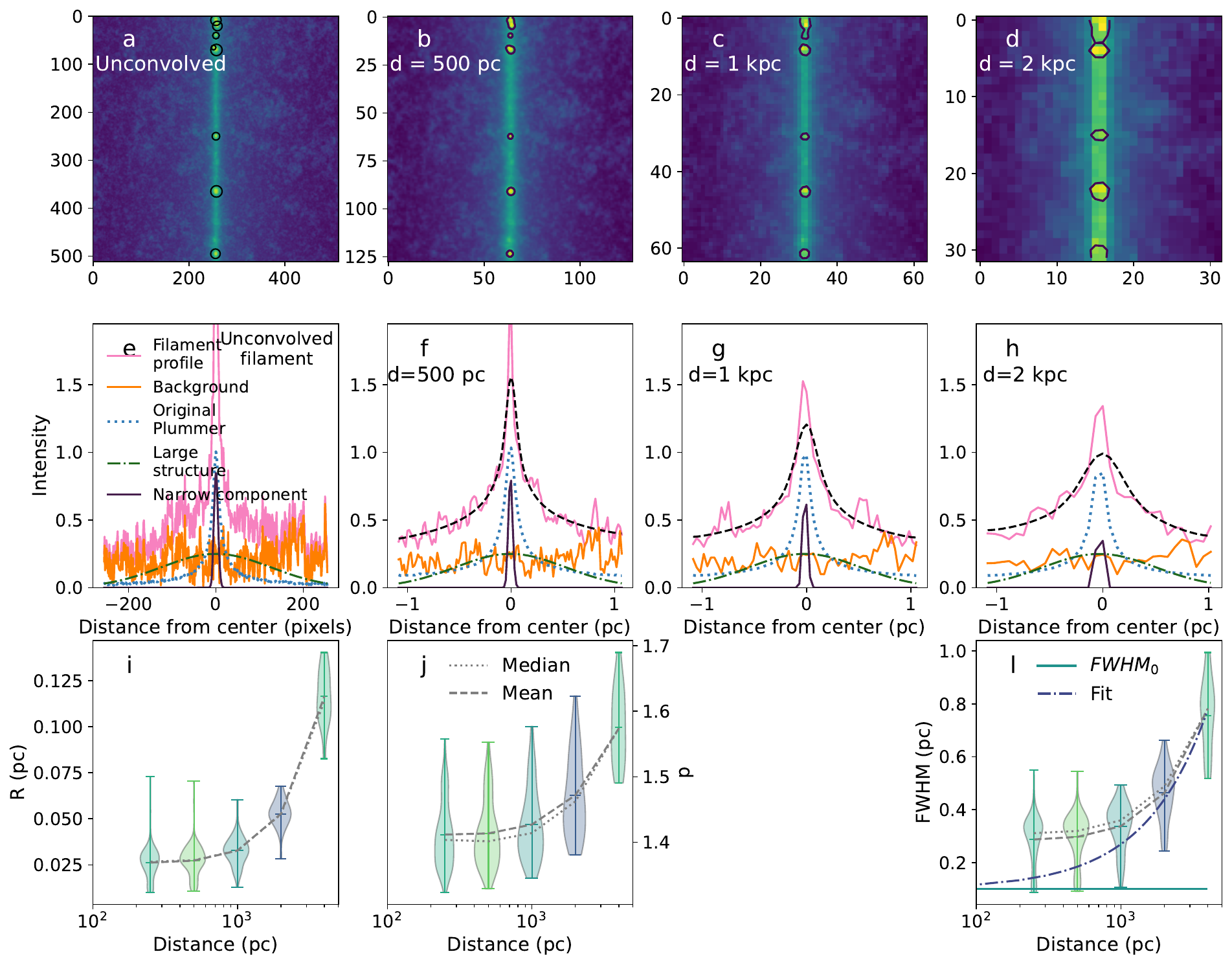}
	\caption{ A simulated ideal filament observed at various distances. (top row): The simulated observation of the filament at the original resolution, and 500, 1000, and 2000\,pc. Added Gaussian clumps are emphasized with black outlines. (Middle row): The components of the observation at different distances. The background (orange solid lines), the original Plummer (blue dotted line), the wide Gaussian structure  to simulate hierarchical structure (green dash-dotted lines), the mean small-scale Gaussian clump profile, and the resulting profile (pink solid lines). Frame e shows the original profile at the initial location of 125\,pc, frames f-h the profiles from \textit{d} = 500\,pc to \textit{d} = 2\,kpc. The Plummer fit to the filament profile is shown with black dashed lines. (Bottom row): Violin plots of derived \R\, (frame i), \textit{p} (frame j), and \textit{FWHM} (frame l). The mean and median values at each distance are plotted with gray lines. On frame l, we plot the intrinsic \textit{FWHM} = 0.1\,pc (teal solid line), as well as the fit where the \textit{FWHM} of the fitted Plummer is fixed to 0.1\,pc (blue dash-dotted line).  \label{fig:plummer_fit_distance_simulation} }
\end{figure*}

\subsection{ Range of offsets from the  filament center}

Previous studies have suggested that fitted parameters can change significantly depending on what range of offsets from the filament center Gaussian fitting is performed \citep{Panopoulou2017}. We tested this for Plummer fits using one of our fields, G208 \herschel. The original data use 100 pixels on either side of the filament center, corresponding to $\sim$ 1.16\,pc in physical space (for a total map width of 2.5\,pc). The map size is decreased to 75, 50, and 25 pixels per side (corresponding to $\sim$ 0.88, 0.59, and 0.30\,pc, respectively). Plummer parameters are calculated with asymmetric fits, and the results are shown in Fig. \ref{fig:distance_from_center}. We do not distinguish between filament quality flag in this test.

Unlike in \cite{Panopoulou2017}, we do not find any difference in derived Plummer parameters with increased offset from filament center. Even when reducing the fitted area to $\sim$0.12\,pc, at which point the full filament width is not sampled, the derived \textit{FWHM} stays remarkably consistent, though individual \textit{p} and \R\, parameters begin to vary. In the case of a more complex, multiple-filament structure such as that in G202, a slight dependence between fitting area and \textit{FWHM} is observed.

\begin{figure*}[h]
	\includegraphics[width=\linewidth]{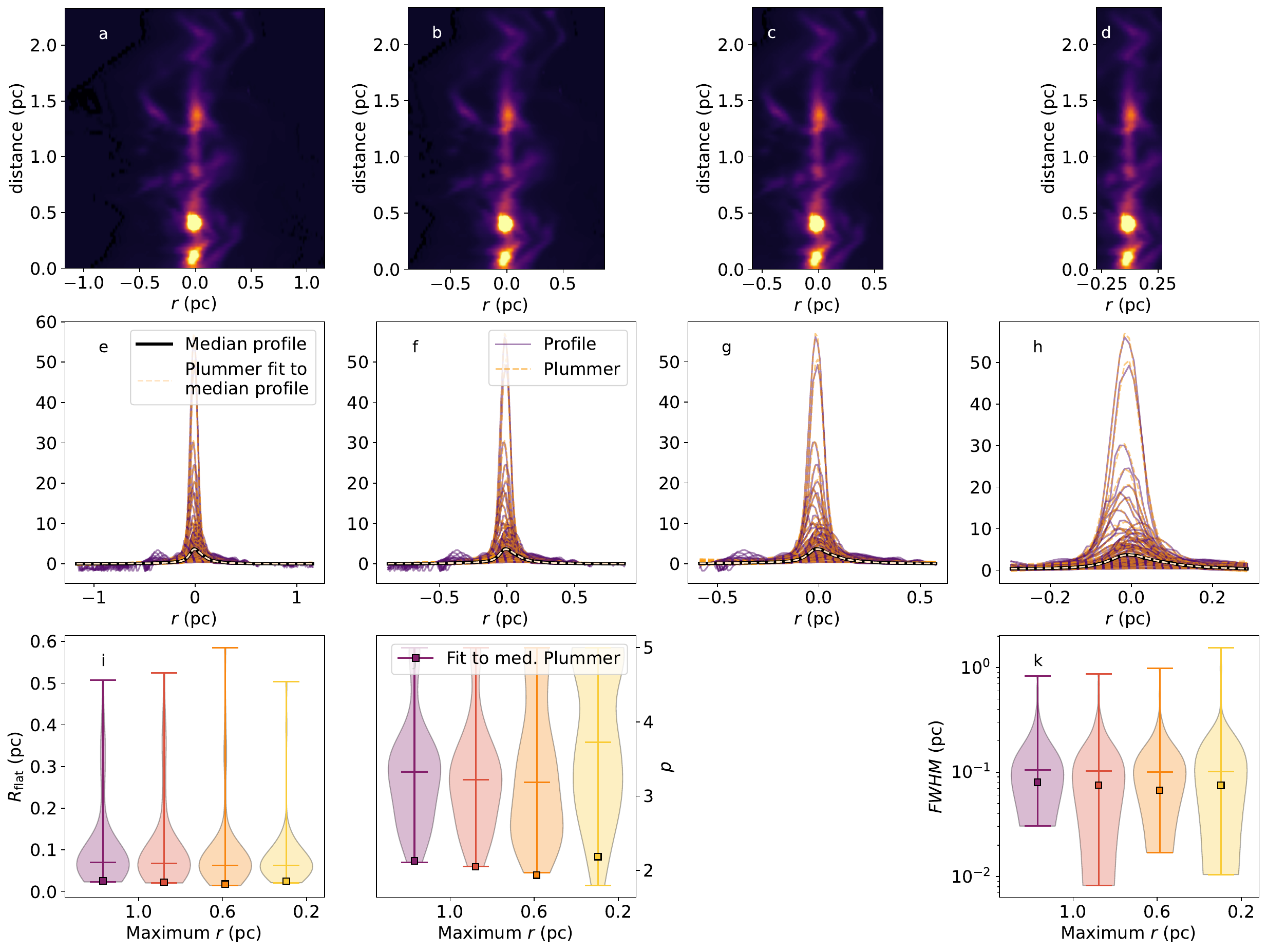}
	\caption{ Simulation of the relation between fitted area from filament center and derived Plummer parameters. Top row (a--d) shows the data, with the original image in frame a. Middle row (e--h) show every profile in purple and the fitted Plummers in orange. The median profile is shown in black solid lines and the Plummer fit to the median Plummer in pale yellow dashed lines. The bottom row (i--k) show violin plots of filament width \R\, (frame i), slope \textit{p} (frame j), and \textit{FWHM} (frame k). The squares show the value for the parameter in the fit to the median profile (yellow line in the middle row). Sharp cut-offs at low values represent physical bounds set by the fitting, and are the same for all runs of this simulation.  \label{fig:distance_from_center} }
\end{figure*}

\subsection{Comparison of 1-D and 2-D convolution \label{sec:app_compasion1D2D}    }

In this paper we have mainly used 1-D convolution, in which each Plummer profile is convolved using a Gaussian. However, it is also possible to perform Plummer fitting by convolving the entire image with a 2-D Gaussian. While there is no difference between a Gaussian convolved by  Gaussian in one or two dimensions, the same is not necessarily true for a Plummer function. We thus perform tests of a simulated filament with \textit{SNR} = 100 and convolution kernel $\sigma$ = 3.0, and fit symmetric Plummer models both with 1-D and 2-D convolution.

We perform this test with three filaments of varying background fluctuation, shown in Fig. \ref{fig:compare_1D2DA}, frames d, h, and l. The median profiles for the original filament components are shown in the first column, the 1-D fit in the second in purple, and the 2-D fit in the third in gray. Derived Plummer parameters are listed in Table \ref{tbl:compare1D2D}. Violin plots of derived \R, \textit{p}, and \textit{FWHM} are shown in Fig. \ref{fig:compare_1D2DB}. 

As seen in the previous section, increasing filament strength results in decreased spread in Plummer parameters. The spread in 1-D convolution is larger, as adjacent rows are not so well correlated. In fields with $SNR_{\rm filament} > $8, the median values are very similar. This is especially true for \textit{FWHM}, in which the median values differ by under 10\%. Due to its faster computing time (Fig. \ref{fig:time_taken}), the 1-D convolution is a viable choice.

\begin{figure*}
	\sidecaption
	\includegraphics[width=12cm]{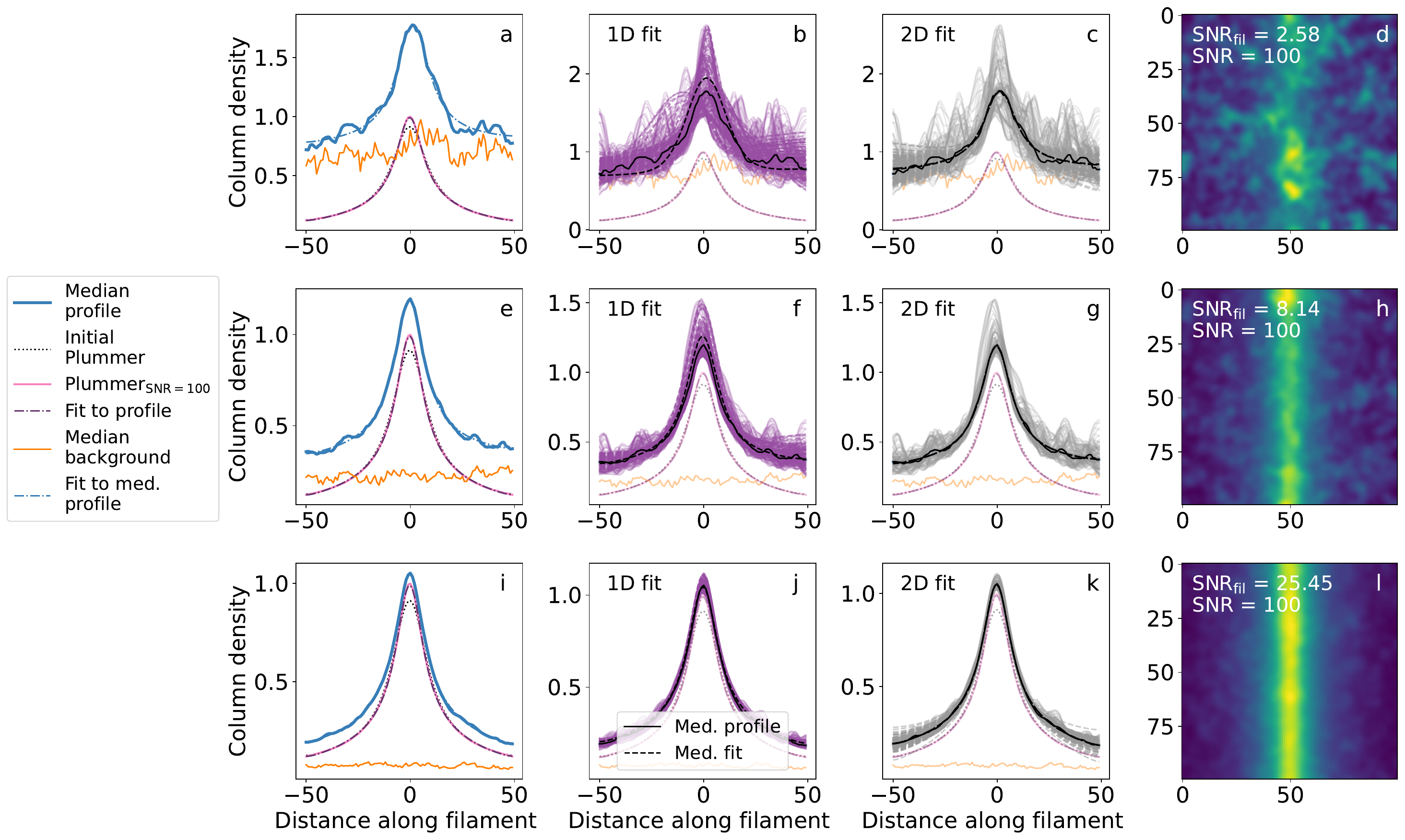}
	\caption{ Plummer fits using 1-D and 2-D convolution to filaments of three different strengths.  (First column)  Components of the simulated filament. Median profile (solid blue line), original Plummer without noise (black dotted line) and with noise (solid pink line), the Plummer fit to only the Plummer with noise (dash-dotted line), the background sky (orange solid line), and the Plummer fit to the full simulated filament (blue dash-dotted line). (Second column) Individual profiles (solid lines) and 1-D Plummer fits (dashed lines) to each row of the profile. Median profile, and the median and mean Plummer fits are shown in black solid, dashed, and dash-dotted lines, respectively. (Third column) Same as the second column, but for the 2-D Plummer fit. (Fourth column) The simulated filament.  \label{fig:compare_1D2DA} }
\end{figure*}

\begin{figure*}
	\sidecaption
	\includegraphics[width=12cm]{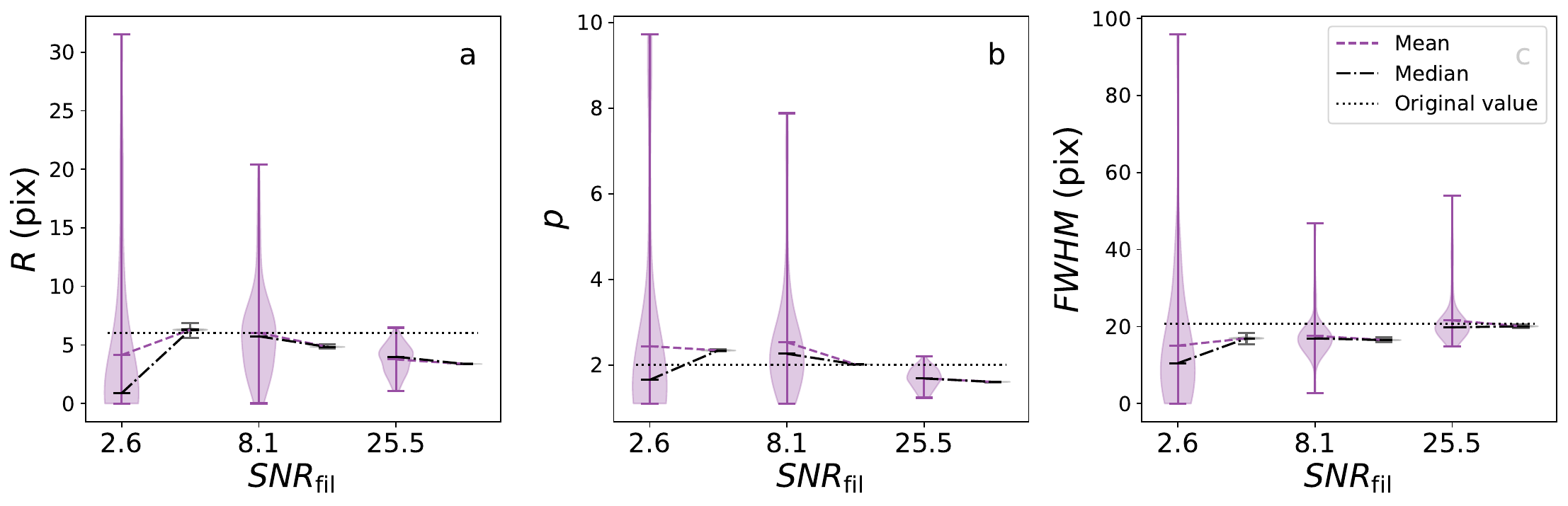}
	\caption{ Comparison of parameters derived for Plummer fits to filaments using 1-D and 2-D convolution. The 1-D Plummer fit is in purple and the 2-D in gray. Mean of each set is shown with purple dashed lines, and median with black dash-dotted lines. The original Plummer parameter is shown with the black dotted line.  \label{fig:compare_1D2DB} }
\end{figure*}

\begin{figure}
	\resizebox{\hsize}{!}{\includegraphics{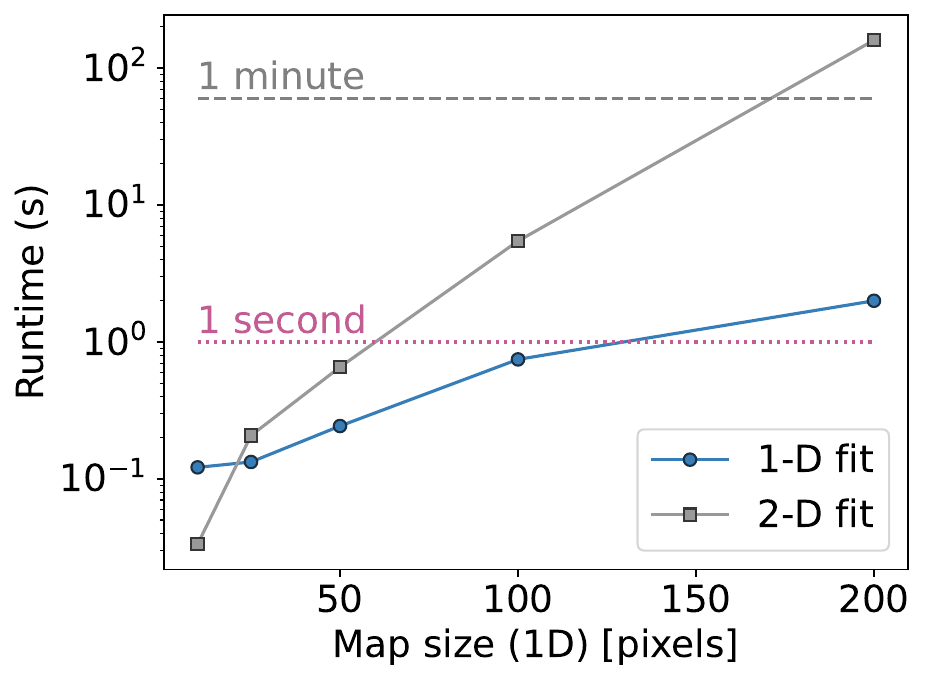}}
	\caption{   Time taken to run 1-D (blue circles) and 2-D (gray squares) Plummer analysis of maps of various sizes. These maps have a high SNR, comparable to the third row in Fig. \ref{fig:compare_1D2DA}. 	  \label{fig:time_taken} }
\end{figure}

\begin{table*} 
	\centering
	\caption{ Results of the simulations comparing 1\,D and 2\,D Plummer fitting to a simulated filament.   }
	\label{tbl:compare1D2D}
	\begin{tabular}{llllllllllllll}
		\hline\hline
		$SNR_{\rm f}$ & Method & $N_{\rm 0}$  & $R_{\rm flat}$  & $p$  &  $a$  & $b$  &  $\Delta r$  & $FWHM$ \\
		\hline
		\hline

		 2.72  &  1D & 1.41  $\pm$  0.91 & 9.85  $\pm$  14.31 & 2.75  $\pm$  4.03 & 0.88  $\pm$  0.43 & 0.0  $\pm$  0.0 & 0.12  $\pm$  5.38 & 14.92  $\pm$  33.11  \\ 
		 2.72  &  2D & 1.41  $\pm$  0.02 & 9.86  $\pm$  0.27 & 2.74  $\pm$  0.03 & 0.88  $\pm$  0.01 & 0.0  $\pm$  0.0 & 0.13  $\pm$  0.05 & 21.69  $\pm$  0.57  \\ 
		 \hline
		 8.49  &  1D & 1.29  $\pm$  0.23 & 4.59  $\pm$  4.39 & 1.67  $\pm$  1.2 & 0.09  $\pm$  0.17 & 0.0  $\pm$  0.0 & 0.01  $\pm$  0.21 & 21.77  $\pm$  7.26  \\ 
		 8.49  &  2D & 1.29  $\pm$  0.0 & 4.59  $\pm$  0.02 & 1.67  $\pm$  0.0 & 0.09  $\pm$  0.0 & 0.0  $\pm$  0.0 & 0.01  $\pm$  0.01 & 24.01  $\pm$  0.22  \\ 
		 \hline
		 26.66  &  1D & 1.22  $\pm$  0.09 & 3.72  $\pm$  0.97 & 1.68  $\pm$  0.23 & 0.03  $\pm$  0.07 & 0.0  $\pm$  0.0 & 0.01  $\pm$  0.07 & 17.35  $\pm$  1.85  \\ 
		 26.66  &  2D & 1.22  $\pm$  0.0 & 3.72  $\pm$  0.01 & 1.68  $\pm$  0.0 & 0.03  $\pm$  0.0 & 0.0  $\pm$  0.0 & 0.01  $\pm$  0.02 & 19.39  $\pm$  0.08  \\ 

		\hline
	\end{tabular}
	\tablefoot{ Median Plummer parameters derived using 1\,D and 2\,D convolution of a simulated filament with $SNR$ = 100. Three $SNR_{\rm filament}$ values are tested.  } 
\end{table*}

\section{Additional tables \label{sec:app_tables}  }
Here we present additional tables and data used in this paper. Fig. \ref{fig:art_fields} shows the three other \artemis\, fields. Centers of the empty reference regions are listed in Table \ref{tbl:centers_of_empty_regions}, uncertainties on symmetric and asymmetric fits to an ideal filament in Tables \ref{tbl:errors_symmetric} and \ref{tbl:errors_asymmetric}, and median $SNR_{\rm filament}$ of each field in Table \ref{tbl:SNR_qualityFlag} for all quality flags. Finally, the results of wavelet analysis are shown in Table \ref{tbl:wavelet}.

\begin{figure*}[h]
	\sidecaption
	\includegraphics[width=12cm]{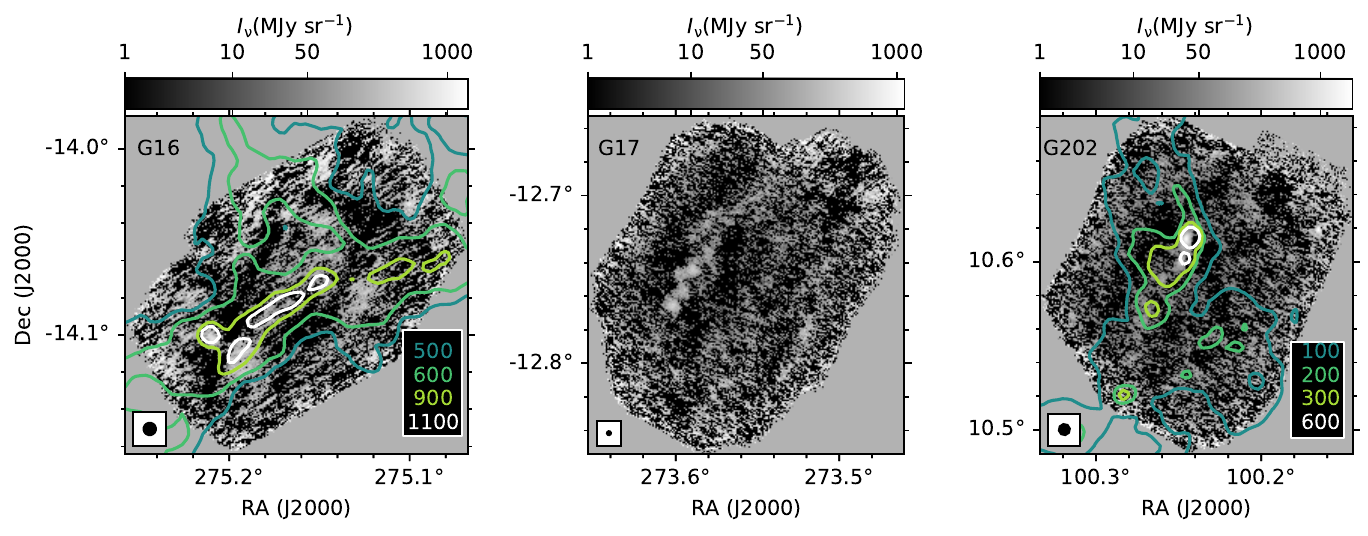}
	\caption{ \artemis\, 350\,\micro\, surface brightness maps (grayscale) of the G16, G17, and G202 fields. For G16 and G202, the contours show \textit{Herschel} 350\,\micro\, surface brightness, with contour levels (in MJy/sr) listed in the frames. The \textit{Herschel} beam is shown in the G16 and G202 frames, and the \artemis\, beam in the G17 frame.  \label{fig:art_fields} }
\end{figure*}

\begin{table}	
	\caption{Parameters of the \artemis\, observations }
	\label{tbl:art_obs_pars}
	\centering
	\begin{tabular}{lllll}
	\hline\hline
	Field & RMS noise & Max. signal & Max. SNR  \\
	& (MJy\,sr$^{-1}$) & (MJy\,sr$^{-1}$) &   \\
	\hline
	G16 & 254.6 & 1174.7	& 4.6 \\
G17 & 133.5 & 612.7	& 4.6 \\
G202 & 140.0 & 1970.6	& 14.1 \\
G208 & 181.7 & 12452.0	& 68.5 \\

	\hline
	\end{tabular}
\end{table}

\begin{table}
	\caption{Coordinates and radii of the reference regions used for noise estimation. }
	\label{tbl:centers_of_empty_regions}
	\centering
	\begin{tabular}{lllll}
	\hline\hline
	Field & ra & dec & radius & instrument \\
	& (\degr) & (\degr) & (\degr) &\\
	\hline
	G017.69-00.15 & 275.26 & -13.92 & 0.03 & \textit{Herschel} \\
	G202.16+02.64 & 100.39 & 10.65 & 0.03 & \textit{Herschel} \\
	G208.63-20.36 & 83.69 & -5.00 & 0.03 & \textit{Herschel} \\
	G16 & 275.16 & -14.07 & 0.02 & \artemis \\
	G17 & 273.53 & -12.73 &  0.02 & \artemis \\
	G202 & 100.22 & 10.59 &  0.02 & \artemis \\
	G208& 83.89 & -5.04 &  0.02 & \artemis \\
	\hline
	\end{tabular}
\end{table}

\begin{table}
	\centering
	\caption{Uncertainties on symmetric fits of a simulated filament.}
	\label{tbl:errors_symmetric}
	\begin{tabular}{lllll}
		\hline
		\hline	
		(1) & (2) & (3) & (4) \\
		Variable & Initial & Mean $\pm$ std & Median  \\
		& value & & \\
		
		\hline
		\R \,(pix) & 6.0 & 5.9 $\pm$ 0.25 & 5.9 \\
		\textit{p} & 1.75 & 1.73 $\pm$ 0.044  & 1.72  \\
		$FWHM$ (pix) & 27.8 & 28.3 $\pm$ 0.83  & 28.2  \\
		$\Delta r$  ($10^{-2}$ pix) & 0.0 & -0.36 $\pm$ 2.0 & -0.41  \\
		\textit{a} & 0.00 & -0.00 $\pm$ 0.020  & -0.00  \\
		\textit{b} ($10^{-3}$) & 0.2 & 0.2 $\pm$ 0.02  & 0.2  \\
		$N_{\rm 0}$ ($10^{22}$\,cm$^{-2}$) & 1.0 & 1.0 $\pm$ 0.02 & 1.0  \\
		\hline
	\end{tabular}
	\tablefoot{Plummer parameters of a simulated filament over a powerlaw background. The initial Plummer parameters are listed in column 2, the mean $\pm$ standard deviation recovered values in column 3, and the median recovered value in column 4.  }
\end{table}

\begin{table}
	\centering
	\caption{Uncertainties on asymmetric fits of a simulated filament.}
	\label{tbl:errors_asymmetric}
	\begin{tabular}{lllll}
		\hline
		\hline
		(1) & (2) & (3) & (4) \\
		Variable & Initial & Mean $\pm$ std & Median  \\
		& value & & \\
		
		\hline
		$R_{\rm L}$ (pix) & 6.0 & 6.0$\pm$  0.40 & 6.1  \\
		$R_{\rm R}$ (pix) & 6.0 & 5.9 $\pm$ 0.41 & 5.9  \\
		$\langle R_{\rm flat}\rangle$ (pix) & 6.0 & 5.9$\pm$  0.24 & 6.0  \\
		$p_{\rm 1}$ & 1.75 & 1.75$\pm$  0.054  & 1.75  \\
		$p_{\rm 2}$ & 1.75 & 1.73$\pm$ 0.061   & 1.73 \\
		$\langle p\rangle$ & 1.75 & 1.74$\pm$ 0.048  & 1.74  \\
		$FWHM$ (pix) & 27.8 & 27.9$\pm$  1.01 & 27.8  \\
		$\Delta r$  (pix) & 0.0 & -0.023$\pm$  0.19 & -0.047  \\
		\textit{a} & 0.00 & 0.00$\pm$  0.023 & 0.00  \\
		\textit{b} ($10^{-3}$) & 0.2 & 0.1$\pm$  0.2 & 0.1  \\
		$N_{\rm 0}$ ($10^{22}$\,cm$^{-2}$) & 1.0 & 1.0 $\pm$ 0.03 & 1.0  \\
		\hline		
	\end{tabular}
	\tablefoot{Same as Table \ref{tbl:errors_symmetric}, but using an asymmetric Plummer fit.   }
\end{table}

\begin{table}
	\centering
	\caption{Median SNR of the filament in each quality flag.   }
	\label{tbl:SNR_qualityFlag}
	\begin{tabular}{lllll}
		\hline
		\hline
		Field & \multicolumn{4}{c}{SNR}\\
		& \textit{q} = 1 & 2 & 3 & 4\\		
		\hline
		G17	  & 12.6  & 9.1  & 6.9  & 4.5  \\ 
		G202	  & 63.8  & 39.7  & 28.8  & 22.1  \\ 
		G208 \herschel   & 31.1  & 14.6  & 8.5  & 4.4  \\ 
		G208 \hcrop   & 122.6  & 67.2  & 51.1  & 28.4  \\ 
		G208 \feather  & 100.9  & 54.1  & 39.4  & 21.0  \\

		\hline	
	\end{tabular}
\end{table}

\begin{table*}
	\centering
	\caption{Results of the wavelet analysis in Sect.~\ref{sec:wavelets}. }
	\label{tbl:wavelet}
	\begin{tabular}{llllllllllll}
		\hline\hline
		Field & Level & Size & $n_{\rm struct}$ & $\langle N$(\MH)$\rangle$  & $\langle n$(H)$\rangle$ & $\sum M$ & $\langle s \rangle$ \\ 
		  &  & (pc) & &  ($\times 10^{22}$\,cm$^{-2}$) &  ($10^{5}$\,cm$^{-3}$) & (\msun) & (pc)   \\ 
		\hline
G017
		& 10 & 3.0  & 2 & 1.01 & 0.06 & 2192.4 & 6.32  \\ 
		& 9 & 2.0  & 3 & 1.46 & 0.17 & 1062.0 & 3.04  \\ 
		& 8 & 1.0  & 4 & 1.70 & 0.28 & 936.8 & 1.79  \\ 
		& 7 & 0.75  & 5 & 1.77 & 0.39 & 618.9 & - \\ 
		& 6 & 0.5  & 8 & 1.84 & 0.55 & 449.5 & - \\ 
		& 5 & 0.25  & 10 & 1.81 & 0.68 & 357.5 & 0.81  \\ 
		& 4 & 0.1  & 9 & 1.82 & 0.75 & 293.2 & - \\ 
		& 3 & 0.075  & 9 & 1.82 & 0.74 & 289.4 & - \\ 
G202
		& 8 & 1.0  & 1 & 0.70 & 0.00 & 409.5 & 1.82  \\ 
		& 7 & 0.75  & 1 & 0.72 & 0.00 & 399.8 & 0.89  \\ 
		& 6 & 0.5  & 3 & 0.80 & 0.15 & 226.7 & - \\ 
		& 5 & 0.25  & 5 & 1.06 & 0.31 & 149.2 & 0.33  \\ 
		& 4 & 0.1  & 3 & 1.58 & 0.55 & 147.7 & - \\ 
		& 3 & 0.075  & 7 & 1.54 & 0.94 & 60.8 & - \\ 
		& 2 & 0.05  & 9 & 1.52 & 1.14 & 63.0 & - \\ 
		& 1 & 0.025  & 9 & 1.53 & 1.16 & 63.0 & 0.30  \\ 
G208 (\herschel)
		& 8 & 1.0  & 1 & 2.53 & 0.00 & 607.2 & 1.07  \\ 
		& 7 & 0.75  & 1 & 2.37 & 0.00 & 641.4 & - \\ 
		& 6 & 0.5  & 2 & 3.27 & 0.68 & 431.7 & 0.55  \\ 
		& 5 & 0.25  & 1 & 2.63 & 0.00 & 110.6 & 0.32  \\ 
		& 4 & 0.1  & 4 & 5.14 & 2.37 & 121.8 & 0.20  \\ 
		& 3 & 0.075  & 4 & 11.03 & 8.69 & 162.1 & - \\ 
		& 2 & 0.05  & 7 & 13.66 & 19.38 & 50.7 & 0.11  \\ 
		& 1 & 0.025  & 7 & 13.52 & 19.85 & 48.5 & - \\ 
G208 (\feather)
		& 6 & 0.5  & 1 & 2.88 & 0.00 & 162.9 & - \\ 
		& 5 & 0.25  & 1 & 3.75 & 0.00 & 134.5 & 0.35  \\ 
		& 4 & 0.1  & 0 & - & - & 0.0 & 0.14  \\ 
		& 3 & 0.075  & 3 & 5.43 & 5.68 & 52.5 & - \\ 
		& 2 & 0.05  & 5 & 6.57 & 14.14 & 22.8 & 0.09  \\ 
		& 1 & 0.025  & 9 & 6.33 & 15.83 & 28.6 & 0.03  \\ 
		\hline
	\end{tabular}
	\tablefoot{ Columns 2-3 list the sizes of the structures in each category, column 4 the number of structures, columns 5-6 the mean column and volume densities, column 7 the total gas mass, and column 8 the separation between adjacent structures.   }
\end{table*}

\section{Testing the flux recovery of uvcombine   \label{sec:app_uvcombineTest}       }


To test whether feathering recovered all flux, we performed simulations of an ideal filament with several Gaussian clumps. A Plummer filament and the clumps are shown in Fig. \ref{fig:feather_test}a. The ideal map was convolved with a beam of 25\arcsec\, and  white noise amounting to 1\% of peak emission added to create the \textit{Herschel} image (Fig. \ref{fig:feather_test}b). The small-scale \artemis\, image  (Fig. \ref{fig:feather_test}c) was created by subtracting an image convolved by 2\arcmin\, from an image convolved by 8.5\arcsec, then adding white noise of 2\% of peak emission. 

The feathered result is shown in frame d, and the difference between the ideal and feathered images in frame e. The difference is under three percent of the original flux, showing that most signal is recovered by feathering. This effect worsens with decreasing SNR, but is not significant especially in \textit{q} = 1--2 filaments. 

\begin{figure}[h]
	\includegraphics[width=\linewidth]{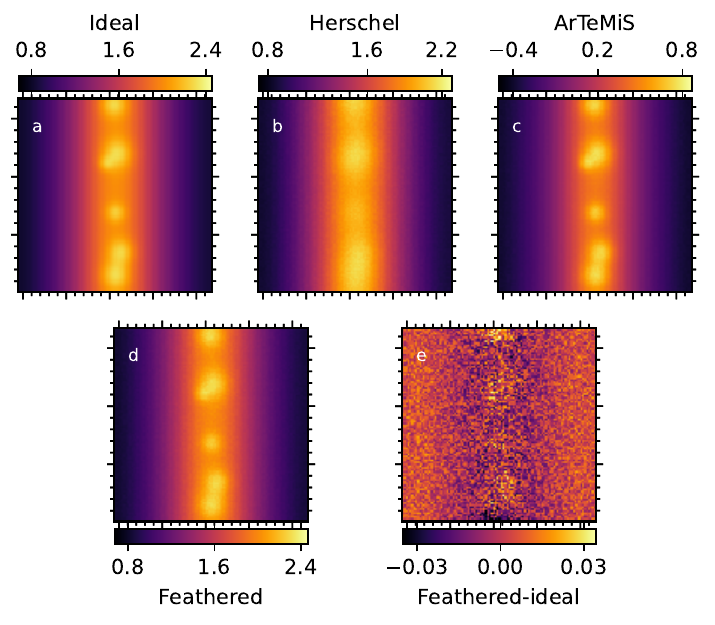}
	\caption{Test of data feathering.  (a) The simulated filament and clumps. (b) Frame a convolved with a 25\arcsec\, beam. (c) Frame a convolved to represent \artemis\, observations.  (d) Frames b and c feathered together with uvcombine. (e) Difference between the feathered and ideal images.  \label{fig:feather_test} }
\end{figure}

\section{Plummer parameters  \label{sec:app_Plummer_pars}  }
The following section presents derived Plummer parameters for the three regions studied in this paper. Tables \ref{tbl:PlummerResults}-\ref{tbl:PlummerResults_sym} list median derived Plummer parameters for asymmetric and symmetric fits. Violin plots of the same results are shown in Fig. \ref{fig:derived_violins} for asymmetric fits.

\begin{sidewaystable*}
	\centering
	\caption{Median asymmetric Plummer parameters for our data.}
	\label{tbl:PlummerResults}
	\begin{tabular}{lllllllllllllll}

				\hline\hline
		Field & Qual & $N_{\rm 0}$ & $R_{\rm flat, L}$ &  $p_{\rm L}$ & $R_{\rm flat, R}$ &  $p_{\rm R}$ & $\Delta r$ & \textit{a} & \textit{b} & \textit{FWHM} \\
		& & ($10^{22}$\,cm$^{-2}$) & (pc) &  & (pc) &  & (pc) & ($10^{22}$\,cm$^{-2}$) & ($10^{22}$\,cm$^{-2}$\,pc$^{-1}$)  & (pc)  \\
		\hline
		G017 &  1 & 1.80 $\pm$ 0.37  & 0.32 $\pm$ 1.51  & 2.24 $\pm$ 2.77  & 0.50 $\pm$ 0.42  & 3.71 $\pm$ 2.61  & -0.35 $\pm$ 0.63  & -0.00 $\pm$ 0.11  & 0.00 $\pm$ 0.03  & 0.89 $\pm$ 0.75   \\
		& 1-2 & 1.58 $\pm$ 0.48  & 0.32 $\pm$ 1.77  & 2.40 $\pm$ 3.42  & 0.50 $\pm$ 1.46  & 3.62 $\pm$ 2.58  & -0.36 $\pm$ 1.04  & -0.00 $\pm$ 0.13  & -0.00 $\pm$ 0.03  & 0.95 $\pm$ 1.94   \\
		& 1-3 & 1.34 $\pm$ 0.51  & 0.44 $\pm$ 1.68  & 3.11 $\pm$ 3.19  & 0.41 $\pm$ 1.46  & 2.88 $\pm$ 2.51  & -0.19 $\pm$ 1.02  & -0.00 $\pm$ 0.11  & -0.00 $\pm$ 0.03  & 0.99 $\pm$ 1.87   \\
		& 1-4 & 1.29 $\pm$ 0.55  & 0.43 $\pm$ 1.71  & 2.95 $\pm$ 3.24  & 0.42 $\pm$ 1.39  & 2.87 $\pm$ 2.42  & -0.13 $\pm$ 0.98  & -0.00 $\pm$ 0.11  & -0.00 $\pm$ 0.03  & 1.06 $\pm$ 1.79  \\
		& \\ 
		G202 &  1 & 1.35 $\pm$ 0.09  & 0.08 $\pm$ 0.04  & 2.34 $\pm$ 0.49  & 0.10 $\pm$ 0.08  & 1.92 $\pm$ 1.23  & 0.02 $\pm$ 0.05  & 0.02 $\pm$ 0.03  & 0.01 $\pm$ 0.04  & 0.28 $\pm$ 0.04   \\
		& 1-2 & 0.94 $\pm$ 0.22  & 0.20 $\pm$ 0.90  & 2.95 $\pm$ 2.97  & 0.08 $\pm$ 0.50  & 2.13 $\pm$ 1.93  & -0.05 $\pm$ 0.46  & 0.00 $\pm$ 0.02  & 0.00 $\pm$ 0.03  & 0.49 $\pm$ 0.45  \\
		& 1-3 & 0.84 $\pm$ 0.27  & 0.23 $\pm$ 0.80  & 2.97 $\pm$ 3.57  & 0.09 $\pm$ 0.79  & 2.18 $\pm$ 3.29  & 0.02 $\pm$ 0.49  & -0.00 $\pm$ 0.02  & 0.00 $\pm$ 0.03  & 0.61 $\pm$ 0.45  \\
		& 1-4 & 0.68 $\pm$ 0.29  & 0.30 $\pm$ 0.76  & 3.17 $\pm$ 3.82  & 0.12 $\pm$ 0.74  & 2.17 $\pm$ 2.87  & 0.03 $\pm$ 0.45  & -0.00 $\pm$ 0.02  & 0.00 $\pm$ 0.03  & 0.78 $\pm$ 0.50  \\
		& \\ 
		G208 \herschel 	&   1 & 11.90 $\pm$ 6.94  & 0.07 $\pm$ 0.11  & 4.18 $\pm$ 3.87  & 0.06 $\pm$ 0.04  & 2.79 $\pm$ 0.80  & 0.02 $\pm$ 0.03  & -0.00 $\pm$ 0.06  & -0.02 $\pm$ 0.06  & 0.11 $\pm$ 0.03   \\
		& 1-2 & 6.26 $\pm$ 6.34  & 0.06 $\pm$ 0.07  & 2.75 $\pm$ 2.80  & 0.07 $\pm$ 0.05  & 2.99 $\pm$ 1.47  & 0.02 $\pm$ 0.02  & -0.00 $\pm$ 0.04  & 0.01 $\pm$ 0.07  & 0.14 $\pm$ 0.03  \\
		& 1-3 & 5.17 $\pm$ 5.84  & 0.06 $\pm$ 0.22  & 2.75 $\pm$ 3.17  & 0.07 $\pm$ 0.06  & 3.01 $\pm$ 1.62  & 0.01 $\pm$ 0.03  & -0.00 $\pm$ 0.03  & 0.01 $\pm$ 0.10  & 0.14 $\pm$ 0.09  \\
		& 1-4 & 3.58 $\pm$ 5.49  & 0.07 $\pm$ 0.20  & 3.08 $\pm$ 3.05  & 0.07 $\pm$ 0.17  & 2.99 $\pm$ 1.77  & 0.02 $\pm$ 0.05  & -0.00 $\pm$ 0.03  & 0.01 $\pm$ 0.09  & 0.15 $\pm$ 0.11  \\
		& \\ 
		G208 \hcrop &   1 & 8.15 $\pm$ 0.09  & 0.10 $\pm$ 0.01  & 9.23 $\pm$ 1.79  & 0.05 $\pm$ 0.00  & 4.58 $\pm$ 0.79  & 0.00 $\pm$ 0.01  & 1.29 $\pm$ 0.05  & 1.89 $\pm$ 0.85  & 0.08 $\pm$ 0.00  \\
		& 1-2 & 5.00 $\pm$ 1.80  & 0.06 $\pm$ 0.02  & 5.19 $\pm$ 2.56  & 0.06 $\pm$ 0.04  & 5.01 $\pm$ 2.85  & 0.00 $\pm$ 0.01  & 0.75 $\pm$ 0.42  & 1.21 $\pm$ 0.71  & 0.08 $\pm$ 0.02   \\
		& 1-3 & 4.05 $\pm$ 1.91  & 0.07 $\pm$ 0.03  & 5.28 $\pm$ 3.27  & 0.05 $\pm$ 0.12  & 4.36 $\pm$ 3.08  & 0.00 $\pm$ 0.01  & 0.49 $\pm$ 0.40  & 0.90 $\pm$ 2.44  & 0.08 $\pm$ 0.14  \\
		& 1-4 & 3.14 $\pm$ 2.11  & 0.11 $\pm$ 0.05  & 6.95 $\pm$ 3.96  & 0.05 $\pm$ 0.12  & 3.10 $\pm$ 2.66  & -0.00 $\pm$ 0.01  & 0.26 $\pm$ 0.39  & 0.49 $\pm$ 1.94  & 0.10 $\pm$ 0.11 \\
		& \\ 
		G208 \feather  &   1 & 12.96 $\pm$ 4.31  & 0.04 $\pm$ 0.03  & 3.67 $\pm$ 2.47  & 0.02 $\pm$ 0.02  & 2.07 $\pm$ 0.86  & 0.00 $\pm$ 0.01  & -0.00 $\pm$ 0.68  & 0.00 $\pm$ 4.82  & 0.05 $\pm$ 0.02  \\
		& 1-2 & 10.36 $\pm$ 4.85  & 0.03 $\pm$ 0.03  & 2.88 $\pm$ 2.84  & 0.02 $\pm$ 0.03  & 2.09 $\pm$ 1.82  & 0.01 $\pm$ 0.01  & -0.00 $\pm$ 0.58  & 1.19 $\pm$ 4.92  & 0.05 $\pm$ 0.02  \\
		& 1-3 & 6.89 $\pm$ 4.90  & 0.04 $\pm$ 0.03  & 3.10 $\pm$ 2.61  & 0.02 $\pm$ 0.04  & 2.05 $\pm$ 2.74  & 0.00 $\pm$ 0.02  & -0.00 $\pm$ 0.50  & 0.70 $\pm$ 4.85  & 0.06 $\pm$ 0.02  \\
		& 1-4 & 6.12 $\pm$ 5.09  & 0.04 $\pm$ 0.04  & 3.10 $\pm$ 2.81  & 0.02 $\pm$ 0.04  & 2.02 $\pm$ 2.59  & 0.00 $\pm$ 0.02  & -0.00 $\pm$ 0.47  & 0.70 $\pm$ 4.51  & 0.07 $\pm$ 0.03  \\
		& \\

		\hline
	\end{tabular}
	\tablefoot{  Median Plummer parameters for asymmetric Plummer fits to \textit{Herschel} SPIRE+PACS 160\,\micro\, column density data (Image \textit{FWHM} = 20\arcsec) and the \feather\, data, only including those fits for which $\langle p_{\rm R}, p_{\rm L} \rangle \leq 10$. Errors are the 1-$\sigma$ standard deviation. Note these are calculated using the median of all derived Plummer parameters, \textit{not} the fit to the median filament profile. $R_{\rm L}$ and $p_{\rm L}$ refer to fits to the left-hand side of the filament, $R_{\rm R}$ and $p_{\rm R}$ to the right-hand side.  }

\end{sidewaystable*}

\begin{sidewaystable*}
	\centering
	\caption{Median symmetric Plummer parameters for our data.}
	\label{tbl:PlummerResults_sym}
	\begin{tabular}{lllllllllllllll}
		\hline\hline

		Field & Qual & $N_{\rm 0}$ & \multicolumn{2}{l}{$R_{\rm flat}$} &  \multicolumn{2}{l}{$p$} &  \textit{a} & \textit{b} & $\Delta r$  & \textit{FWHM} \\ 
		& & ($10^{22}$\,cm$^{-2}$) & \multicolumn{2}{l}{(pc)}  & \multicolumn{2}{l}{ }  & ($10^{22}$\,cm$^{-2}$) & ($10^{22}$\,cm$^{-2}$\,pc$^{-1}$)   & (pc)   & (pc)  \\ 
		\hline

		G017 &  1 & 1.87 $\pm$ 0.40  & \multicolumn{2}{l}{0.43 $\pm$ 1.73}  & \multicolumn{2}{l}{2.66 $\pm$ 9.35}  & 0.02 $\pm$ 0.14  & -0.01 $\pm$ 0.01  & -0.06 $\pm$ 0.70  & 0.98 $\pm$ 0.70   \\ 
		& 1-2 & 1.45 $\pm$ 0.50  & \multicolumn{2}{l}{0.48 $\pm$ 2.04}  & \multicolumn{2}{l}{3.28 $\pm$ 9.17}  & 0.04 $\pm$ 0.16  & -0.01 $\pm$ 0.01  & -0.11 $\pm$ 0.74  & 1.11 $\pm$ 1.24   \\ 
		& 1-3 & 1.19 $\pm$ 0.52  & \multicolumn{2}{l}{0.58 $\pm$ 1.94}  & \multicolumn{2}{l}{3.64 $\pm$ 9.56}  & 0.08 $\pm$ 0.14  & -0.00 $\pm$ 0.02  & -0.04 $\pm$ 0.66  & 1.11 $\pm$ 1.17   \\ 
		& 1-4 & 0.98 $\pm$ 0.57  & \multicolumn{2}{l}{1.25 $\pm$ 1.92}  & \multicolumn{2}{l}{4.64 $\pm$ 10.26}  & 0.08 $\pm$ 0.13  & -0.00 $\pm$ 0.02  & -0.02 $\pm$ 0.62  & 1.22 $\pm$ 1.08   \\

		& \\ 
		G202 &  1 & 1.55 $\pm$ 0.53  & \multicolumn{2}{l}{0.10 $\pm$ 0.47}  & \multicolumn{2}{l}{2.43 $\pm$ 10.43}  & 0.06 $\pm$ 0.03  & 0.02 $\pm$ 0.01  & 0.03 $\pm$ 0.08  & 0.29 $\pm$ 0.14   \\ 
		& 1-2 & 1.13 $\pm$ 0.58  & \multicolumn{2}{l}{0.19 $\pm$ 0.77}  & \multicolumn{2}{l}{2.70 $\pm$ 9.61}  & 0.05 $\pm$ 0.03  & 0.02 $\pm$ 0.02  & 0.03 $\pm$ 0.28  & 0.32 $\pm$ 0.34   \\ 
		& 1-3 & 0.83 $\pm$ 0.58  & \multicolumn{2}{l}{0.27 $\pm$ 0.80}  & \multicolumn{2}{l}{4.78 $\pm$ 8.82}  & 0.04 $\pm$ 0.03  & 0.02 $\pm$ 0.02  & 0.03 $\pm$ 0.31  & 0.43 $\pm$ 0.39   \\ 
		& 1-4 & 0.77 $\pm$ 0.57  & \multicolumn{2}{l}{0.28 $\pm$ 0.79}  & \multicolumn{2}{l}{3.71 $\pm$ 8.63}  & 0.04 $\pm$ 0.04  & 0.02 $\pm$ 0.03  & 0.04 $\pm$ 0.32  & 0.56 $\pm$ 0.44   \\

		& \\ 

		G208 \herschel 	&   1 & 11.81 $\pm$ 13.51  & \multicolumn{2}{l}{0.07 $\pm$ 0.07}  & \multicolumn{2}{l}{3.72 $\pm$ 9.51}  & 0.07 $\pm$ 0.12  & 0.03 $\pm$ 0.04  & 0.01 $\pm$ 0.02  & 0.11 $\pm$ 0.02   \\ 
		& 1-2 & 7.48 $\pm$ 11.45  & \multicolumn{2}{l}{0.07 $\pm$ 0.12}  & \multicolumn{2}{l}{3.45 $\pm$ 8.21}  & -0.00 $\pm$ 0.10  & 0.03 $\pm$ 0.06  & 0.00 $\pm$ 0.02  & 0.13 $\pm$ 0.05   \\ 
		& 1-3 & 5.57 $\pm$ 10.26  & \multicolumn{2}{l}{0.07 $\pm$ 0.23}  & \multicolumn{2}{l}{3.24 $\pm$ 8.17}  & -0.00 $\pm$ 0.09  & 0.03 $\pm$ 0.07  & 0.00 $\pm$ 0.03  & 0.13 $\pm$ 0.10   \\ 
		& 1-4 & 3.71 $\pm$ 9.41  & \multicolumn{2}{l}{0.07 $\pm$ 0.31}  & \multicolumn{2}{l}{3.13 $\pm$ 8.45}  & -0.00 $\pm$ 0.08  & 0.03 $\pm$ 0.07  & -0.00 $\pm$ 0.04  & 0.15 $\pm$ 0.13   \\ 
		& \\ 

		G208\hcrop &  1 & 7.90 $\pm$ 1.42  & \multicolumn{2}{l}{0.16 $\pm$ 0.03}  & \multicolumn{2}{l}{15.39 $\pm$ 5.17}  & 1.08 $\pm$ 0.50  & 2.19 $\pm$ 2.59  & -0.01 $\pm$ 0.01  & 0.10 $\pm$ 0.01   \\ 
		& 1-2 & 5.13 $\pm$ 2.17  & \multicolumn{2}{l}{0.15 $\pm$ 0.03}  & \multicolumn{2}{l}{15.58 $\pm$ 6.85}  & 0.99 $\pm$ 0.57  & 1.96 $\pm$ 2.65  & -0.01 $\pm$ 0.01  & 0.09 $\pm$ 0.01   \\ 
		& 1-3 & 3.94 $\pm$ 2.26  & \multicolumn{2}{l}{0.14 $\pm$ 0.04}  & \multicolumn{2}{l}{15.39 $\pm$ 7.76}  & 0.63 $\pm$ 0.55  & 1.79 $\pm$ 2.42  & -0.00 $\pm$ 0.02  & 0.08 $\pm$ 0.02   \\ 
		& 1-4 & 3.43 $\pm$ 2.42  & \multicolumn{2}{l}{0.14 $\pm$ 0.04}  & \multicolumn{2}{l}{11.95 $\pm$ 8.20}  & 0.52 $\pm$ 0.55  & 1.42 $\pm$ 2.22  & -0.00 $\pm$ 0.02  & 0.10 $\pm$ 0.03   \\ 
		& \\

		G208 \feather  &   1 & 12.90 $\pm$ 4.30  & \multicolumn{2}{l}{0.02 $\pm$ 0.02}  & \multicolumn{2}{l}{2.72 $\pm$ 6.60}  & 0.35 $\pm$ 0.89  & 4.62 $\pm$ 5.02  & 0.00 $\pm$ 0.01  & 0.04 $\pm$ 0.02   \\ 
		& 1-2 & 9.03 $\pm$ 4.88  & \multicolumn{2}{l}{0.02 $\pm$ 0.04}  & \multicolumn{2}{l}{2.74 $\pm$ 7.60}  & 0.82 $\pm$ 0.88  & 2.85 $\pm$ 4.64  & 0.00 $\pm$ 0.01  & 0.05 $\pm$ 0.02   \\ 
		& 1-3 & 6.64 $\pm$ 4.77  & \multicolumn{2}{l}{0.02 $\pm$ 0.04}  & \multicolumn{2}{l}{2.58 $\pm$ 7.63}  & 0.53 $\pm$ 0.83  & 2.47 $\pm$ 4.08  & 0.01 $\pm$ 0.02  & 0.06 $\pm$ 0.03   \\ 
		& 1-4 & 5.32 $\pm$ 4.87  & \multicolumn{2}{l}{0.03 $\pm$ 0.04}  & \multicolumn{2}{l}{2.50 $\pm$ 7.98}  & 0.32 $\pm$ 0.79  & 2.16 $\pm$ 3.60  & 0.00 $\pm$ 0.02  & 0.07 $\pm$ 0.04   \\ 

%
%
%

		\hline
	\end{tabular}
	\tablefoot{As in Table~\ref{tbl:PlummerResults} but for symmetric fits.   }

\end{sidewaystable*}

\begin{figure*}
	\centering
	\includegraphics[width=\textwidth]{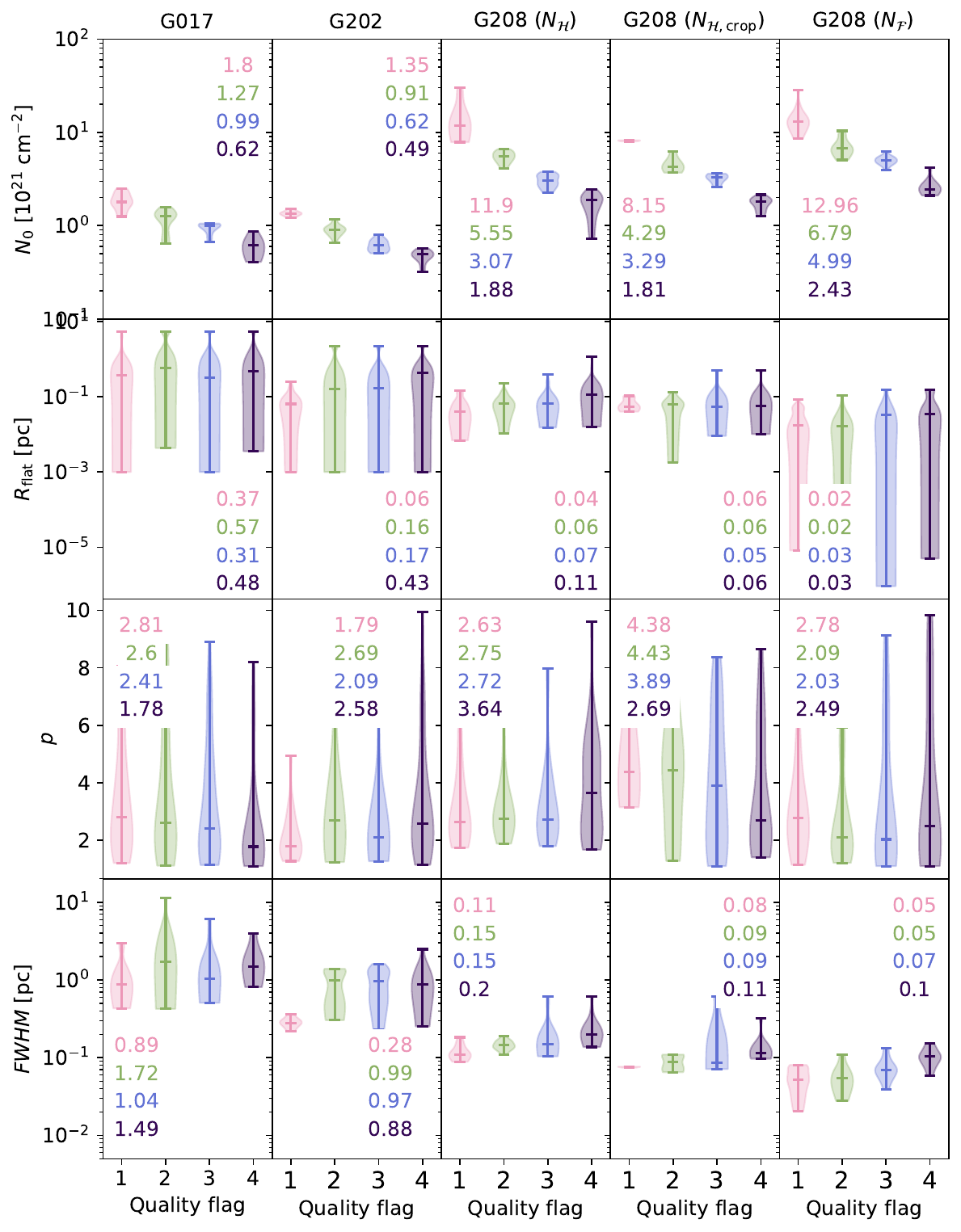}
	\caption{Violin plots of derived Plummer parameters for each quality bin for all fields used in this paper, for those fields with $p \leq 10$.   \label{fig:derived_violins}  }
\end{figure*}

\end{appendix}

\end{document}